\documentclass[5p]{elsarticle}

\biboptions{numbers,sort&compress}

\usepackage{geometry}
\usepackage{todonotes}
\usepackage[utf8]{inputenc}
\usepackage[T1]{fontenc}
\usepackage{placeins}
\usepackage{libertine}
\usepackage{libertinust1math}
\usepackage{amsmath}
\usepackage{amsfonts}
\usepackage{amssymb}
\usepackage{float}
\usepackage{eurosym}
\usepackage[normalem]{ulem}
\usepackage{booktabs}
\usepackage{tabularx}
\usepackage{multirow}
\usepackage[hyphens]{url}
\usepackage[colorlinks=true]{hyperref}
\usepackage[nameinlink,sort&compress,capitalise]{cleveref}
\usepackage{lineno}
\usepackage{pdflscape}
\usepackage{adjustbox}
\usepackage{fixltx2e}

\usepackage[final]{microtype}
\graphicspath{{graphics/}}

\DeclareGraphicsExtensions{.pdf, .png}

\newcommand{\ubar}[1]{\text{\b{$#1$}}}

\newcommand{\tablelegend}{(PF = perfect
foresight; myopic X/Y = myopic foresight with X hours of foresight and Y
hours of overlap; VOLL = perfectly inelastic demand up to value of lost load of 2000~\euro/MWh; PWL-elastic =
Piecewise-linear elastic; C$\pm$X\% = capacity perturbation of $\pm$X\%;
LT = long-term model; ST = short-term model; STD = Standard Deviation)}

\def\l{\lambda}  \def\m{\mu}  
\def\d{\partial} \def\cL{\mathcal{L}}

\journal{Energy Economics}

\begin{document}
\begin{frontmatter}

  \title{Price formation without fuel costs:\\the interaction of demand elasticity with storage bidding}

  \author[tub]{Tom~Brown\corref{cor1}}
  \ead{t.brown@tu-berlin.de}
  \author[tub]{Fabian~Neumann}
  \author[tub]{Iegor~Riepin}

  \cortext[cor1]{Corresponding author}
  \address[tub]{Department of Digital Transformation in Energy Systems, Institute
    of Energy Technology,\\
    Technische Universität Berlin, Fakultät III, Einsteinufer 25 (TA 8), 10587
    Berlin, Germany}

  \begin{abstract}
    Studies looking at electricity market designs for very high shares of wind
    and solar often conclude that the energy-only market will break down.
    Without fuel costs, it is said that there is nothing to set prices. Symptoms
    of breakdown include long phases of zero prices, scarcity prices too high to
    be politically acceptable, prices that collapse under small perturbations of
    capacities from the long-term equilibrium, cost recovery that is impossible
    due to low market values, high variability of revenue between different
    weather years, and difficulty operating long-term storage with limited
    foresight. We argue that all these problems are an artefact of modelling with
    perfectly inelastic demand. If short-term elasticity to reflect today's
    flexible demand (-5\%) is implemented in the model, these problems are
    significantly reduced. The combined interaction of demand willingness to pay
    and storage opportunity costs is enough to produce stable pricing. This
    behavior is illustrated by a model with wind, solar, batteries, and
    hydrogen-based storage, where the price duration curve is significantly
    smoothed with a piecewise linear demand curve. This removes high price
    peaks, reduces the fraction of zero-price hours from 90\% to around 30\%,
    and guarantees more price stability for perturbations of capacity and
    different weather years. Fuels derived from green hydrogen take over the
    role of fossil fuels as the backup of final resort. Furthermore, we show
    that with demand elasticity, the long-term optimisation model exactly reproduces the prices
    of the short-term model with the same capacities. We then use insights from
    the long-term model to derive simple bidding strategies for storage so that
    we can also run the short-term model with limited operational foresight. We
    demonstrate this short-term operation in a model optimised using 35 years of
    weather data and then tested on another 35 years of unseen data. We conclude that
    the energy-only market can still play a key role in coordinating dispatch
    and investment in the future.
  \end{abstract}


  \begin{keyword}
    electricity markets, price formation, capacity expansion, variable renewables, demand elasticity, storage bidding, energy-only market

    \JEL Q400 \sep Q410 \sep Q420 \sep C610 \sep D410 \sep D470
  \end{keyword}

\end{frontmatter}

\section{Introduction}

\subsection{Problem statement}

In electricity markets dominated by fossil fuel generators, prices are set in
most hours by the variable costs of the power plant on the margin. In hours of
scarcity, when all generation is at full capacity, prices are determined by the
demand's willingness to pay.

As more zero-marginal-cost generation from wind and solar enters the market, the
supply curve shifts to the right, which reduces the market price, sometimes to
zero or even negative values \cite{sensfuss2008}.  This so-called \textit{merit
  order effect} puts pressure on the revenues of conventional generators as well
as on variable renewables. If conventional generators leave the market, scarcity
pricing could become more pervasive.

There is a widespread concern in the literature that if shares of wind and solar
rise even higher, providing more than 80\% of yearly electricity, prices might
become so singular that the energy-only market might no longer function in a
meaningful way \cite{zhou2025}. Taylor et al.~\cite{taylor2016} express the concern `because
there are no fuel costs, employing nodal pricing as is would simply result in
all prices being equal to zero'. Some simulations indicate that prices are not
always zero, but zero most of the time, with very high prices in other hours.
Mallapragada et al.~\cite{mallapragada2023} find that in `VRE-dominant
energy-only wholesale power markets \dots generators and storage facilities
would earn the bulk of their annual energy market revenues in relatively few
hours'. Similarly, Levin et al.~\cite{Levin2023} raise the concern that `\dots a
large fraction of the value of [energy storage] may be realised during a
relatively small number of periods'. A study for the Royal Society~\cite{royalsociety2023}
suggests that `if paid only on the basis of short-run costs, the large-scale
long-term storage \dots could never recover its capital costs' and that
`traditional spot markets, which were developed to deal with gas and coal
powered generation, are not automatically \dots adaptable to \dots wind, solar,
and storage.'

Additional concerns include an expectation of wide swings in prices between
different years reflecting changing weather conditions \cite{green2010,brown2020,thomassen2022}, raising
risks around cost recovery \cite{hirth2013,mays2022,egli2020,barroso2021}, and concerns
about how storage dispatch and prices behave when transitioning from long-term
capacity expansion models with perfect foresight to short-term operational
models with myopic foresight \cite{bistline2020,guerra2024}. The apparent
challenge to secure sufficient revenue for assets needed for system reliability
\cite{frew2016} has led to discussions about alternative market designs oriented
towards capacity rather than energy provision
\cite{cramton2013,hogan2017,wolak2021,zhou2025}.

\subsection{Other literature}

Other modelling studies incorporating newly-electrified sectors and demand
flexibility in systems with high renewable shares have seen different behavior
\cite{boettger2021,haertel2021,helisto2023,neumann2023}. These studies see more
moderate price duration curves with fewer zero and scarcity prices in capacity
expansion models that incorporated flexible demands from heating, transport, and
coupling to hydrogen production, which frequently becomes price-setting. Similar
effects have been observed for systems with storage \cite{aaslid2021}. Even
without the dominance of renewables, it has been noted that a small amount of
demand elasticity can improve market operation, screening away price spikes
\cite{stoft2002,kirschen2003,chua2009}. Others point out that systems without
fuel costs might not undermine the working principles of energy-only markets if
taking into account storage operation, scarcity prices and demand response
\cite{leslie2020}. 

In power systems dominated by hydroelectricity, which also has near-zero
marginal costs, it has been understood for decades how the opportunity costs for
dispatch give the storage medium water a value
\cite{masse1946,bellman1957,little1955,koopmans1957,hveding1968,lederer1984overall,Rotting1992,fosso1999},
thereby determining the bidding behavior of hydroelectric generators. These
insights have been recently transferred to the dispatch of long-duration energy
storage (like hydrogen storage or thermal energy storage in district heating),
where marginal storage values have been used in operational dispatch settings to
develop bidding rules \cite{crampes2019} and deal with myopic foresight
\cite{dupre2023}. Bidding strategies for storage have also been investigated by
Ward et al.~\cite{ward2018,ward2019} without optimisation methods, looking at
how price variability affects the arbitrage earnings of storage.

A series of papers \cite{korpas2020,tarel2024} examines the price structure in a
stylised setup with renewables and storage only and find that prices set by
storage play an important role in the cost recovery of all assets. However, the
analysis is simplified to a residual load duration curve setup without
addressing intertemporal storage dynamics or storage capacity constraints.
Furthermore, the authors assume perfectly inelastic demand and only one type of renewable
generation and storage technology.

Antweiler and Müsgens~\cite{antweiler2024} explore price formation and market equilibria using analytical and empirical models with wind, solar, and storage technologies and demand response. The authors show that energy-only markets remain functional even when the system only includes generators with near-zero marginal costs, as long as free entry and competition ensure effective price setting. However, the analysis is limited to a stochastic state space covering only one hour, so that the intertemporal dynamics of storage operation is not modeled explicitly.

Ekholm and Virasjoki~\cite{ekholm2021} analyse pricing mechanisms in fully
renewable systems with demand elasticity (-25\%) under perfect and
Cournot-type imperfect competition and identify the key role of storage
operation and demand elasticity for price setting. They consider a short-term
operational setting with weekly time frames without capacity expansion,
multi-decadal time series, or myopic foresight.

De Jonghe et al.~\cite{dejonghe2012} describe different approaches to
incorporate demand elasticity into capacity expansion models with perfect
foresight, including cross-price elasticities between different hours. In a model with high
conventional capacity and no storage, they find that demand elasticity increases
the share of wind and reduces peak capacity required as demand response clears
the market in those peak hours. However, their analysis does not focus on the
price formation mechanisms.

Adachi et al.~\cite{adachi2024} include price-responsive demand in a
model with renewables and storage, and find that `stable average
prices can be maintained throughout the transition to 100\% VRE'.
However, they did not consider multiple weather years, long-duration
energy storage, or analyse the resulting price duration curves.

\subsection{Main thesis}

In this paper, we provide analytic insights and modelling results to provide a
unifying framework for understanding price formation at high penetrations of
variable renewables like wind and solar. We show how storage bidding and demand
flexibility combine to stabilise price formation. We use a model complex enough
to illustrate the salient points, yet simple enough to trace cause and effect.

Our novelty lies in comparing price formation in systems with high shares of
variable zero-marginal cost generation under different demand modelling
paradigms, various levels of operational foresight, and across an order of
magnitude more weather conditions than used previously in the literature. This
allows us to disentangle the contradictory results from the literature and
explain them as particular corners of the solution space. In addition, we use
state-of-the-art recent measurements of demand elasticity
\cite{hirth2024,arnold2023} to enhance the realism of our simulations.

Our main contention is that much of the singular price behavior observed in the
literature stems from the assumption of perfectly inelastic demand used in capacity
expansion models for electricity. If more realistic demand behavior is used,
with, for example, the -5\% short-term price elasticity of demand at the average
price that was recently observed from today's industrial consumers in Germany
\cite{hirth2024,arnold2023}, then many artefacts resulting from assuming
perfectly inelastic demand disappear. Prices no longer jump between zero and
scarcity; they remain stable over many weather years and do not collapse with
small capacity perturbations. This explains the smoother price results observed
in sector-coupling studies that include demand flexibility (although this is often modelled with virtual storage units rather than explicit demand curves) \cite{boettger2021,haertel2021,neumann2023}.
Furthermore, demand elasticity also smooths the marginal storage values for
long-duration energy storage, which allows us to deduce heuristics for storage
bidding based on these marginal values and dispatch the systems with myopic
foresight similar to \cite{dupre2023}.

In addition to developing an understanding of these features theoretically
through the interaction of shadow prices in the optimisation model, we also
provide hourly modelling simulations for three European countries featuring
different renewable potentials (Germany, Spain, and the United Kingdom) with
over 70 years of weather data. The present study is the first study we are aware
of that considers demand elasticity in such capacity expansion models with
multi-decadal time series. Solving such models is enabled by innovations in
interior-point solvers for such convex problems \cite{gurobi}. We illustrate
price formation for various demand types and examine which demand modelling
approaches cause the price formation mechanisms to diverge or align between
long-term models with capacity expansion and short-term models with only
dispatch variables. We then show how prices behave under capacity perturbations
from the long-term equilibrium, how well simple storage bidding heuristics can
reproduce the price formation with realistic myopic operational foresight, and
how this affects cost recovery.

\section{Theory}
\label{sec:theory}

\subsection{Introduction}

\begin{figure}[!t]
  \centering
  \includegraphics[trim=0 0cm 0 0cm,width=\linewidth,clip=true]{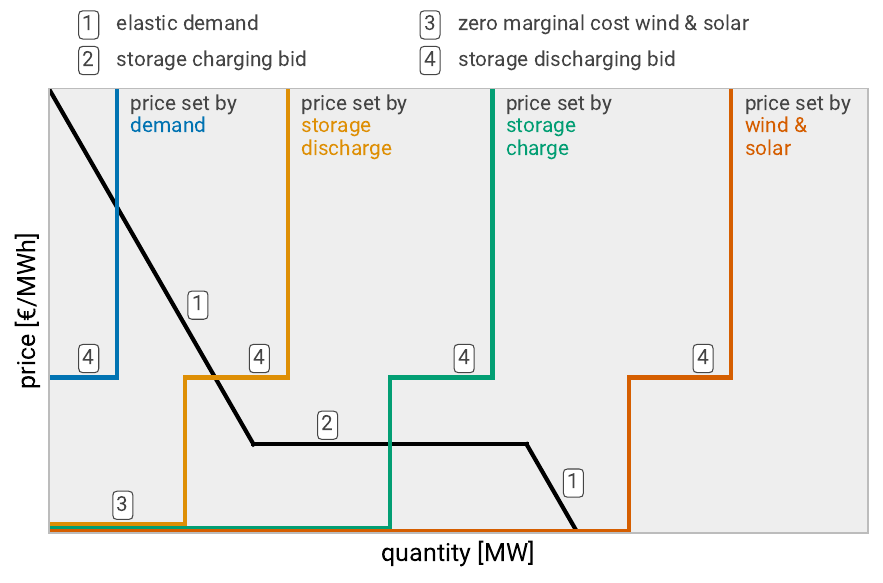}
  \caption{Example market situations with demand curve ((1) and (2)) and supply curves ((3) and (4)) for four different VRE feed-in situations. Depending on the VRE availability, either the demand (1), storage (2) and (4) or wind and solar production (3) sets the price.}
  \label{fig:price_setting}
\end{figure}

We first provide some intuition by examining hourly supply and demand curves for
a system with zero-marginal-cost variable renewable energy (VRE), a single
infinite\footnote{Scarcity and cost in energy storage capacity would introduce
  multiple price levels for charging and discharging bids over time
  (\cref{fig:si:example}).} storage technology and demand elasticity.
\cref{fig:price_setting} shows how the price setting works with rising availability of VRE.

In the aggregated demand curve, part of the demand curve (1) is determined by
the regular electricity demand, which is assumed to have elasticity with a low
magnitude (e.g. below -5\% at the average price). The other part is the willingness to
pay for the storage charging (2). The storage willingness to pay is non-zero
because the storage medium has a value. For hydrogen storage, this would
correspond to the option to sell the hydrogen for a non-zero price.

The supply curve has two price levels: zero for VRE (3) and non-zero for the
storage discharging (4). The storage bids with a non-zero price into the market
because there is always an opportunity cost of saving energy to dispatch later
in higher-price hours. This translates to a marginal storage value (MSV) of the
storage medium. This MSV is also called the \textit{water value} or \textit{Bellman
  value} in the water resource management and hydroelectricity literature
\cite{bellman1957,little1955,koopmans1957}. For a storage medium like hydrogen,
it would correspond to the market price of the hydrogen.

As a result of these bidding curves, the electricity price is only zero in the
case that the VRE availability exceeds both the peak of the regular demand and
the storage charging capacity. When demand exceeds VRE feed-in but not the
available storage discharging capacity, then storage discharging sets the price
based on the MSV. When VRE feed-in exceeds demand but not the storage charging
capacity, then storage charging is price-setting. Finally, when demand exceeds
VRE and storage, prices are set by the willingness to pay for the regular
demand, resulting in some scarcity pricing.

This scarcity pricing is important because it helps to set the opportunity costs
of the storage, which can hold back energy for these high-priced hours. Because
the demand curve is sloped, the storage constantly makes trade-offs for how
long to hold its energy, store it, or discharge it. The opportunity costs set
the MSV, which in turn determines the electricity prices in most hours through
the storage bidding.

\subsection{Mathematical theory}

This section provides a mathematical representation of the price formation by
mapping the prices to dual variables (or \textit{shadow prices}) of an
optimisation problem. We first start with the long-term (LT) capacity expansion
problem, in which the capacities of all assets are co-optimised with their dispatch.
In the next section we look at the consequences of moving to a short-term (ST)
model, where we fix all the capacities and optimise only the dispatch.

The long-run welfare maximisation runs over electricity consumers $c$,
generators $r$, and storage units $s$, each of which can have different costs and
efficiencies. The yearly welfare is represented by hourly weather and load conditions $t$.
The optimisation variables are the dispatch for each period
$t$  of demand $d_{c,t}$, generation $g_{r,t}$, storage discharging $f_{s,t}$,
charging $h_{s,t}$ and state of charge $e_{s,t}$ as well as capacities for each
generator $G_r$, storage discharge $F_s$, charging $H_s$ and energy $E_s$.
\begin{align}
  \max_{d_{c,t}, g_{r,t}, G_r,f_{s,t},F_s,h_{s,t},H_s,e_{s,t}, E_s} & \left[\sum_{c,t} U_{c,t}(d_{c,t}) - \sum_{r}c_r G_r -  \sum_{r,t} o_{r} g_{r,t}\right. \nonumber \\
                                                                    & \left. - \sum_s \left( c^f_s F_s + c^h_s H_s + c^e_s E_s \right)\right] \label{eq:objective}
\end{align}
Specific annual fixed costs $c_*$, made up of annualised investment costs and fixed operation and maintenance costs, are applied for all assets, while
linearised variable costs $o_r$ are given only for generators. We assume
non-linear utility for the demand and linear bidding for generation, assuming
constant short-term marginal generation costs.

The constraints are split into hourly energy conservation for electricity and each
storage medium $s$, and capacity constraints:
\begin{align}
  \sum_c d_{c,t} + \sum_s h_{s,t} - \sum_r g_{r,t} - \sum_s f_{s,t}   =  0 & \hspace{.4cm}\leftrightarrow\hspace{.4cm} \l_t \in \mathbb{R} \nonumber                   \\
  e_{s,t} -   e_{s,t-1} - \eta_s^h h_{s,t} + ( \eta_s^f )^{-1} f_{s,t} = 0 & \hspace{.4cm}\leftrightarrow\hspace{.4cm} \l^s_t \in \mathbb{R} \nonumber                 \\
  - d_{c,t}   \leq  0                                                      & \hspace{.4cm}\leftrightarrow\hspace{.4cm} \ubar{\m}^d_{c,t} \geq 0 \nonumber              \\
  d_{c,t} - D_{c,t} \leq  0                                                & \hspace{.4cm}\leftrightarrow\hspace{.4cm} \bar{\m}^d_{c,t} \geq 0 \nonumber               \\
  - g_{r,t}   \leq  0                                                      & \hspace{.4cm}\leftrightarrow\hspace{.4cm} \ubar{\m}^g_{r,t} \geq 0  \nonumber             \\
  g_{r,t} - G_{r,t} G_r  \leq 0                                            & \hspace{.4cm}\leftrightarrow\hspace{.4cm} \bar{\m}^g_{r,t} \geq 0 \nonumber               \\
  - f_{s,t}   \leq  0                                                      & \hspace{.4cm}\leftrightarrow\hspace{.4cm} \ubar{\m}^f_{s,t} \geq 0  \nonumber             \\
  f_{s,t} - F_s  \leq 0                                                    & \hspace{.4cm}\leftrightarrow\hspace{.4cm} \bar{\m}^f_{s,t} \geq 0 \nonumber               \\
  - h_{s,t}   \leq  0                                                      & \hspace{.4cm}\leftrightarrow\hspace{.4cm} \ubar{\m}^h_{s,t}  \geq 0 \nonumber             \\
  h_{s,t} - H_s  \leq 0                                                    & \hspace{.4cm}\leftrightarrow\hspace{.4cm} \bar{\m}^h_{s,t} \geq 0 \nonumber               \\
  - e_{s,t}   \leq  0                                                      & \hspace{.4cm}\leftrightarrow\hspace{.4cm} \ubar{\m}^e_{s,t}  \geq 0 \nonumber             \\
  e_{s,t} - E_s  \leq 0                                                    & \hspace{.4cm}\leftrightarrow\hspace{.4cm} \bar{\m}^e_{s,t} \geq 0  \label{eq:constraints}
\end{align}
$D_{c,t}$ is the constant upper limit on the demand, while $G_{r,t}$ is the
time-dependent availability factor for each generator. The electricity price
corresponds to the shadow price of the energy balance constraint $\l_t$, while
the MSV for each storage unit is given by $\l^s_t$ of
the storage balance constraint.


The price formation can be deduced from the stationarity equations of the
Karush-Kuhn-Tucker (KKT) conditions.\footnote{The KKT conditions are the
  first-order conditions necessary for an optimal solution. When the objective
  function to maximise is concave, and the constraints are affine, these conditions
  are also sufficient for optimality \cite{boyd2004}.} For the generation, we have
\begin{equation}
  0 =  \frac{\d \cL}{\d g_{r,t}} = \l_t - o_r + \ubar{\m}^g_{r,t} - \bar{\m}^g_{r,t} \quad \forall r,t.  \label{eq:gstat}
\end{equation}
The generator's variable cost $o_r$ and available capacity determine its bid
into the market. When a generator $r$ is on the margin, so that neither capacity
limit is binding, $\bar{\m}^g_{r,t} = \ubar{\m}^g_{r,t}  = 0$, the generator's
variable cost is price-setting $\l_t = o_r$. This corresponds to part (3) of the
supply curve in \cref{fig:price_setting}.

If we now consider the storage dispatch,
\begin{equation}
  0 =  \frac{\d \cL}{\d f_{s,t}} =  \l_t - ( \eta_s^f )^{-1} \l^s_t + \ubar{\m}^f_{s,t} - \bar{\m}^f_{s,t} \quad \forall s,t, \label{eq:fstat}
\end{equation}
it has the same form as a generator bidding with variable cost $( \eta_s^f
  )^{-1} \l^s_t$, similar to a generator with a fuel cost of the value of the
storage medium $\l^s_t$ with conversion efficiency $ \eta_s^f $. The term $(
  \eta_s^f )^{-1} \l^s_t$ corresponds to the \textit{effective bid} of the storage
dispatch variable $f_{s,t}$ obtained from Lagrangian relaxation
\cite{brown2021}. If it is on the margin, it is price setting. We see here how
the MSV $\l^s_t$ sets the prices like fuel costs. This
corresponds to part (4) of the supply curve in \cref{fig:price_setting}.

On the demand side, each demand satisfies
\begin{equation}
  0 =  \frac{\d \cL}{\d d_{c,t}} = U'_{c,t} (d_{c,t}) - \l_t +  \ubar{\m}^d_{c,t} - \bar{\m}^d_{c,t} \quad \forall c,t,  \label{eq:kktdemand}
\end{equation}
where the derivative of the utility determines its demand curve. This
corresponds to part (1) of the demand curve in \cref{fig:price_setting},
where the demand is price setting.

If we now consider the storage charging,
\begin{equation}
  0 =  \frac{\d \cL}{\d h_{s,t}} =  \eta_s^h \l^s_t  - \l_t + \ubar{\m}^h_{s,t} - \bar{\m}^h_{s,t} \quad \forall s,t,  \label{eq:hstat}
\end{equation}
it has the same form as a demand bidding with a willingness to pay $\eta_s^h   \l^s_t$,
i.e.~it needs an electricity price low enough that it can sell storage
medium at its going price $\l^s_t$ considering its conversion efficiency
$\eta_s^h$.  This corresponds to part (2) of the demand curve in
\cref{fig:price_setting}.

From this, we see how relevant the marginal storage values $\lambda^s_t$ are in
setting prices, since parts (2) and (4) both depend on them.

\subsection{Conditions for matching prices in long-term and short-term
  models}\label{sec:proofpricessame}

In the long-term model (LT), storage and generation capacities are endogenous
variables of the optimisation problem. In this case, fixed costs play a
role in determining capacity investments that satisfy demand in a way that
maximises social welfare. In the short-term model (ST), the capacities
$G_r,F_s,H_s,E_s$ are fixed from the beginning and not part of the optimisation.

It is not obvious that there is a relationship between the hourly
prices in the LT and ST optimisation models. Economic theory suggests
they should converge, since if the ST model deviates from the LT
equilibrium, asset owners should notice that they are making either
losses or excess profits and then (in the long-term) exit or enter the market
respectively. However, from the perspective of the optimisation
models, they have different sets of variables and objective functions.
The ST model doesn't see the fixed costs of the assets at all, and the
ST objective function only involves dispatch variables:
\begin{align}
  \max_{d_{c,t}, g_{r,t}, f_{s,t},h_{s,t},e_{s,t}} & \left[\sum_{c,t} U_{c,t}(d_{c,t}) -  \sum_{r,t} o_{r} g_{r,t}\right].
\end{align}
If all variable costs are zero ($o_r = 0$), only the demand utility has non-zero
coefficients in the objective function.

By taking the optimal capacities $G_r,F_s,H_s,E_s$ from the long-term model and
fixing them as constants in the short-term model, we can compare the resulting
electricity prices $\l_t$. Economic theory says the ST and LT should have the same prices,
but we will show that in the optimisation models, this only occurs under certain
conditions. As a result, prices from LT optimisation models should be treated with care.

The key point is that in the presence of storage the fixed costs from
the LT objective, i.e. the terms multiplied with $c_*$ in
\eqref{eq:objective}, can mix themselves into the prices. This causes
price deviations versus the ST model, where these fixed cost terms are
not present. Without storage, the fixed costs are usually screened away
in the screening curve analysis by a combination of the variable costs
of the generators $o_*$ and the cost for load shedding in the form of
the value of lost load (VOLL) \cite{biggar2014}.  With storage, the fixed costs
for the storage play a critical role in setting the marginal storage
values (MSV) $\l^s_t$, since the MSV enable the storage to recover their fixed costs
from price arbitrage. The MSV then leach into the market prices
whenever the storage is price setting.\footnote{On a mathematical level, the fixed costs interact with the prices via the $\bar{\m}$ variables in equations \eqref{eq:gstat}-\eqref{eq:hstat}, which are then linked to the fixed costs via the stationarity equations for the capacity variables in the LT optimisation problem, e.g. for storage energy capacity $0 =  \frac{\d \cL}{\d E_{s}} = -c^e_s + \sum_t \bar{\m}^e_{s,t}$ \cite{brown2021}.}

We will now demonstrate that, in the presence of storage, the LT and ST model prices are identical as
long as the aggregated inverse demand curve (aggregate of $U'_{c,t}$)
provides a unique mapping from the demand to the price through a
strictly monotonically decreasing demand curve. If there are
steps\footnote{Steps would represent segments of perfectly elastic
demand, e.g.~load shedding for a fixed value of lost load (VOLL).} in
the shape of the demand curve, the prices in the LT and ST model can
diverge.

Suppose that there are no constraints on $d_{c,t}$, so that $ \ubar{\m}^d_{c,t}
  = \bar{\m}^d_{c,t} = 0$ for all $c$ and $t$. Furthermore, suppose that the
generation and load conditions in each hour are different so that there is a
unique long-term solution for the demands $d_{c,t}$, i.e.~there are no
degeneracies between hours in the objective, so that load cannot be moved from
one hour to another without a cost impact.

It follows that the optimal demands $d_{c,t}$ and generation and storage
dispatch in the LT model must be identical to those in the ST model. This is
because generation capacities are the same in both models, and any improvement
to the objective function by altering dispatch variables in one model would also
benefit the other.

If there is a unique mapping from the demand curve to prices from stationarity
for $d_{c,t}$ (cf.~\cref{eq:kktdemand}), and the demand curves are identical,
then it follows that the prices must also be identical in both models in all
hours. In other words, if there is a smooth elastic demand curve with a
different price for each demand, prices will be identical. This holds even for
very low elasticity values.

However, if the demand curve is a step function, such as with a VOLL at
2000~\euro/MWh, then prices are only identical during load shedding or excess
supply hours in the absence of other generators with non-zero marginal cost. In
the hours with load shedding, prices reach 2000~\euro/MWh in both the ST and LT
models. In hours with excess supply beyond the storage charging capacity, prices
are zero in both models. The constant demand for varying prices during
non-shedding periods causes a demand-side degeneracy, which means the
supply-side sets the price unless storage charging does. As fixed costs can
set the price in the LT model but not in the ST model, we no longer have
identical prices.

Symmetrically, if the supply curve were strictly monotonically increasing, e.g., with a variety
of generators with different efficiencies and fuel costs, rather than in linear
steps, this would also be sufficient to resolve the prices and guarantee
identical prices between ST and LT models. Some structure is needed either on
the supply or the demand to eliminate the degeneracy of prices.

\section{Methods for simulations}

\subsection{Implementation details}

To demonstrate how price formation works, modelling is done for three European
countries (Germany, Spain, and the United Kingdom) using 70 years of historical
hourly weather data from 1951 to 2020 \cite{bloomfield2022}. We use 70 years of
data for different countries to capture a wide variety of weather conditions,
rather than illustrating a 70-year pathway of system transformation. Systems are
modeled with only a limited selection of technologies: onshore wind,
utility-scale solar photovoltaics, batteries for short-term storage, and a
hydrogen-based chemical storage medium for long-term storage (e.g.~underground
hydrogen storage). The techno-economic parameters, such as investment costs and
efficiencies, are provided in \cref{tab:technology-data}. The figures in the
main body show results for Germany, while the results for Spain and the United
Kingdom are provided in \ref{sec:si:countries}.

The modelling framework used is Python for Power System Analysis (PyPSA)
\cite{PyPSA}, and the open source code to reproduce our experiments is available
on GitHub;\footnote{\url{https://github.com/fneum/price-formation} (v0.2.0)}
output data were deposited on
Zenodo.\footnote{\url{https://doi.org/10.5281/zenodo.12759247} (v0.2.0)} The
experiments were run on a high-performance computing cluster with PyPSA v0.27.1
\cite{PyPSA}, linopy v0.3.8 \cite{linopy}, and Gurobi v11.0.2 \cite{gurobi}. For
the resulting linear (LP) and quadratic (QP) problems, we used the interior
point method without crossover, a barrier convergence tolerance of $10^{-7}$,
and 32 threads per model. The maximum memory requirement per scenario was 22 GB.

In this setup, we then vary the number of weather years in the simulation, the
choice of the demand curve (\cref{sec:demand-method}), long-term optimisation
with capacity expansion versus short-term optimisation with fixed capacities,
capacity perturbations of $\pm5\%$ around the long-term equilibrium, and myopic
versus perfect foresight dispatch (\cref{sec:myopic-method}).

\subsection{Demand modelling}
\label{sec:demand-method}

\begin{figure}[!t]
  \centering
  \includegraphics[trim=0 0cm 0 0cm,width=\linewidth,clip=true]{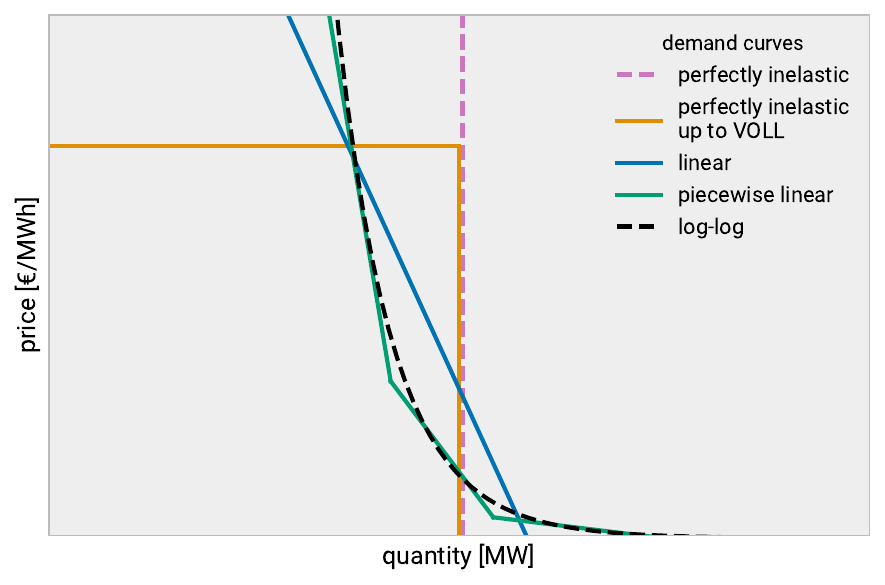}
  \caption{Illustration of different demand curves considered in the scenarios.}
  \label{fig:demand_curves}
\end{figure}

A single demand curve is modelled to represent all aggregated consumers.
The demand curve is varied between a perfectly inelastic demand, an inelastic
step function up to VOLL, a linear demand curve, and a piecewise-linear (PWL)
approximation of a log-log demand curve, as shown in
\cref{fig:demand_curves}. For most of the modelling analysis we focus on the
case of inelastic demand up to VOLL, since this is the common modelling approach and also
reflects the price caps enforced in many markets, and the PWL log-log elasticity case, since
this was empirically measured in the German market \cite{arnold2023}.
Non-linear curves are not considered because solving
becomes inefficient once we go beyond quadratic terms in the objective function.
The same demand curve is used for each hour to isolate effects of supply from
demand. The level of demand is chosen to represent a token average of 100~MW in
each country to simplify case comparisons.

The simplest case of perfectly inelastic demand is modelled by removing the utility from the
objective function, since it is effectively infinite, and replacing the demand variable by
a demand constant.

The next simplest case is that of demand that is completely inelastic up to a given value of
lost load (VOLL). For a given demand $0 \leq d \leq D$ with value of lost
load $V$, the utility function is
\begin{equation}
  U(d) = Vd
\end{equation}
The demand curve is a step function. It is perfectly inelastic up to a price of
$p = U'(d) = V$, at which point it is perfectly elastic. Implementing
this demand function in the optimisation problem
\cref{eq:objective,eq:constraints} results in a linear problem (LP). For the modelling
we assume a VOLL of $V=2000$~\euro/MWh and a peak load of $D=100$~MW.

For a linear demand curve, the utility is quadratic:
\begin{equation}
  U(d) = a d - \frac{b}{2} d^2
\end{equation}
giving a linear demand curve
\begin{equation}
  p = U'(d) = a - b d
\end{equation}
This results in a quadratic problem (QP), similar to one of the approaches in De
Jonghe et al.~\cite{dejonghe2012}. For a linear demand curve, the elasticity
varies along it, becoming more elastic at higher prices. Using parameters $a =
2000$ and $b=20$ results in an elasticity of $-5\%$ near the approximate
average system price $p = 100$~\euro/MWh.

To approximate a demand curve of log-log form $\ln(p) = a - b \ln(d)$ with
constant price elasticity, we use an aggregated demand curve given by several
demands $d\in\left[0,D\right]$ with different constants ($a, b$), so that we get
a piecewise-linear curve with different slopes. In \cref{fig:demand_curves}, an
example approximation is shown with three linear segments. Here, we use three
segments with parameters $a = (8000, 400, 200)^\top$, $b = (80, 40, 20)^\top$,
and $D = (95, 5, 10)^\top$, which results in an elasticity of -5\% at a price of 100~\euro{}/MWh and a demand
around 100~MW. The elasticity of $-5\%$ reflects the empirical analysis
of the German market made in \cite{arnold2023}.
For sensitivity cases with higher and
lower elasticity at the same price levels, coefficients $a$ and $b$ are halved
or doubled which results in elasticities of $-10\%$ and $-2.5\%$ respectively at
a price of 100~\euro{}/MWh.

In our main scenarios, we assume for simplicity that there are no inter-temporal dependencies
for the demand elasticity, i.e. the demand level $d_t$ at time $t$
only depends on the price $p_t$. This is consistent with how the empirical
measurements of German elasticity were made, since the inter-temporal effects
were already present in the data \cite{arnold2023}.
In reality there may be complex dependencies
on the hours before or after. In particular, some loads may not be able
to reduce their demand for multiple consecutive hours, or they may
have to increase the demand later to recover the lost demand. This can
be represented in the model with cross-price elasticities between
different hours \cite{dejonghe2012}, which adds bilinear terms between
the hourly demands in the objective and does not affect our
theoretical analysis, or by representing load shifts using a virtual
storage for the demand, which is a common approach for e.g. flexible
electric vehicle charging or heat pump operation \cite{Brown2018}.
In \ref{sec:cross-elasticity}, we show in a sensitivity analysis that
cross-elastic terms have only a limited effect on our results.

To enhance numerical stability, we substitute demand variables with
load-shedding generators as described in \ref{sec:numerics}, yielding identical
results. Notably, the QPs with demand elasticity were observed to solve much
faster than the LPs with perfectly inelastic demand up to the VOLL (6-16 times; 8-13~min vs.~72-144~min).

\subsection{Myopic dispatch}
\label{sec:myopic-method}

Assuming perfect foresight over multiple years, where future periods
of scarcity and abundance are fully anticipated, is unrealistic.
While forecasts may be reliable 1-3 days ahead, and hydroelectric
planners are accustomed to making seasonal forecasts for water levels
in their dams, the multi-week time scales on which wind power vary are
harder to anticipate in advance \cite{bauer2015}. 
In particular this creates challenges for the dispatch of long-duration storage so that it charges during days of excess in order to discharge during periods of scarcity outside the myopic foresight horizon.

To evaluate the impact of myopic foresight on price formation for different
demand modelling cases, we initially derive optimised capacities from an LT model
optimised using 35 randomly selected years from the 70-year dataset (\ref{sec:year-selection}). Subsequently, we
run the dispatch in an ST model using the remaining 35 years, comparing cases
with perfect foresight with those with myopic foresight. We also apply
perturbations of $\pm5\%$ to all generation, conversion and storage capacities,
to show the impacts of imperfectly anticipating the system's needs.

For cases with myopic foresight, we apply a rolling-horizon dispatch
optimisation with a 96-hour look-ahead horizon and a 48-hour overlap between windows.
This approach aligns with the horizon of current forecast skill and provides sufficient
information for the short-term battery operation \cite{bauer2015}. Since
there are already good weather forecasts up to 5 days ahead \cite{bauer2015,magnusson2022},
this represents a conservative worst case for the system dispatcher.

For the chemical storage, we dispatch assuming a constant fuel value equal to
the mean of $\lambda^s_t$ from the LT model. For instance, if the hydrogen has
an average value $\langle \lambda^s_t \rangle = 100$~\euro/MWh, the electrolysis
with efficiency 70\% bids 70~\euro/MWh for consuming electricity and the
hydrogen turbine with efficiency 50\% offers it at 200~\euro/MWh. By
implementing these approximate bids as marginal costs on the storage, we use the
average MSV as an orientation to operate long-term storage reasonably without
foresight.\footnote{Knowing the marginal storage value $\lambda^s_t$ of the ST
  model in advance would be sufficient to reproduce the optimal dispatch with
  perfect foresight, as illustrated by the effective bids of Lagrangian relaxation
  \cite{brown2021}.}

\begin{figure*}[!t]
  \centering
  \includegraphics[trim=0 0cm 0 0cm,width=0.92\linewidth,clip=true]{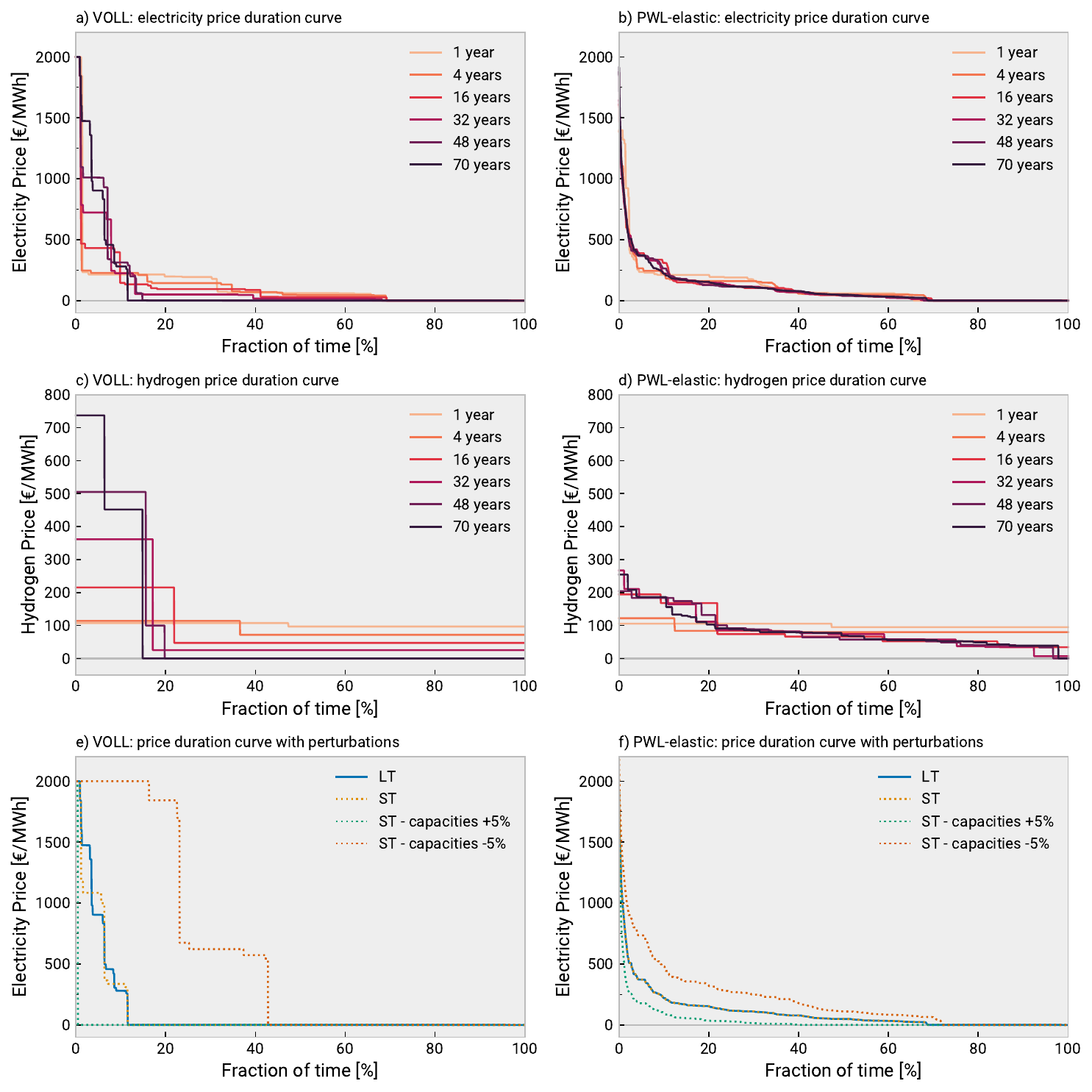}
  \caption{Different price duration curves: Panels a) and b) show electricity price duration curves as the number of years used in the optimisation rises from 1 to 70. Panels c) and d) show the corresponding hydrogen price duration curves for the same simulations. Panels e) and f) show for the 70-year optimisation how the prices change when going from the long-term (LT) to short-term (ST) model with the LT capacities, and then perturbing the asset capacities in the ST model up and down by 5\%.}
  \label{fig:price_duration}
\end{figure*}

However, although $\lambda^s_t$ becomes relatively stable when demand elasticity
is modeled (\cref{fig:price_duration}),\footnote{In order for the chemical
  storage itself to recover its capital costs from arbitrage, the marginal storage
  value $\lambda^s_t$ has to vary. It would only be constant if the storage
  capacity were free and unconstrained (\cref{fig:si:example}).} assuming a
constant value is not perfect for short-run dispatch and will cause price
deviations from perfect foresight, which are quantified in
\cref{sec:results-myopic}.

A similar methodology was employed in \cite{dupre2023}, utilising a
state-of-charge-dependent storage value, as also noted in \cite{aaslid2021}. In
contrast, we maintain a simpler method with a constant MSV to demonstrate that
reasonable results can be achieved even with this simplification. This
assumption could be justified if industrial offtake users with long-term
contracts and producers on global markets contribute to stabilising the hydrogen
price.

\section{Results}

\subsection{Results from the long-term model}
\label{sec:results-lt}

\cref{fig:price_duration} shows different price duration curves from the model
with perfectly inelastic demand up to a value of lost load (VOLL) of 2000~\euro/MWh (left column) and with a piecewise-linear (PWL) approximation of -5\% price elasticity of
demand (right column). The results assume hydrogen is used for long duration
storage.

Panels a) and b) in \cref{fig:price_duration} reveal how the electricity price
duration curves from the long-term model vary as the number of years in the
simulation is increased from one year up to the full 70 years (\ref{sec:year-selection}). For a duration
curve to be meaningful, it should not change significantly as the number of
years increases, but for the perfectly inelastic case, it becomes more and more singular
as more years are considered. This is because high prices are clustered into the
wind-scarce years, which drive the investment (\cref{fig:msv-cf}). As the high prices
cluster in the worst weather years, prices at other times collapse, since at these
times there is over-capacity.
For the
case with demand elasticity, the elastic demand curve spreads the scarcity pricing
out over the years, resulting in price duration curves that have a more
consistent shape as the number of years in the model is varied. Furthermore, going from the perfectly inelastic to
the price-responsive case for 70 years of data, the fraction of zero-price hours reduces
from 89\% to 31\%, and the fraction of hours with prices over 400~\euro/MWh
reduces from 8.5\% to 3.9\% so that the extreme variability of prices is
substantially damped. This means that revenues for most technologies are more
evenly distributed across different price bands (\cref{fig:price-bands}) instead
of being concentrated in just a few high-priced hours. These results reproduce
observations from sector-coupled models \cite{boettger2021,neumann2023}, where
demand elasticity plays the same role as demand flexibility to reduce and stabilise
prices.

We conclude from this that price duration curves from models with perfectly inelastic
demand are not robust enough between simulations that anything can be deduced
from them. Strong price bifurcations seen in previous studies
(e.g.~\cite{mallapragada2023}) are modelling artefacts that cannot be taken at
face value, since they are not consistent as the simulation length is changed
and they depend strongly on how demand is modelled.

Panels c) and d) in \cref{fig:price_duration} show hydrogen's corresponding
price duration curves. Since hydrogen can be inexpensively stored in underground
caverns ($\approx$~0.15~\euro/kWh, cf.~\cref{tab:technology-data}), its price is
more stable between different hours. This is because between periods of scarcity
there is no system benefit in moving hydrogen between hours without binding
capacity constraints. However, it still shows variations between wind-scarce and
windy years (\cref{fig:msv-cf}). The price variability is much more extreme for
the perfectly inelastic case, mirroring the tight coupling between electricity prices and
marginal storage values.

Per unit of electricity supplied, the system cost in the elastic case is also 9.3\%
lower than in the inelastic case, as higher demand curtailment replaces some
storage capacity (\cref{tab:metrics-70a-de}, \cref{fig:load-duration}). For the
inelastic case, we observe 70~MW of firm capacity compared to 46~MW for the
elastic case, which results in more regular demand curtailment over 70 years
which, in effect, spreads the prices more evenly. In the inelastic case, there
is less frequent load shedding concentrated in bad weather years that causes
very high prices, while we trade this in the elastic case for more regular load
reductions that cause much milder price increases.

\subsection{Results from short-term model with perfect foresight}
\label{sec:results-st}

Panels e) and f) in \cref{fig:price_duration} show the results of taking the optimal
capacities from the long-term (LT) model, fixing them, and rerunning the
operational short-term (ST) optimisation of the model with the same weather
years and perfect operational foresight. We then perturb all capacities of
generation and storage up and down by 5\% to reveal how sensitive the prices are
to capacities away from the long-term equilibrium.

Going from the LT to the ST model, we see the result of the theoretical argument in
\cref{sec:proofpricessame} confirmed: the prices are identical for
demand elasticity, where we have a unique mapping from demand to price, but for the
perfectly inelastic case the lack of price regularisation causes the prices of the ST
model to diverge from the LT prices. Furthermore, prices are susceptible to
small capacity perturbations, either becoming very high in many hours if capacity is
tight or collapsing to zero if capacities are slightly over-dimensioned. In the
elastic case, the price changes follow the same direction but are milder in
magnitude and more regular over time. Average prices rise by 104\% with 5\% less
capacity and go down by 63\% with 5\% more capacity (\cref{tab:metrics-70a-de}),
incentivising market exits or entries with investments towards the long-term
equilibrium.

\subsection{Forcing dispatchable capacity to limit load reduction}
\label{sec:forcing}

Demand elasticity means that there are more regular reductions in
demand, as can be seen from the load duration curves in
\cref{fig:load-duration}. This can be a problem for some consumers if
the reductions last more than a few hours. In our results there are
reductions of 20\% lasting 1-2 days roughly every other year during
dark wind lulls, while there are 2 events in the 70 year sample
lasting 10 days. To mitigate these high price events, we explore
in \cref{fig:si:reserve} how forcing dispatchable capacity affects the model.

If some conventional dispatchable capacity is retained (\cref{fig:si:dispatchable}) or
power capacity for re-electrification of hydrogen is procured outside the
energy-only market (\cref{fig:si:reserve}), our results remain largely
unchanged. Some high-price episodes gradually disappear, both for the inelastic and
elastic cases, as scarcity costs of power deficits are replaced by the cost of
reserve capacities or dispatchable power plants. For the elastic case, we still
see identical prices between the LT and ST models.

\subsection{Inter-annual variability}
\label{sec:results-interannual}

\begin{figure}[!t]
  \centering
  \includegraphics[trim=0 0cm 0 0cm,width=0.95\linewidth,clip=true]{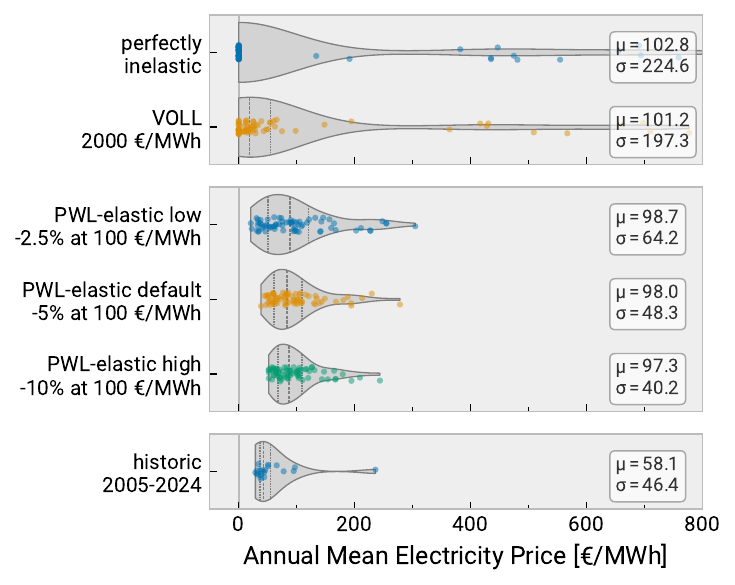}
  \caption{ Distribution of average annual baseload prices in different
    scenarios. Violin plots show the kernel density estimate of the data.
    Vertical lines show the quartiles of the data. The $\mu$ signifies the
    mean, $\sigma$ the standard deviation.}
  \label{fig:interannual}
\end{figure}

\begin{figure}[!t]
  \centering
  \includegraphics[trim=0 0cm 0 0cm,width=\linewidth,clip=true]{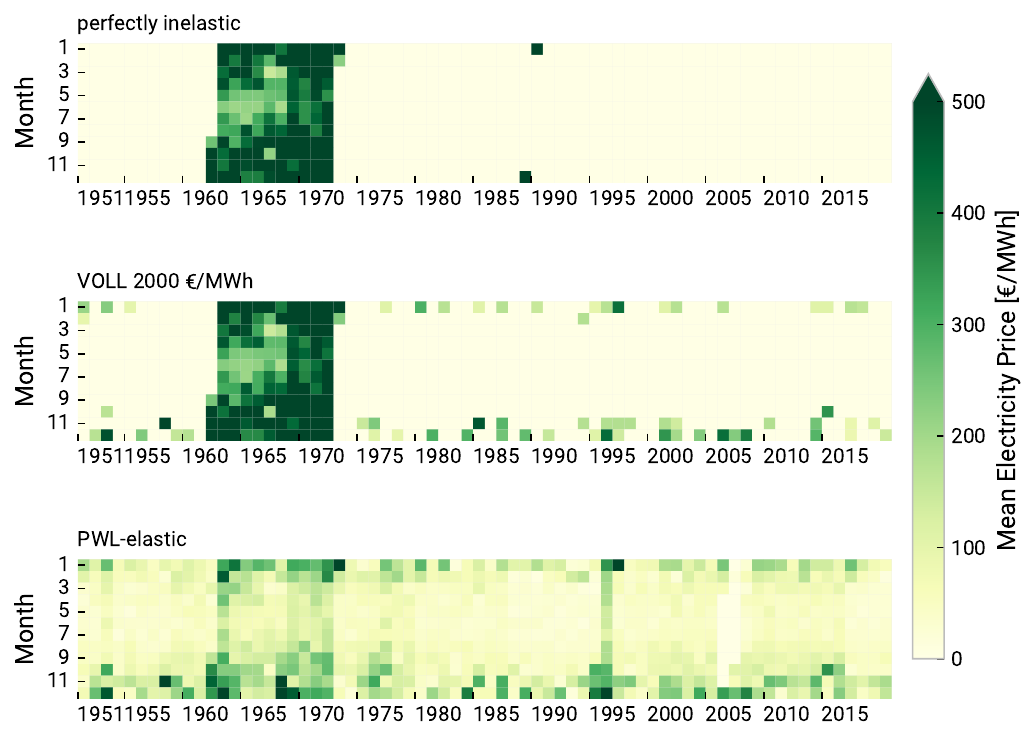}
  \caption{Heatmap of average monthly baseload prices over 70 years in different scenarios.}
  \label{fig:heatmaps}
\end{figure}

One of the concerns raised about systems with high shares of variable generation
is that prices might vary strongly between years as weather conditions change.
The wind, in particular, is known to exhibit strong inter-annual variability
(\cref{fig:msv-cf}) \cite{pryor2006}. Widely varying prices would make asset
cost recovery uncertain and raise the risk premium for financing the assets
\cite{mays2022,stoft2002}.

To explore this, we show in \cref{fig:interannual,fig:heatmaps} the distribution
of average annual and monthly baseload prices from the optimised 70-year period.
The perfectly inelastic demand cases show a very high variance with a strong
clustering of high prices in the 1960s. In contrast, prices cluster ever tighter
for increasing elasticity and are more evenly distributed. In all cases except
for fully inelastic demand, we see a seasonal pattern with higher prices in
winter and broadly lower prices in summer.  If we compare the results to the
distribution of prices over the last 19 years \cite{opsd2020,entsoe2024}, we see
a comparable range as seen in the conventional system due to fossil fuel price
variability.\footnote{It should be noted that the historical wholesale market
prices do not yield full cost recovery for the assets in the system because
aspects like renewable subsidies are not included.} Such stable prices would
increase certainty and reduce risk for investors, as annual revenues would also be
less volatile (\cref{fig:annual-revenue}).

\subsection{Results from short-term model with myopic foresight}
\label{sec:results-myopic}

\begin{figure*}[!t]
  \centering
  \includegraphics[trim=0 0cm 0 0cm,width=0.8\linewidth,clip=true]{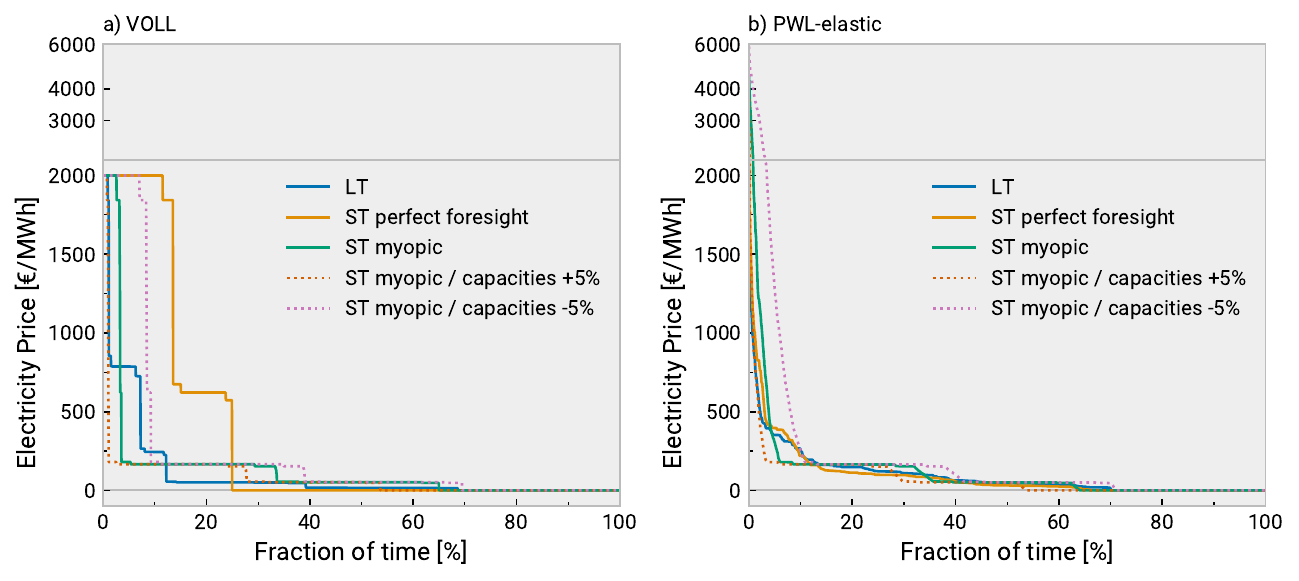}
  \caption{Price duration curves comparing short-term models with myopic
    foresight on unknown years, i.e.~the LT and ST model take two disjunct sets of
    35 weather years as input data. Blue is from the original LT model based on 35
    years of data. Orange is from a ST model with optimised capacities from the LT
    model and perfect foresight on 35 unseen test years. Green is another ST model
    with myopic foresight of 96 hours and a rolling horizon overlap of 48 hours
    for the same unseen test years. Dotted lines add capacity perturbations of
    $\pm 5\%$ to that.}
  \label{fig:myopic}
\end{figure*}

Until now, the short-term models have been run with perfect foresight over the
same weather years used for capacity expansion planning. Now, we show results
with myopic operational foresight and heuristic storage bidding as described in
\cref{sec:myopic-method}. Capacities are determined based on 35 randomly
selected years in the LT model, and then operations are optimised on 35 other
unseen years in the ST model. As hydrogen storage bids for ST model the we take
the mean MSV of the LT model.

\cref{fig:myopic} compares price duration curves from myopic model runs (96-hour
foresight, 48-hour overlap) with the LT and ST models with perfect foresight.
Since the MSV is set to a constant value, there are now only two large non-zero
steps in the price duration curves corresponding to the charging and discharging
bids for hydrogen storage. The small intermediary steps are caused by
interactions with the bids of the battery storage and the demand curve, if
present (\cref{fig:si:example}).

In the perfectly inelastic case, there is a large increase in hours with load-shedding
going from the LT capacity expansion model to the ST operational model, since
the weather years of the ST model were different to those in the LT model. With
myopic foresight the heuristic storage bidding based on a constant MSV acts to
regularise prices, resulting in two main non-extreme price levels besides
load-shedding and zero-price hours. A constant MSV was not something found in
the LT inelastic model (see \cref{fig:price_duration}, panel c)) but exogenously set in
the model inspired by observations from the elastic case (see
\cref{fig:price_duration}, panel d)). As a result, the monthly average baseload
price distribution has a similar regularity to the case with demand elasticity
(\cref{fig:si:heatmap-myopic}). Taking the constant averaged MSV blends out the
volatility of the MSV formation in the absence of demand elasticity. With myopic
foresight, the peak load shedding increases by 13.9\% from 77.6~MW to 88.3~MW,
while the average load served is just reduced by 0.5\% from 99.96~MW to
99.47~MW. To compare the system operation between perfect and myopic foresight
with and without demand elasticity, we refer to \cref{fig:operation}.

In the elastic case, the price duration curves are similar. Unlike in
\cref{fig:price_duration}, the price duration curves between the LT and ST
models are close but not identical. This is again because the wind and solar
capacity factors are taken from different weather years. The minor differences
observed between the ST models with myopic and perfect foresight outline where
the heuristic storage bids under or overestimate the MSV. With myopic foresight,
the average electricity price increases from 95.4~\euro/MWh to 126.2~\euro/MWh.
This reflects the cost of lacking foresight. Peak demand curtailment increases
by 23.2\%, whereas the average load served is reduced by only 0.8\%. The drop in
total welfare is just 0.1\%, since the welfare is dominated by the demand utility (see Table \ref{tab:metrics-35a-35a-de}).

In either case, the capacity perturbations of $\pm5\%$ have a similar effect as
seen previously with perfect foresight in \cref{fig:price_duration}. An
additional sensitivity analysis run with a shorter foresight window of 48 hours
exhibits only slight differences between 96 hours and 48 hours of foresight.
With longer foresight, the average load served is increased by less than 0.1\%,
peak load shedding is reduced by 3-7\%, and average electricity prices are
reduced by around 4\% (\cref{tab:metrics-35a-35a-de}).

These results highlight that long foresight horizons are not required for
operating a highly renewable energy system with storage. A horizon of a few days
corresponding to current forecasting skill suffices for battery scheduling and
simple bidding heuristics for long-duration energy storage yield acceptable
dispatch decisions. Any estimates of weather conditions further ahead will
only help to improve the economic dispatch. A live website was developed by one
of the authors to demonstrate how 24-hour-ahead forecasts combined with fixed hydrogen MSV
are able to dispatch a fully renewable German power system over 10 years of weather data \cite{modelenergy}.

\subsection{Cost recovery}
\label{sec:results-crf}

\begin{figure}[!t]
  \centering
  \includegraphics[trim=0 0cm 0 0cm,width=\linewidth,clip=true]{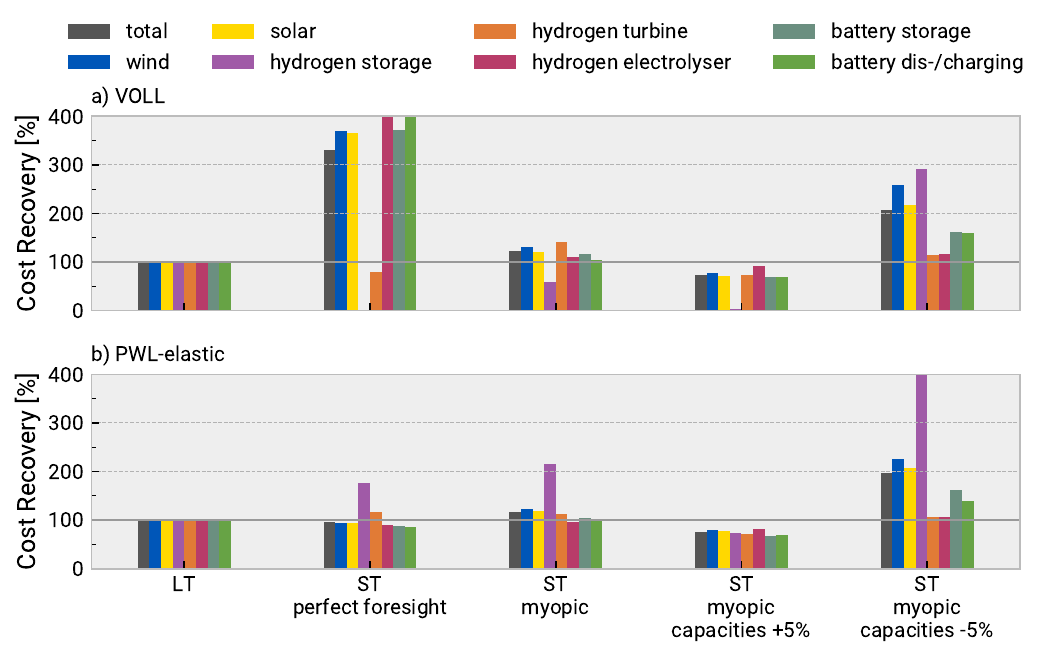}
  \caption{Cost recovery of each component as a fraction of investment, depending on demand modelling approach, operational foresight and capacity perturbations for 35-year LT plus 35-year ST optimisations.}
  \label{fig:cost_recovery}
\end{figure}

Cost recovery refers to the fraction of the investment and operation costs that
are recovered by the revenues from the operation of the assets.
\cref{fig:cost_recovery} shows the cost recovery for each asset in different
scenarios. For the LT model, the cost recovery is perfect for all components,
following the zero-profit rule shown in Brown and Reichenberg~\cite{brown2021}.
The variability of revenue from year to year is, however, much higher for the
VOLL case than the PWL-elastic case (\cref{fig:annual-revenue}).

Transitioning from the LT to the ST model with perfect foresight, drastic
differences are observed between demand modelling cases. The high overshoot of
cost recovery rates for the perfectly inelastic case until VOLL, due to the high prices seen in
\cref{fig:myopic}, is not observed in the elastic case and highlights the
inadequacy of modelling with perfectly inelastic demand.

For scenarios with myopic operational foresight in unseen years, the differences
between inelastic and elastic demand cases are pruned by the heuristic storage
bids. Especially for the elastic case, all components exceed or come close to full
cost recovery. The lowest cost recovery is 97.2\% for the electrolyzer. With
over-capacities of 5\%, cost recovery falls short and varies between 67\% and
82\%, incentivising market exits. The opposite holds for deficient capacities,
where profits between 14\% and 190\% incentivise certain market entries.\\~\\

All results shown in preceding
\cref{sec:results-lt,sec:results-st,sec:forcing,sec:results-interannual,sec:results-myopic,sec:results-crf}
for Germany are consistent with those observed for a more solar-dominated system
in Spain and a more wind-dominated system in the United Kingdom, as shown in
\ref{sec:si:countries}.

\section{Discussion}

\subsection{Consquences for interpreting prices in energy system models}

Our results establish that the strong price bifurcations seen in many capacity
expansion models are due to modelling demand without price elasticity. The
combination of step-like inelastic demand curves with step-like supply curves
from VRE together cause the price bifurcations. Given that today's electricity
markets already feature price-sensitive demand \cite{hirth2024}, price duration
curves from these models are unrealistic and should not be taken at face value.

While average prices and market values might be indicative in models with
perfectly inelastic demand, the shape of the price duration curves is not stable as the
number of optimised periods is changed, nor under transition from long-term to
short-term modelling, nor under small perturbations of capacity, nor between
different weather years. Prices either collapse to zero for extended periods or
spike dramatically. The concerns in the literature, for instance, that revenues
of generation and storage would be realised in just very few hours
\cite{taylor2016,mallapragada2023,Levin2023,royalsociety2023}, are rooted in
these observations but are, in fact, interpretations of a modelling artifact.

The short-term elasticity level observed in the German wholesale electricity
market today (cf.~\cite{hirth2024}) would already be sufficient to stabilise
price formation and yield a completely different picture about when revenues are
made. The price formation with demand elasticity is then identical between long-
and short-term models, stable under capacity perturbations, stable between
between weather years, and allows for more consistent and evenly spread cost
recovery.

The reason demand elasticity helps to stabilise prices is that it
gives the storage a wide range of willingness-to-pay (WTP) values
against which to calibrate its opportunity costs and arbitrage. With
perfectly inelastic demand, the WTP jumps from zero to the value of
lost load with nothing in between for the storage to bid against.

The prospect of more cross-sectoral integration further opens new opportunities
for demand-side management (DSM) of electric vehicles, heat pumps, and industrial
processes, beyond today's elasticity level, although inter-temporal dependencies
need to be treated with care. Modelling DSM with virtual storage units is not
sufficient to resolve pricing problems, as the examples above with storage
show; sloped demand curves with smoothly varying WTP are what is needed for stable price formation.

\subsection{Relating the modelling results to real electricity markets}
\label{sec:relating-to-reality}

Until now, we have focused mainly on explaining model results and comparing them
to observations in the literature, but we believe many of our insights are
transferable to real electricity markets. Our model's operation with slightly
elastic demand and myopic foresight is close to how real markets function, and
price formation remains stable under these conditions, as implied in Stoft's
textbook \cite{stoft2002}.

Cost recovery is naturally imperfect with imperfect information. For instance,
with capacities below or above the long-term equilibrium, there would be
incentives for market entries or exits as asset owners would adjust their
capacities to prevailing market conditions. The more regular and lower high
price episodes in the elastic case would provide investors with a much more
stable investment environment than the infrequent extreme high price episodes
seen with the perfectly inelastic case, thus helping to avoid boom and bust cycles. The
political tension between allowing scarcity pricing and remunerating
dispatchable generators via capacity markets to guarantee security of supply is
present in our high-VRE scenarios, just as it is in today's markets.

Our results suggest that the energy-only market remains functional in future energy systems with high VRE penetration.
However, it may not be sufficient on
its own to attain a low-cost renewable system.
Investor risk remains thanks to uncertainties surrounding inter-annual revenue variability,
changing weather patterns due to climate change,
other factors that can confound price forecasts (price cannibalisation
from other renewables, CO$_2$ price uncertainty) as well as regulatory risks
and political risks from changing climate targets.
To lower investor risk and reduce financing
costs, support guarantees such as investment subsidies, contracts for difference
or feed-in tariffs may still be advisable \cite{neuhoff2022}. The strike prices
of these contracts could use the energy-only market for orientation towards
efficient prices.

The assumption of a constant marginal storage value for the myopic dispatch is
strong, and in reality, storage operators would adjust their bidding based on
their filling state and medium to long-range weather forecasts \cite{dupre2023}.
Moreover, a flat price would mean storage operators could not perform arbitrage
and recover their costs. Potentially, long-term contracts for industrial offtake
or global supply (e.g.~H2Global~\cite{h2global}) could stabilise the hydrogen
price.

We also assume perfect forecast skill within the 96-hour foresight horizon. In
reality, stochastic weather forecast ensembles for the next 1-2 days would guide
storage owners in their bidding strategy under uncertainty \cite{aaslid2021}.
However, this neglect of short-range uncertainty does not invalidate the use of
bidding heuristics for long-term storage dispatch.

\subsection{Comparison of chemical storage to conventional generation}

In terms of their impact on price formation in electricity markets, we can
compare chemical storage to conventional generation technologies. For instance,
hydroelectricity shares many similarities with chemical storage. Both can
provide long-duration storage that stabilises prices through long-term opportunity costs.
However, unlike chemical storage, hydroelectric storage is constrained by
natural inflow, self-discharge, and the fact that the storage medium cannot be traded on global
markets. Hydroelectric inflow is easier to forecast on a seasonal basis
that wind power fluctuations, thanks to predictable snow and rainfall patterns.
On the other hand, chemical storage expansion is not geologically limited and offers more
control over dimensioning and scaling.

The price of fossil fuels like gas and coal also stems from a mix of production
and opportunity costs, which is similar to how the value of chemical storage is
determined in our model. For hydrogen, production costs include wind and solar
generation and power-to-X elements. For fossil fuels, it is land rents as well as the energy and labor
costs for extraction, which may be further impacted by emission prices or
scarcity prices due to supply disruptions. Similarly, prices are linked to
feedstock availability for electricity production from biomass and the mining,
enrichment, and fabrication of uranium into fuel rods for nuclear power. The
connection to currently used fuels becomes even more apparent when considering
green fuel imports for power backup, where the price is decoupled from the
domestic electricity system. The price formation mechanism is then similar to that
of a fossil-based system.

\section{Conclusion}

Our study addresses several contradictions seen in the literature about the
viability of electricity markets at high penetrations of wind and solar. Our
results demonstrate that price structures consisting of many zero- and high-price hours
are a symptom of capacity expansion models using perfectly inelastic demand. Already a
slightly elastic demand alleviates these symptoms and reduces the fraction of
zero-price hours from 90\% to around 30\%. Besides eliminating price
bifurcations, modelling demand elasticity stabilises prices between weather
years, which reduces revenue risks, and aligns prices observed in long- and
short-term models. Therefore, we conclude that electricity models using
perfectly inelastic demand are inadequate for analysing price formation at high
penetrations of variable renewable energy.

Demand elasticity not only stabilises price formation but also
facilitates the operation of long-term storage with myopic foresight. The
stability of the marginal values of chemical storage, which are closely linked
to electricity price formation, enables the development of storage bidding
strategies. Even with the simple assumption of constant storage bids, we can
achieve adequate system dispatch with similar prices, reliability metrics, and
cost recovery levels, with only a few days of foresight, and even if capacities
do not exactly match the long-term equilibrium.

With these insights, we address many of the concerns about price formation in
energy-only markets with high penetrations of wind and solar. We conclude that
the energy-only market can still play a key role in coordinating future dispatch
and investment given current price elasticity levels. If demand can be further
flexibilised in future, this will only help to stabilise prices even more.

\section*{Code and data availability}

The open source code to reproduce our experiments, including the download and
installation of data and software dependencies, is available on GitHub licensed
under the \href{https://opensource.org/license/mit}{MIT} license at
\url{https://github.com/fneum/price-formation} (v0.2.0). The results data was
deposited on Zenodo under the
\href{https://creativecommons.org/licenses/by/4.0/}{CC-BY-4.0} license at
\url{https://doi.org/10.5281/zenodo.12759247} (v0.2.0).

\section*{Author contributions}


\textbf{T.B.}:
Conceptualization --
Formal Analysis --
Funding acquisition --
Investigation --
Methodology --
Project administration --
Software --
Validation --
Visualization --
Writing - original draft --
Writing - review \& editing
\textbf{F.N.}:
Formal Analysis --
Funding acquisition --
Investigation --
Methodology --
Software --
Validation --
Visualization --
Writing - original draft --
\textbf{I.R.}:
Formal Analysis --
Investigation --
Validation --
Writing - review \& editing

\section*{Declaration of interests}

The authors declare no competing interests.

\section*{Acknowledgements}

I.R.~acknowledges support by the German Federal Ministry for Economic Affairs
and Climate Action (BMWK) under Grant No. 03EI4083A (RESILIENT) jointly with the
CETPartnership (\url{https://cetpartnership.eu/}) through the Joint Call 2022.
As such, I.R.~further acknowledges funding from the European Union's Horizon
Europe research and innovation programme under grant agreement no. 101069750.

We thank Lina Reichenberg, Wolf-Peter Schill, Falko Ueckerdt, Robert Pietzcker,
Gunnar Luderer, Chris Gong, Oliver Ruhnau, Richard Schmalensee, Dharik
Mallapragada, Emil Dimanchev, Richard Green and Tim Schittekatte for fruitful discussions.

\renewcommand{\ttdefault}{\sfdefault}
\bibliography{elasticity}

\onecolumn

\appendix

\renewcommand{\thefigure}{S\arabic{figure}}
\renewcommand{\thetable}{S\arabic{table}}

\setcounter{figure}{0}
\setcounter{table}{0}

\section{Substitutions to improve numerics}
\label{sec:numerics}

Many of our initial runs with VOLL or demand elasticity represented as outlined in
\cref{eq:objective} were numerically unstable and did not converge when
many weather years were included. We found that making the substitutions $d_c
  \rightarrow D_c - g_c$ for VOLL and $d_c \rightarrow a_c / b_c - g_c$ for
demand elasticity drastically improved the convergence of the model while yielding
the identical results, as they remove large constants from the objective.

The suggested substitutions switch from modelling the demand as a decision
variable to modelling load shedding from fixed peak consumption ($D_c$ or $a_c
  /b_c$) as a decision variable. In terms of the objective function, it yields for
VOLL
\begin{align}
  U_c(d_c) & = V_c d_c \nonumber                    \\
  U_c(g_c) & = V_c (D_c - g_c) = V_c D_c - V_c g_c,
\end{align}
and for the elastic case
\begin{align}
  U_c(d_c) & = a_c d_c - \frac{b_c}{2} d_c^2 \nonumber                                                                                   \\
  U_c(g_c) & = a_c \left(\frac{a_c}{b_c} - g_c\right) - \frac{b_c}{2} \left(\frac{a_c}{b_c} - g_c\right)^2 \nonumber                     \\
           & = \frac{a_c^2}{b_c} - a_c g_c - \frac{b_c}{2} \left( \frac{a_c^2}{b_c^2} - 2 \frac{a_c}{b_c} g_c + g_c^2 \right)  \nonumber \\
           & = \frac{a_c^2}{b_c} - a_c g_c - \frac{a_c^2}{2b_c} + a_c g_c - \frac{b_c}{2} g_c^2 \nonumber                                \\
           & = \frac{a_c^2}{2b_c} - \frac{b_c}{2}g_c^2.
\end{align}
Since the constant terms $V_c D_c$ and $a_c^2 / 2b_c$, which represent the
welfare under the demand curve, do not affect the optimisation, we can drop them
from the objective function. What remains is a load-shedding generator with
the cost curve $C_c(g_c) = V_c g_c$ or $C_c(g_c) = \frac{b_c}{2} g_c^2$ for the
elastic case. In addition, a fixed demand of $D_c$ or $a_c/b_c$, respectively,
needs to be added to the model.

This substitution trick can also be generalised for piecewise-linear demand
curves with segments $c$, where the demand curves are not zero at $a_c/b_c$. We now apply the more general substitution $d_c
  \rightarrow D_c - g_c$ for each segment $c$.
\begin{align}
  U_c(d_c) & = a_c d_c - \frac{b_c}{2} d_c^2 \nonumber                                                                          \\
  U_c(g_c) & = a_c \left(D_c - g_c\right) - \frac{b_c}{2} \left(D_c - g_c\right)^2 \nonumber                                    \\
           & = a_c D_c - a_c g_c - \frac{b_c}{2} \left( D_c^2 - 2 D_c g_c + g_c^2 \right)  \nonumber                            \\
           & = a_d D_c - a_c g_c - \frac{b_c D_c}{2} + b_c D_c g_c - \frac{b_c}{2} g_c^2 \nonumber                              \\
           & = \left(a_c D_c - \frac{b_c D_c}{2} \right) + \left(b_c D_c - a_c\right) g_c  + \left(\frac{-b_c}{2}\right) g_c^2,
\end{align}
resulting in load-shedding generators with a cost curve
\begin{equation}
  C_c(g_c) = \left(a_c - b_c D_c\right) g_c + \frac{b_c}{2} g_c^2,
\end{equation}
again dropping the constant term from the objective function.

\newpage
\section{Including cross-elasticity terms}
\label{sec:cross-elasticity}

To account for how demand changes in one hour affect the demand curve in other
hours, we extend the quadratic utility term in the objective function to include
cross-elasticity terms:
\begin{equation}
  U_c(d_{c,t}) = \sum_t \left(a_c d_{c,t} - \frac{b_c}{2} d_{c,t}^2 + \sum_k \frac{\gamma_{c,t,k}}{2} d_{c,t} d_{c,k}\right)
\end{equation}
where $\gamma_{c,t,k}$ represents the cross-elasticity between hours $t$ and $k$
for segment $c$. The factor $\frac{1}{2}$
appears because each cross-term appears twice in the summation. The
cross-elasticity coefficient is symmetric ($\gamma_{c,t,k} = \gamma_{c,k,t}$)
and is only non-zero for adjacent hours within a specified time window:
\begin{equation}
  \gamma_{c,t,k} = \begin{cases}
    \gamma_{c} & \text{if } |t-k| \leq \hat{T} \text{ and } t \neq k \\
    0 & \text{otherwise}
  \end{cases}
\end{equation}
where $\hat{T}$ is the maximum difference in time steps for considering
cross-elasticity effects and $\gamma_{c}$ is chosen proportionally to $b_c$. For
each demand segment $c$, we use $\gamma_{c} = \frac{b_c}{16}$ in the default
case to ensure cross-effects do not dominate own-price elasticity. We also
choose as default $\hat{T} = 4$ hours. The diagonal elements ($t = k$) are zero
since own-price effects are already captured by the quadratic term with
coefficient $b_c$.

This approach means that if demand in hour $t$ decreases by 1~MW, the
willingness to pay in all hours $k$ where $|t-k| \leq \hat{T}$ increases by
$\gamma_{c}$ €/MWh, reflecting the increased need to consume in those hours to
maintain the same total consumption level. The cross-elasticity does not change
the price elasticity itself; the demand curve shifts up in those hours, but its
slope (elasticity) stays the same. The objective function remains concave with
the inclusion of cross-elasticity terms as long as $\gamma_{c}$ is small enough,
such that we do not expect a strong computational impact.

For numerical stability, we apply the same substitution as in
\ref{sec:numerics}, replacing $d_{c,t} \rightarrow D_c - g_{c,t}$ where $D_c =
a_c/b_c$. For the cross-elastic terms, this yields:
\begin{align}
  \sum_k \left( \frac{\gamma_{c,t,k}}{2} d_{c,t} d_{c,k} \right) & = \sum_k \left( \frac{\gamma_{c,t,k}}{2} (D_c - g_{c,t})(D_c - g_{c,k}) \right) \nonumber \\
  & = \sum_k \left( \frac{\gamma_{c,t,k}}{2} (D_c^2 - D_c g_{c,t} - D_c g_{c,k} + g_{c,t} g_{c,k}) \right) \nonumber \\
  & = \sum_k \left( \frac{D_c^2 \gamma_{c,t,k}}{2} - \frac{D_c \gamma_{c,t,k}}{2} g_{c,t} - \frac{D_c \gamma_{c,t,k}}{2} g_{c,k} + \frac{\gamma_{c,t,k}}{2} g_{c,t} g_{c,k} \right) \\
\end{align}
The constant terms $\sum_k D_c^2 \gamma_{c,t,k} / 2$ can be dropped from the objective
function. The remaining terms are added to the objective function as cost terms of load shedding generators $g_{c,t}$ for cost minimisation, analogous to \ref{sec:numerics}:
\begin{equation}
  \sum_k \left( \frac{D_c \gamma_{c,t,k}}{2} g_{c,t} + \frac{D_c \gamma_{c,t,k}}{2} g_{c,k} - \frac{\gamma_{c,t,k}}{2} g_{c,t} g_{c,k} \right).
\end{equation}

\begin{figure}
  \includegraphics[width=\linewidth]{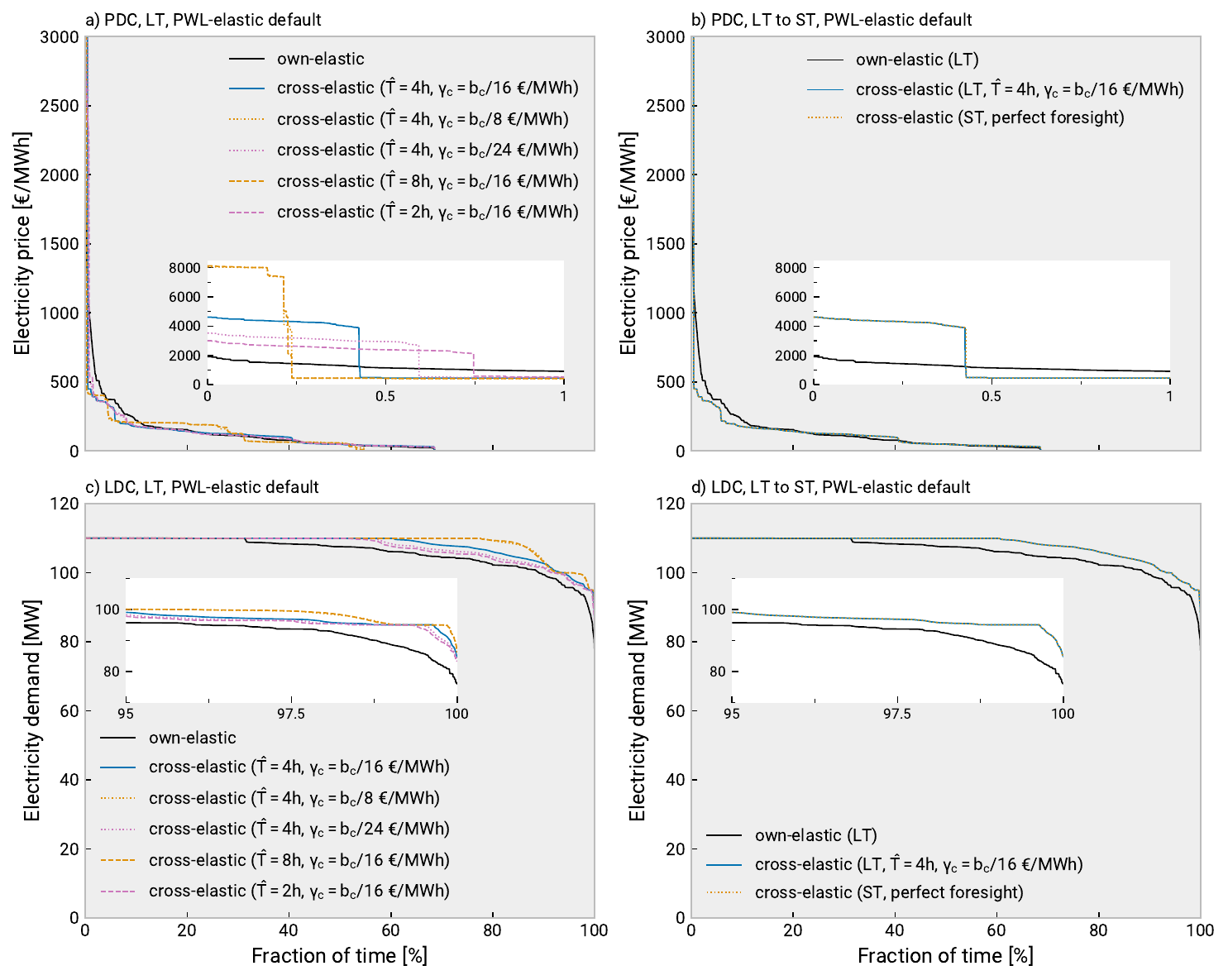}
  \caption{Price and load duration curves for LT and ST models with
    cross-elasticity terms alongside piecewise-linear own-elastic demand. Panel
    a) shows the effect of varyign cross-elasticity coefficients $\gamma_c$ and
    $\hat{T}$ on the price duration curve (PDC). Panel b) shows the effect of
    moving from LT to ST models with cross-elasticity terms on the prices.
    Panels c) and d) show the load duration curves (LDC) corresponding to the
    price duration curves shown in panels a) and b), respectively.}
  \label{fig:cross-elasticity}
\end{figure}

In \cref{fig:cross-elasticity}a, we compare price duration curves for different
cases with and without cross-elasticity terms. We see that across a range of
different parameterisations, the cross-elasticity terms lead to higher peak
prices and more price discontinuities. With only own-price elasticity, a demand
reduction in a high price hour directly reduces the price. There is no
interaction with other hours that could counteract this price-moderating effect.
With cross-elastic terms, demand shifts can create new price pressures in
adjacent hours. This results in higher peak price levels, lower overall demand
reduction, and larger backup capacities. For $\gamma_c = b_c / 16$ and
$\hat{T}=4$h, the capacity of hydrogen turbine increases by 20\% to 56~MW$_\text{el.}$
compared to the 46~MW$_\text{el.}$ without cross-elasticity. Larger cross-elastic terms
further intensify the price spikes and discontinuities. The high price hours are those when
the output of the hydrogen turbines is at full capacity.
Because it is unclear
what values for cross-elasticity are realistic, we show results for a selection of
cross-elastic terms while still maintaining a concave objective function.

In \cref{fig:cross-elasticity}b, we show that cross-elastic terms can also
affect the price identity between LT and ST models with identical capacities.
With only own-price elasticity, we have a strict monotonic relationship between
price and demand in each hour. As demonstrated in \cref{sec:proofpricessame},
this leads to a unique solution and identical prices between LT and ST models.
When we add cross-price elasticities, demand in each hour depends on prices in
multiple hours. This creates, in principle, multiple valid ways to distribute
demand across hours while achieving the same total welfare. The LT and ST models
can choose different temporal distributions of demand/prices that are equally
optimal (i.e.~they achieve exactly the same level of utility and supply the same total
demand). The results in \cref{fig:cross-elasticity}b for Germany and also
\cref{fig:cross-elasticity_es,fig:cross-elasticity_uk} for Spain and the UK show
that, while prices align in the vast majority of hours, the price duration
curves are no longer completely identical between LT and ST models.

\clearpage
\section{Technology data}

\cref{tab:technology-data} shows the techno-economic assumptions used in the model.

\begin{table*}[!htb]
  \caption{Overview of techno-economic assumptions, mostly based on projections
    for the year 2030 from the Danish Energy Agency (DEA) \cite{dea2024}. The currency year is 2020.
    A uniform discount rate of 7\% is assumed. Efficiencies are shown as one-way
    efficiencies. Fixed operation and maintenance cost are given as a percentage
    of the overnight investment cost per year.}
  \label{tab:technology-data}
  \centering
  \vspace*{0.5cm}
  \begin{tabular}{lrlrrr}
    \toprule
    Technology                & Investment Cost &                 & Fixed Operation \& Maintenance & Efficiency
                              & Lifetime                                                                                \\ \midrule Onshore wind & 1095.9 &\euro/kW  & 1.22 \%/a & -- &
    30 a                                                                                                                \\
    Solar photovoltaics       & 543.3           & \euro/kW        & 1.95 \%/a                      & --         & 40 a  \\
    Battery inverter          & 169.3           & \euro/kW        & 0.34 \%/a                      & 96.0\%     & 10 a  \\
    Battery storage           & 150.3           & \euro/kWh       & --                             & --         & 25 a  \\
    Electrolysis              & 1500.0          & \euro/kW$_{el}$ & 4.00 \%/a                      & 62.2\%     & 25 a  \\
    H$_2$ turbine & 1164.0          & \euro/kW$_{el}$ & 5.00 \%/a                      & 50.0\%     &
    10 a                                                                                                                \\
    H$_2$ cavern storage      & 0.15            & \euro/kWh       & --                             & --         & 100 a \\
    \bottomrule
  \end{tabular}
\end{table*}

\section{Selection of years in random sequence}
\label{sec:year-selection}

The 70 years (1951-2020) we consider are first put in a random sequence using
the Fischer-Yates shuffle. The underlying Mersenne Twister random number
generator is initialised with the seed 123.
For our myopic dispatch analysis in \cref{sec:myopic-method}, where an LT model
is first optimised on 35 years and then a ST model is run on the remaining 35
years, we take the last 35 years from the shuffled list for the LT model and the
first 35 years for the ST model. The resulting years for each model are given in
\cref{tab:lt-st-years}.
For \cref{fig:price_duration}, where the number of years $N$ in the simulation
is increased step by step, we use the $N$ last entries in the shuffled list of
weather years. That means, for instance, that the years in the 4-year run are
included in the 16-year run. The shuffled sequence is not put in chronological
order. The resulting years for each case are given in
\cref{tab:inc-years}.
All other 70-year runs use the weather years 1951 to 2020 in chronological
order. All other single-year runs use 2020 as the weather year.

\begin{table}[htb!]
  \centering
  \small
  \begin{tabular}{p{3cm}p{14cm}}
    \toprule
    \textbf{Model} & \textbf{Years (in order)} \\
    \midrule
    LT model (35 years) & 1960, 1996, 1953, 2020, 1979, 1971, 1998, 2014, 2013, 1989, 1956, 1978, 1951, 2006, 1966, 1995, 2004, 2011, 2009, 1959, 1961, 1954, 2005, 2010, 1972, 1986, 2016, 1975, 1955, 1964, 2019, 2003, 1962, 1985, 1957 \\
    ST model (35 years) & 2007, 1987, 1974, 1976, 1981, 1993, 1988, 2015, 1958, 2018, 1970, 1990, 1968, 1991, 1965, 1963, 1992, 1973, 2002, 2001, 1982, 1967, 1999, 2017, 1994, 1984, 1977, 1980, 2012, 2000, 1983, 1997, 1969, 1952, 2008 \\
    \bottomrule
  \end{tabular}
  \caption{Years used in the 35-year LT and ST models.}
  \label{tab:lt-st-years}
\end{table}

\begin{table}[htb!]
  \centering
  \small
  \begin{tabular}{p{3cm}p{14cm}}
    \toprule
    \textbf{Number of years} & \textbf{Years (in order)} \\
    \midrule
    1 year & 1957 \\
    4 years & 2003, 1962, 1985, 1957 \\
    16 years & 1959, 1961, 1954, 2005, 2010, 1972, 1986, 2016, 1975, 1955, 1964, 2019, 2003, 1962, 1985, 1957 \\
    32 years & 2020, 1979, 1971, 1998, 2014, 2013, 1989, 1956, 1978, 1951, 2006, 1966, 1995, 2004, 2011, 2009, 1959, 1961, 1954, 2005, 2010, 1972, 1986, 2016, 1975, 1955, 1964, 2019, 2003, 1962, 1985, 1957 \\
    48 years & 1999, 2017, 1994, 1984, 1977, 1980, 2012, 2000, 1983, 1997, 1969, 1952, 2008, 1960, 1996, 1953, 2020, 1979, 1971, 1998, 2014, 2013, 1989, 1956, 1978, 1951, 2006, 1966, 1995, 2004, 2011, 2009, 1959, 1961, 1954, 2005, 2010, 1972, 1986, 2016, 1975, 1955, 1964, 2019, 2003, 1962, 1985, 1957 \\
    70 years & 2007, 1987, 1974, 1976, 1981, 1993, 1988, 2015, 1958, 2018, 1970, 1990, 1968, 1991, 1965, 1963, 1992, 1973, 2002, 2001, 1982, 1967, 1999, 2017, 1994, 1984, 1977, 1980, 2012, 2000, 1983, 1997, 1969, 1952, 2008, 1960, 1996, 1953, 2020, 1979, 1971, 1998, 2014, 2013, 1989, 1956, 1978, 1951, 2006, 1966, 1995, 2004, 2011, 2009, 1959, 1961, 1954, 2005, 2010, 1972, 1986, 2016, 1975, 1955, 1964, 2019, 2003, 1962, 1985, 1957 \\
    \bottomrule
  \end{tabular}
  \caption{Weather years for scenarios with incremental number of years represented.}
  \label{tab:inc-years}
\end{table}

\newpage
\section{Additional figures}
\FloatBarrier

\begin{figure*}[!htb]
  \centering
  \includegraphics[width=\linewidth]{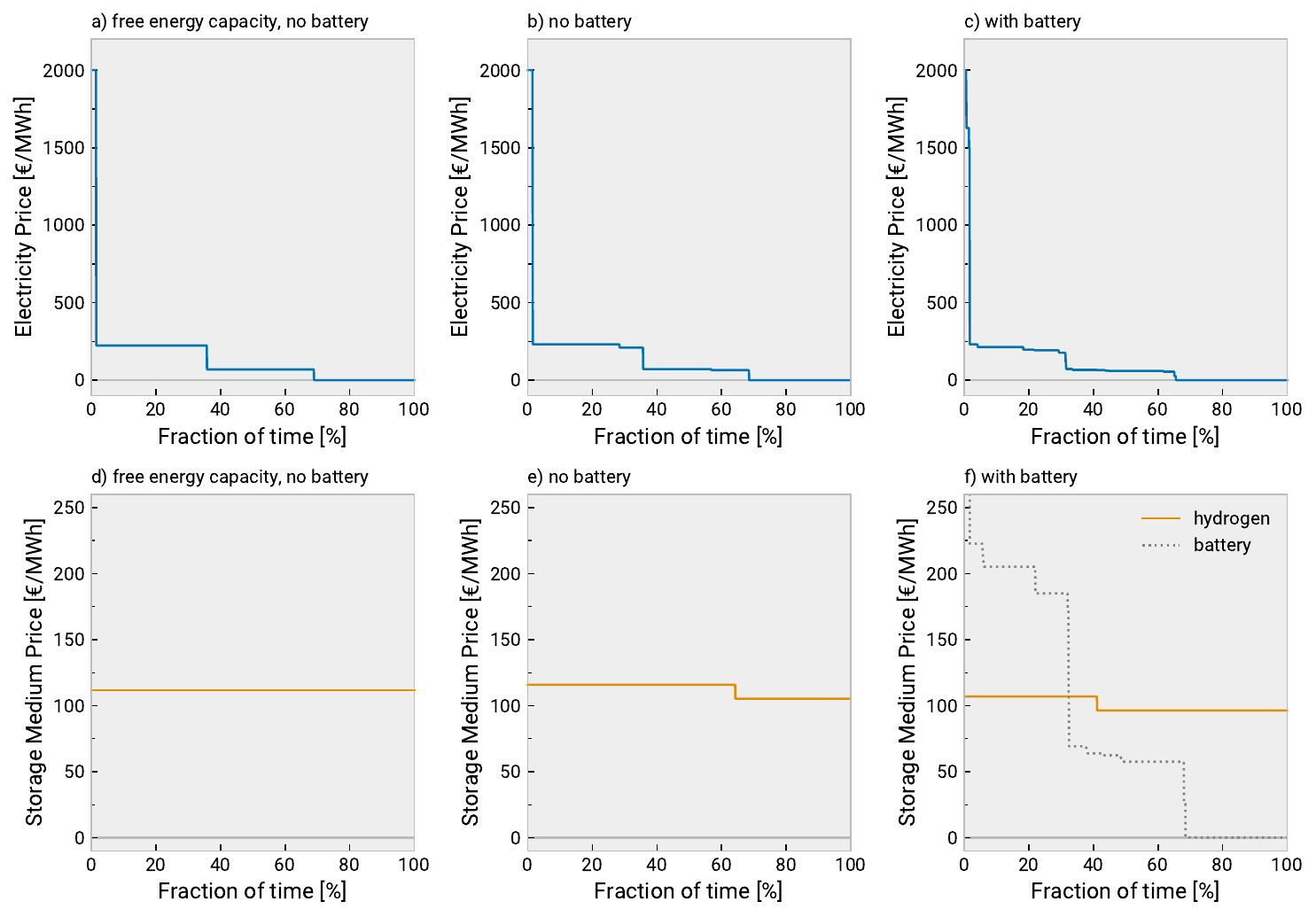}
  \caption{ Price duration and marginal storage value duration curves for a
    single year (2020) with perfectly inelastic demand up to a VOLL of 2000~\euro/MWh.
    Case a)/d) only has VRE (wind and solar) and unconstrained hydrogen storage
    without energy capacity cost. Since the storage itself does not need to
    recover any investment costs through price arbitrage, we see a uniform
    hydrogen price. This translates to four price levels in the electricity
    price duration curve; for VRE, VOLL, storage charging and discharging. Case
    b)/e) includes energy capacity costs for the hydrogen storage, creating two
    price levels for hydrogen such that the storage energy capacity can earn a
    revenue to recover its costs. This carries over to the electricity price
    duration curve, where there are now six price levels; two levels each for
    storage charging and discharging. Case c)/f) additional includes battery
    storage, which leads to a more complex price duration curve with the
    previous six major price levels and small steps at their boundary, resulting
    from the battery's arbitrage operations.}
  \label{fig:si:example}
\end{figure*}

\begin{figure*}[!htb]
  \centering
  \includegraphics[width=\linewidth]{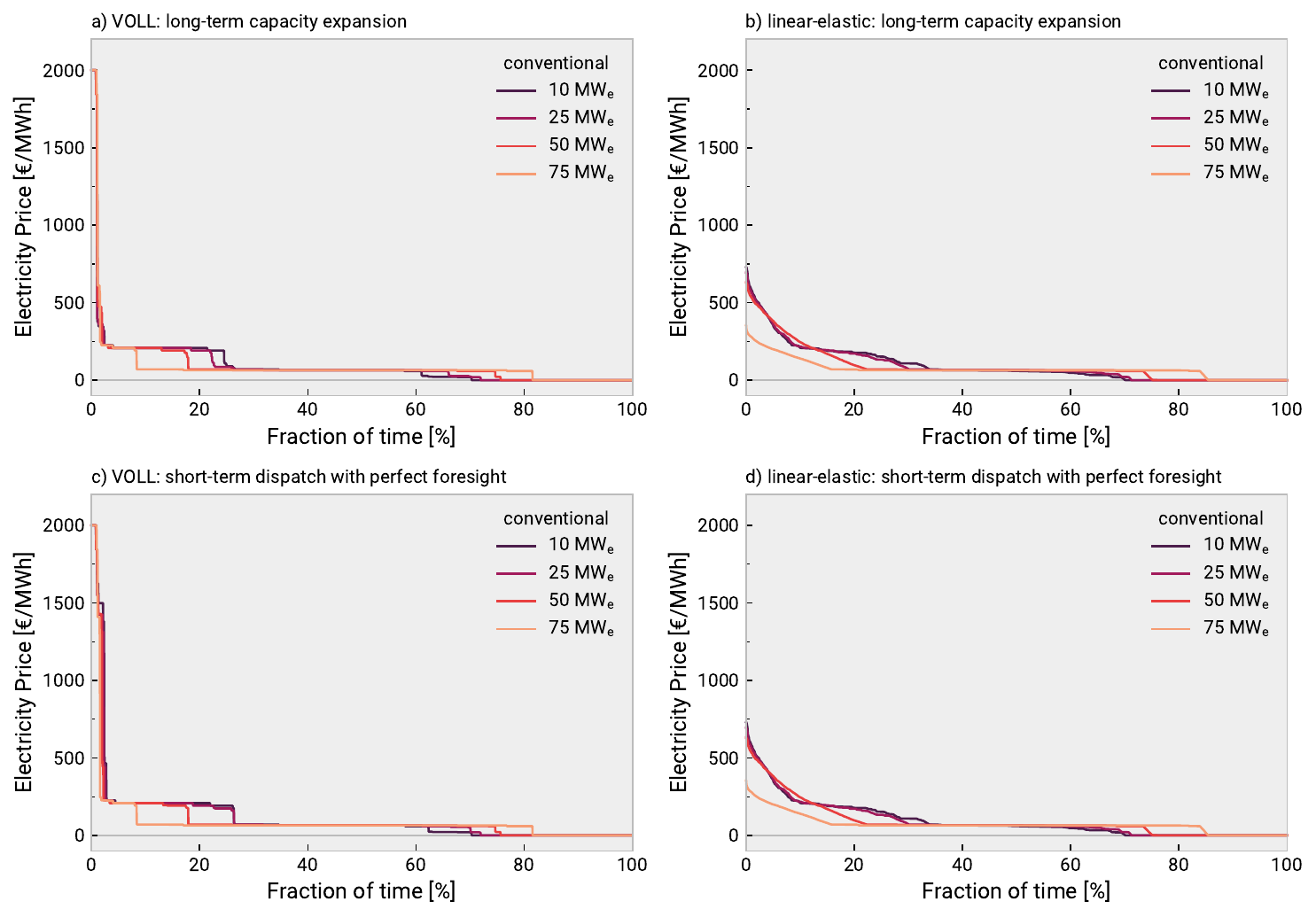}
  \caption{Sensitivity of price duration curves to dispatchable conventional
    capacity in LT and ST model with 20 weather years from 2001-2020.
    Dispatchable electricity cost is 64.7~\euro/MWh creating an additional price
    level in the centre of the price duration curve that widens with increasing
    dispatchable capacity. VOLL is 2000~\euro/MWh. Linear elastic demand run
    with parameters $a=2000$ and $b=10$. Overall, our findings remain unchanged
    with dispatchable conventional capacity in the system.}
  \label{fig:si:dispatchable}
\end{figure*}

\begin{figure*}[!htb]
  \centering
  \includegraphics[width=\linewidth]{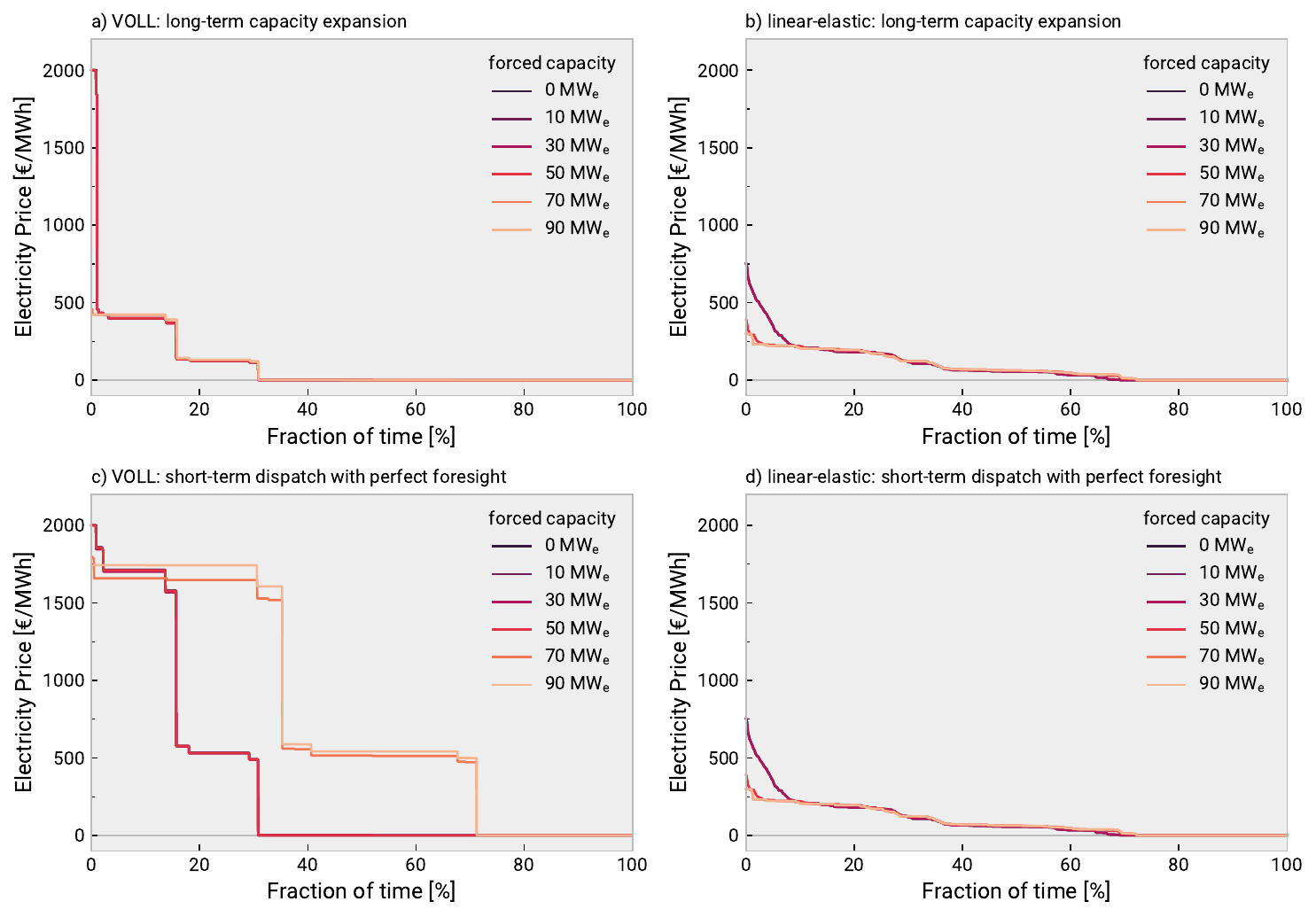}
  \caption{Sensitivity of price duration curves to reserve capacity in LT and ST
    model with 20 weather years from 2001-2020. Reserve hydrogen storage discharging capacity is forced into
    the system by setting a minimum capacity for the hydrogen turbine to
    electricity. VOLL is 2000~\euro/MWh. Linear elastic demand run with parameters
    $a=2000$ and $b=10$. Overall, our findings remain unchanged with reserve
    capacity procured outside the market. For the elastic case, we have price
    identity between LT and ST model. For the perfectly inelastic case, the prices rise
    drastically in the ST model compared to the LT model as the reserve capacity
    is increased. The forced additional dispatch capacity adds scarcity prices to
    the storage medium, which is then reflected in the electricity prices. As this
    setup is outside the model equilibrium, total welfare is lower. These price
    increases are not seen in the elastic case, where the prices remain stable
    because the demand willingness to pay sets the price rather than the storage
    bid.}
  \label{fig:si:reserve}
\end{figure*}

\begin{figure*}[!htb]
  \centering
  \includegraphics[trim=0 0cm 0 0cm,width=0.75\linewidth,clip=true]{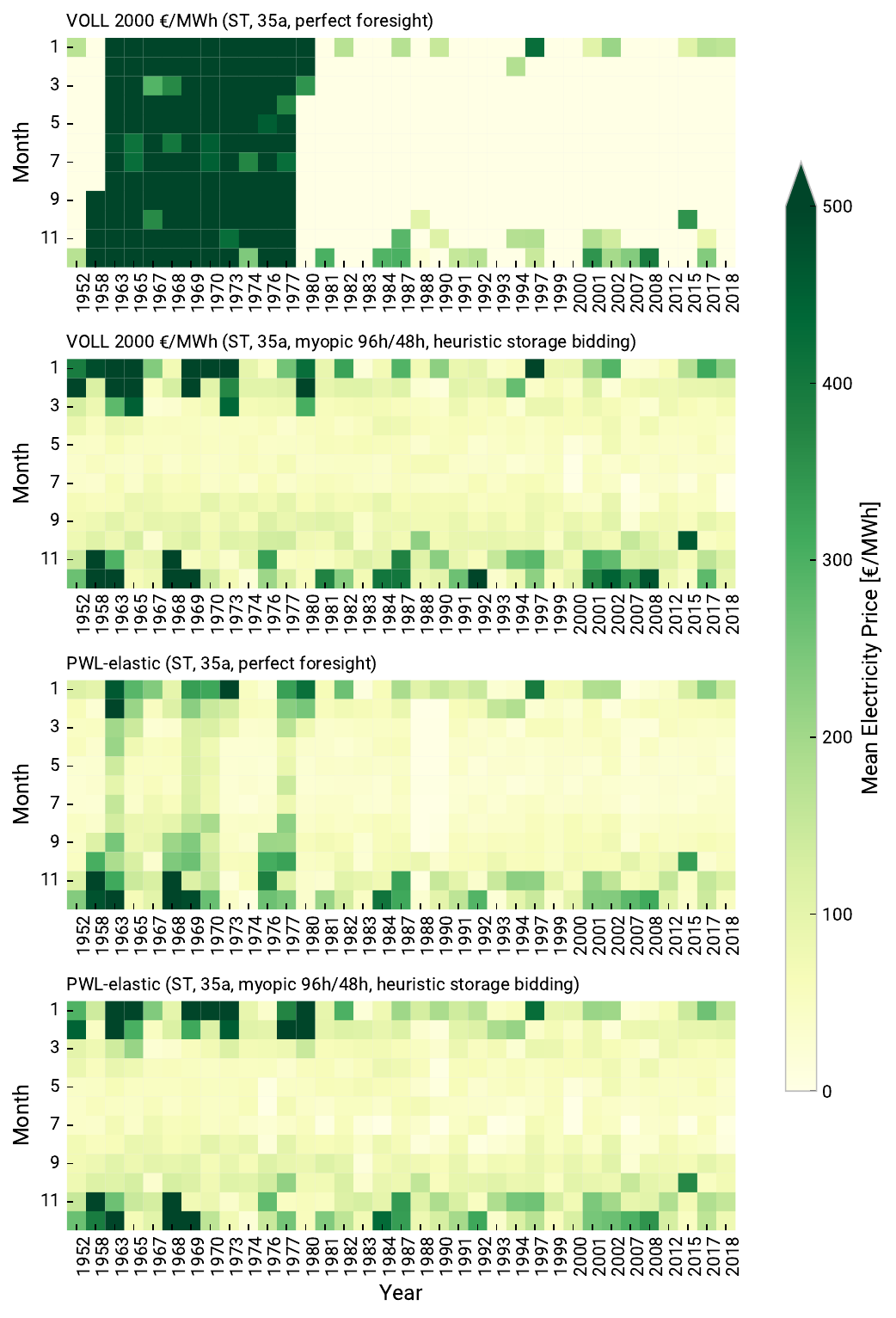}
  \caption{Heatmap of average monthly baseload prices over 35 unseen weather years for different demand modelling scenarios with perfect and myopic foresight with heuristic storage bidding. Years between 1951 and 2020 not indexed were used for the corresponding LT model.}
  \label{fig:si:heatmap-myopic}
\end{figure*}

\begin{figure*}[!htb]
  \centering
  \includegraphics[trim=0 0cm 0 0cm,width=0.8\linewidth,clip=true]{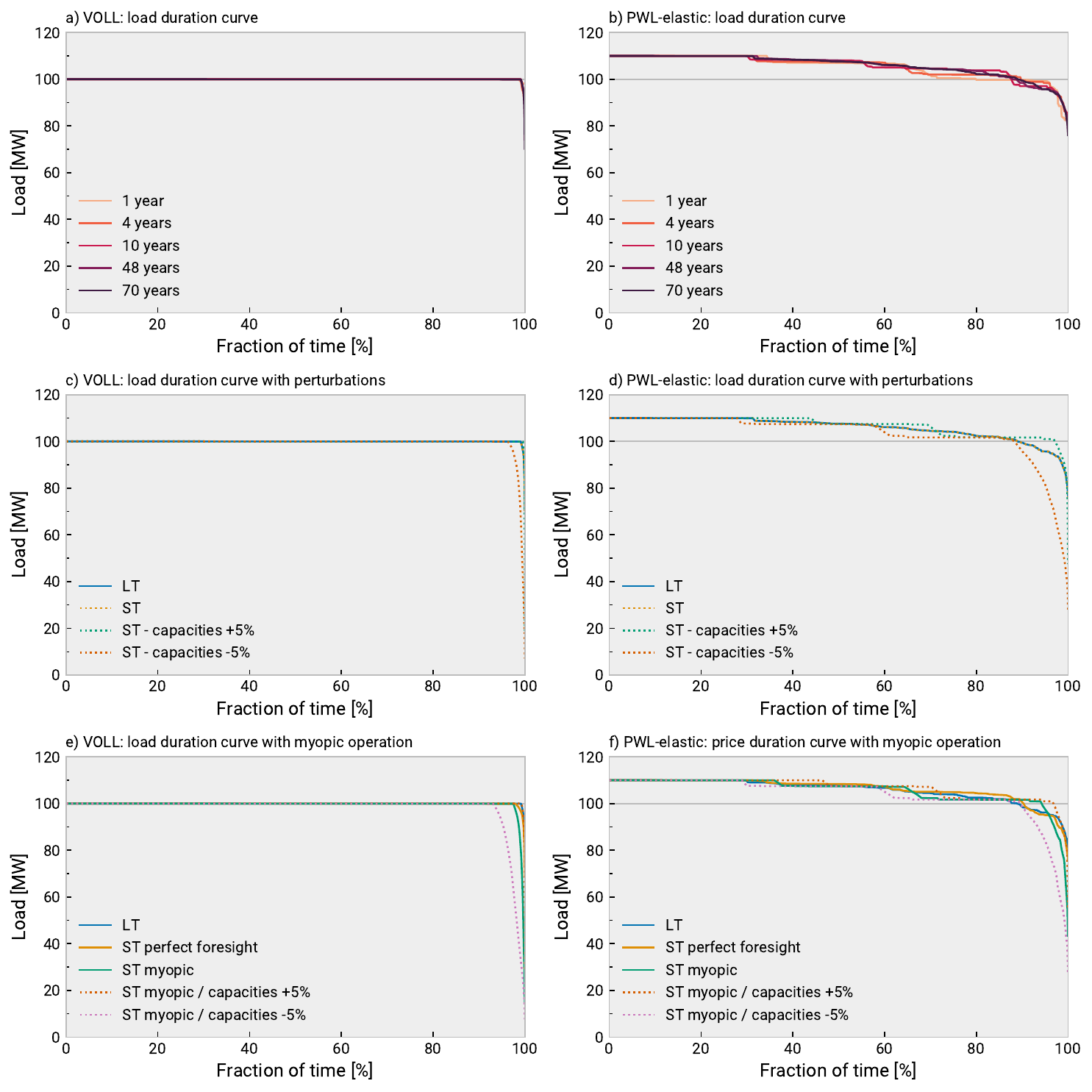}
  \caption{Load duration curves for Germany for cases analogous to Fig.~3.}
  \label{fig:load-duration}
\end{figure*}

\begin{figure*}[!htb]
  \centering
  \includegraphics[trim=0 0cm 0 0cm,width=0.8\linewidth,clip=true]{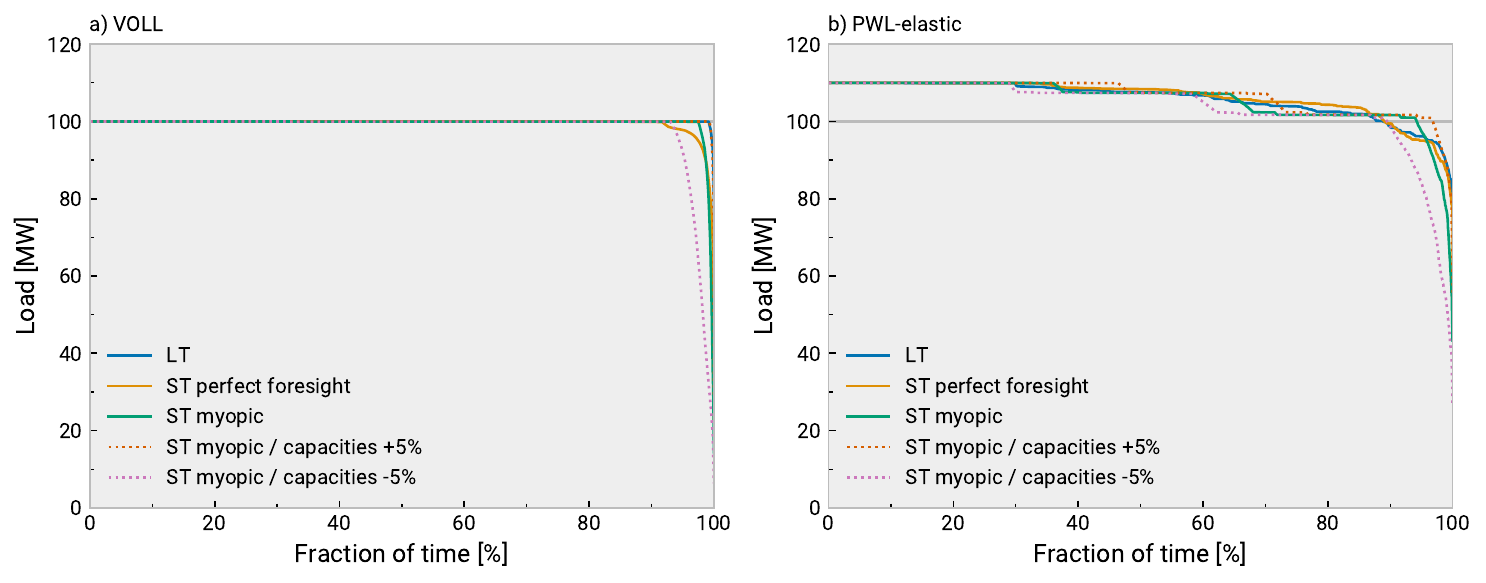}
  \caption{Load duration curves for Germany for cases with myopic foresight analogous to Fig.~6.}
  \label{fig:load-duration-myopia}
\end{figure*}

\begin{figure*}[!htb]
  \includegraphics[width=\textwidth]{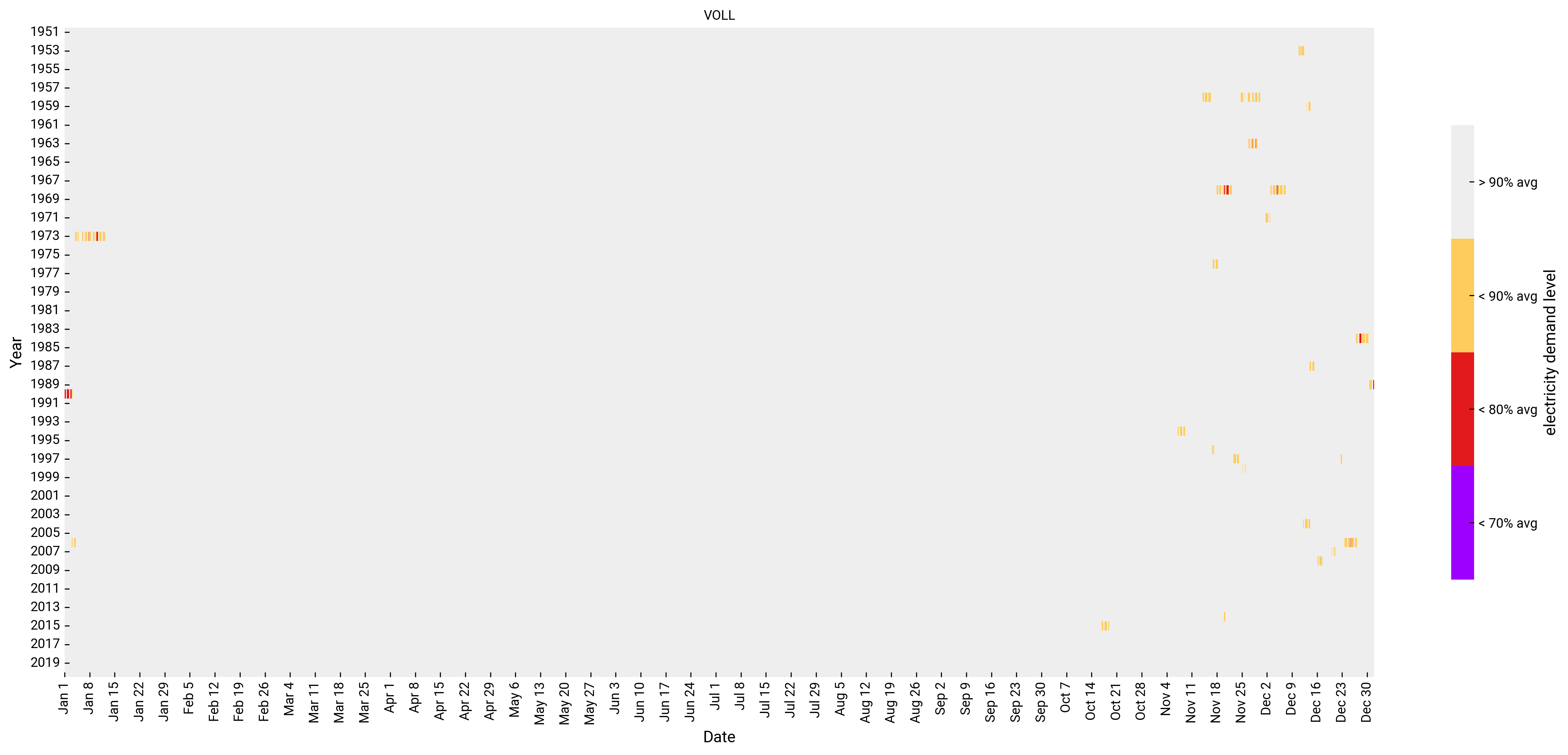}\\
  \includegraphics[width=\textwidth]{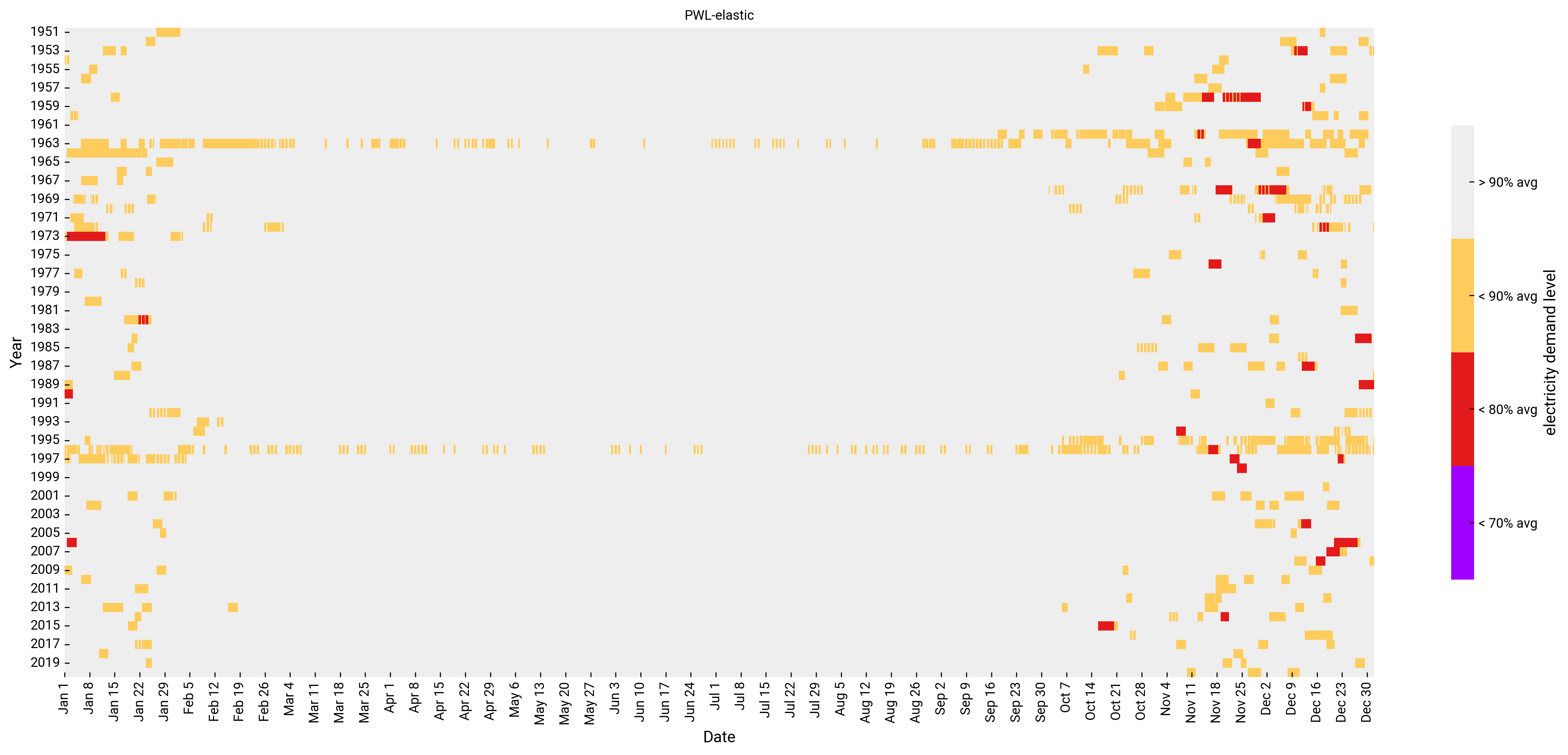}\\
  \caption{Time periods with demand reductions of 90\%, 80\%, 70\% and higher, shown for different demand modelling scenarios.}
  \label{fig:load-heatmap}
\end{figure*}

\begin{figure*}[!htb]
  \small
  \begin{tabular}{cc}
    a) VOLL & b) PWL-elastic \\
    \includegraphics[width=0.49\textwidth]{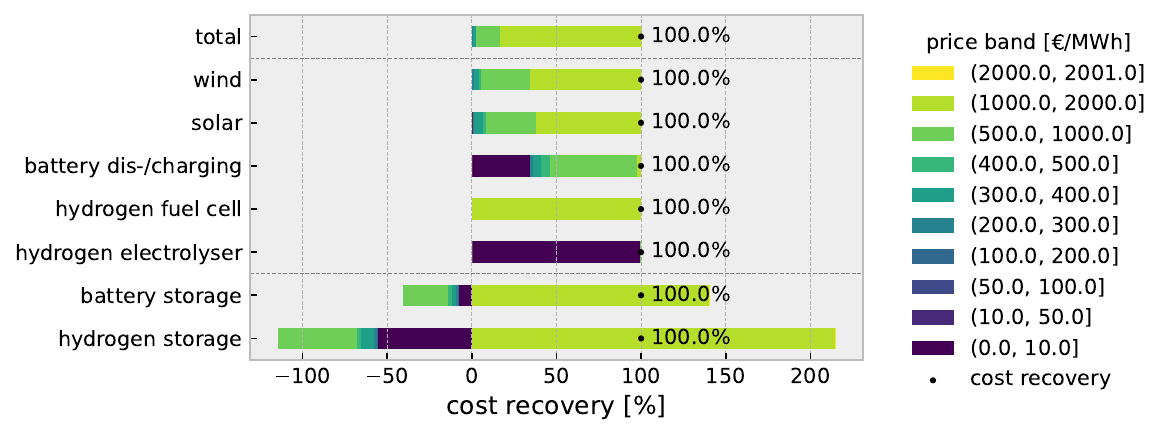}
            &
    \includegraphics[width=0.49\textwidth]{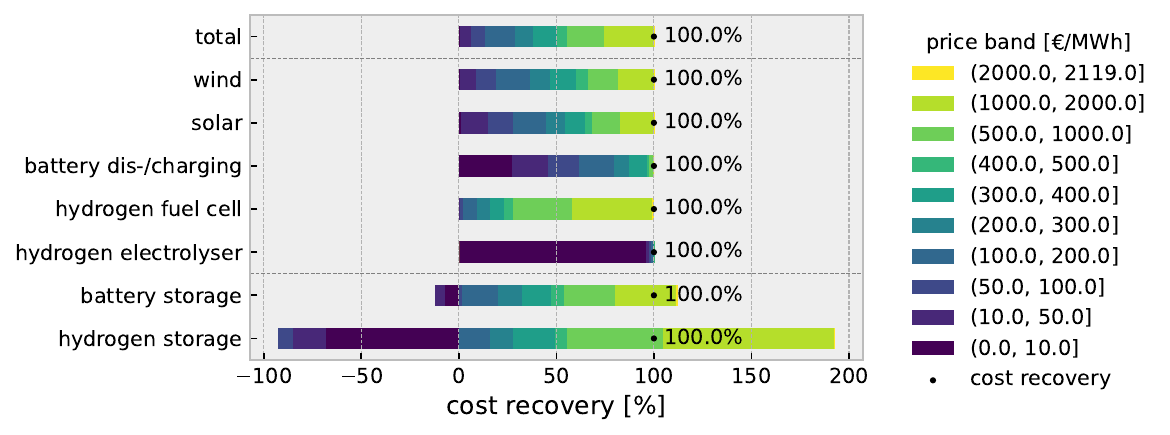}
    \\
  \end{tabular}
  \caption{Cost recovery of each component by price band. In the perfectly inelastic case, most of the revenue is made in the price band of 1000-2000~\euro/MWh, while in the elastic case the revenue is more evenly distributed across different price bands. In both cases, storage buys electricity in low-priced hours and sells it in high-priced hours.}
  \label{fig:price-bands}
\end{figure*}

\begin{figure*}[!htb]
    \includegraphics[width=\linewidth]{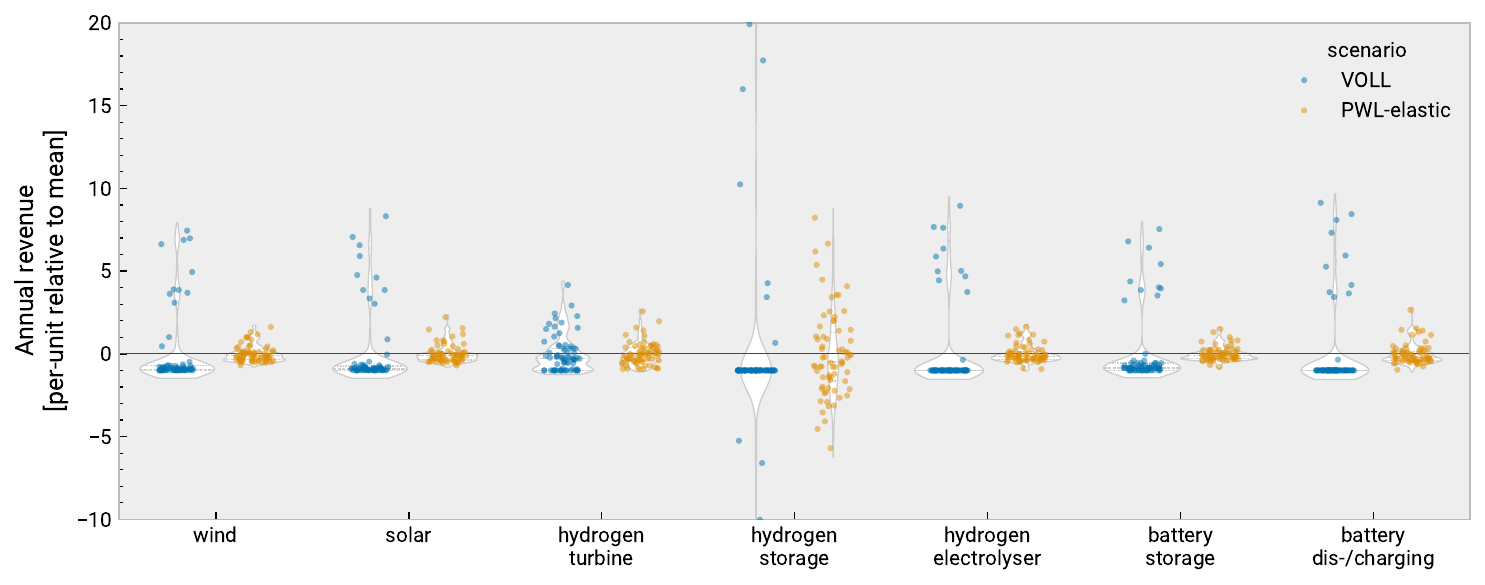}
  \caption{ Distribution of annual revenues per technology normalised to the
    long-term average revenue. Revenues shown are derived from the 70-year LT
    models with VOLL and PWL-elastic demand. Annual revenues show less
    year-to-year variation with demand elasticity, similar to average annual
    baseload prices in Fig.~4. Wind and solar revenues benefit from a natural
    volume-price hedge: higher availability lowers prices while lower
    availability raises them, resulting in even tighter revenue clustering.}
  \label{fig:annual-revenue}
\end{figure*}

\begin{figure*}[!htb]
  \small
  a) VOLL: perfect foresight \\
  \includegraphics[width=\textwidth]{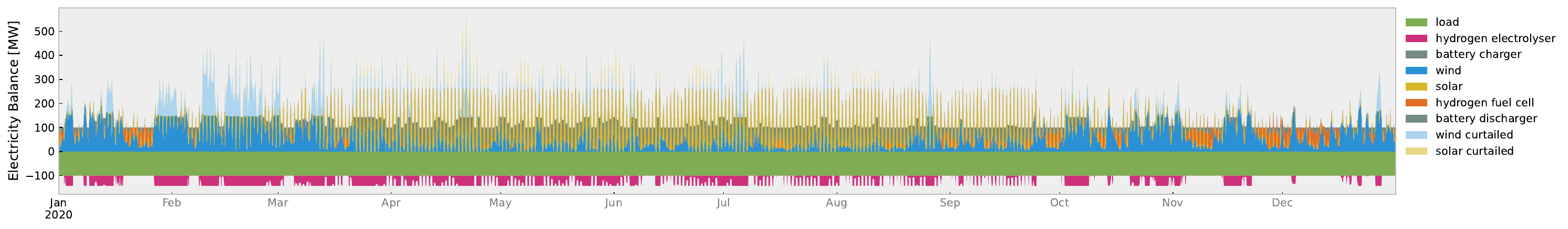}\\
  b) VOLL: myopic foresight of 96 hours with overlap of 48 hours and heuristic
  storage bidding \\
  \includegraphics[width=\textwidth]{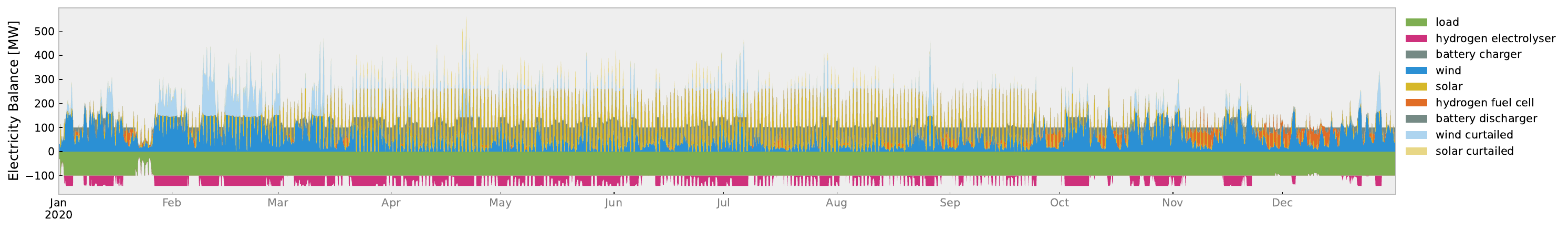}\\
  c) PWL-elastic: perfect foresight \\
  \includegraphics[width=\textwidth]{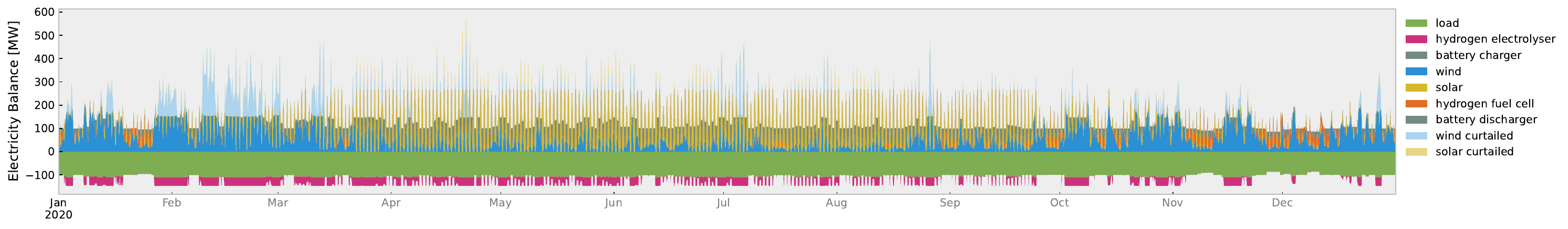}\\
  d) PWL-elastic: myopic foresight of 96 hours with overlap of 48 hours and
  heuristic storage bidding \\
  \includegraphics[width=\textwidth]{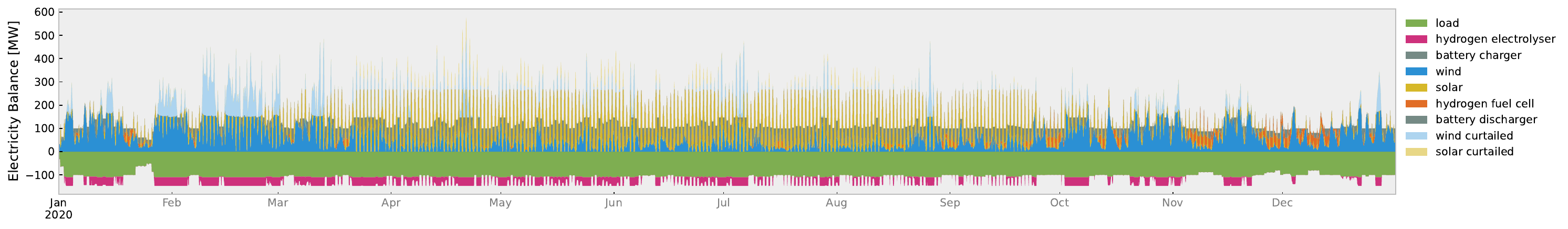}\\
  \caption{Electricity system balance time series for myopic and perfect operational foresight runs with elastic or inelastic demand for single example year 2020.}
  \label{fig:operation}
\end{figure*}

\begin{figure*}[!htb]
  \includegraphics[width=\textwidth]{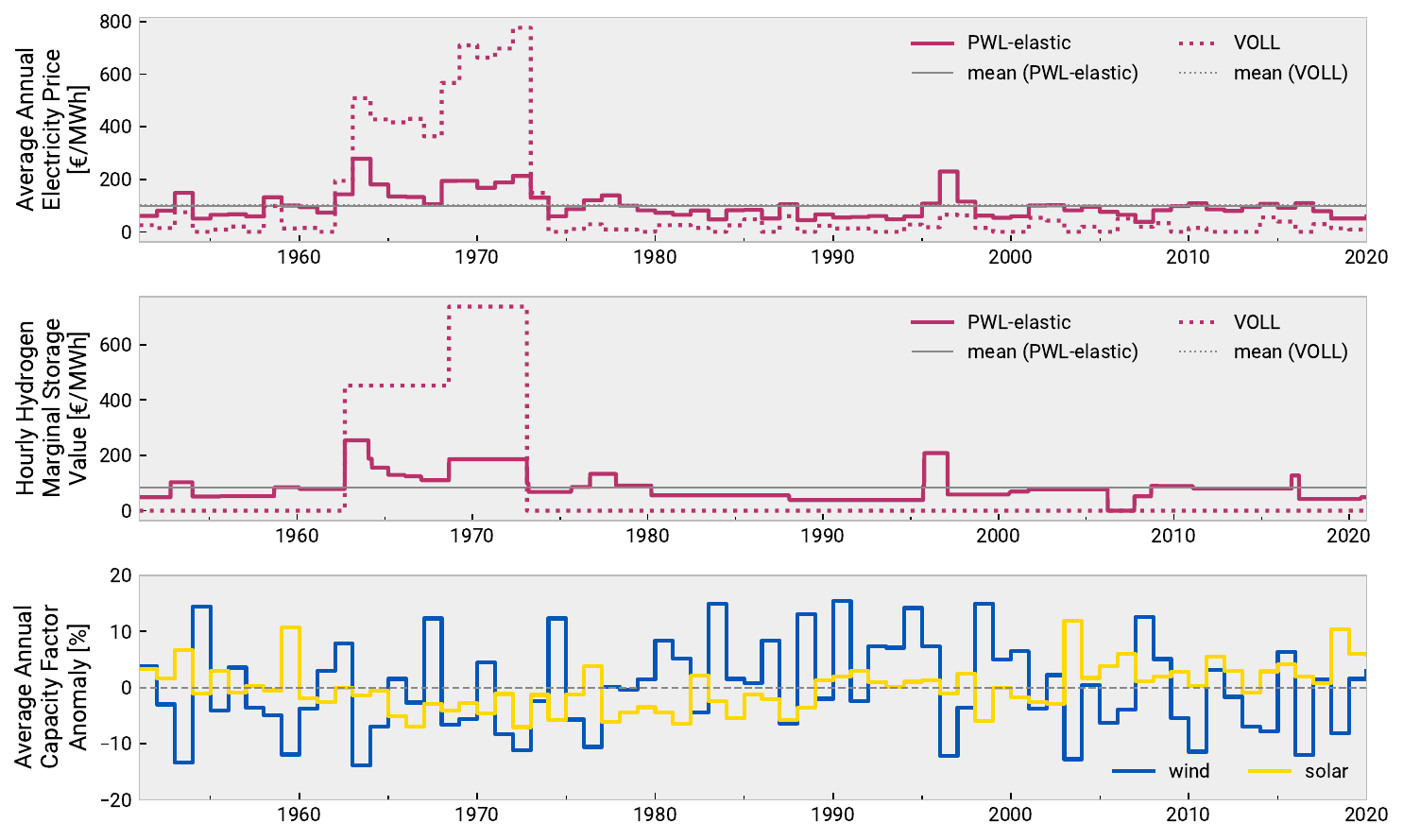}\\
  \caption{ Relation between average annual baseload price, hourly marginal
    storage values and average annual capacity factor anomaly of wind and solar
    (i.e.~relative deviation from long-term mean) for elastic and inelastic
    case. High marginal storage values are associated in particular with
    multi-annual low wind capacity factors, indicating scarcity.}
  \label{fig:msv-cf}
\end{figure*}

In this section we show in a simplified case for a single year how prices
duration curves are built up (\cref{fig:si:example}), a sensitivity analysis for
how price duration curves change if some dispatchable conventional is retained
in the market (\cref{fig:si:dispatchable}) or when some reserve storage
discharging capacity is procured out-of-market (\cref{fig:si:reserve}). We also
show the heatmaps of average monthly baseload prices for different demand
modelling scenarios with perfect and myopic foresight on 35 unseen test weather
years (\cref{fig:si:heatmap-myopic}). Load duration curves corresponding to the
scenarios shown in \cref{fig:price_duration,fig:myopic} respectively are shown
in \cref{fig:load-duration,fig:load-duration-myopia}. In
\cref{fig:load-heatmap}, we show in which time periods demand reductions of
varying magnitude are located. In \cref{fig:price-bands}, we show the revenues
of different system components by price band, similar to Mallapragada et
al.~\cite{mallapragada2023}. In \cref{fig:annual-revenue}, we show the
distribution of annual revenues for different system components. In
\cref{fig:operation}, we illustrate the operation of the electricity system as
balance time series for different demand modelling and operational foresight
cases. Finally, \cref{fig:msv-cf} shows the relation between hourly marginal
storage values and the average annual capacity factor anomaly of wind and solar.

\FloatBarrier
\newpage
\section{Main figures for different countries}
\label{sec:si:countries}

\begin{figure*}[!htb]
  \centering
  \includegraphics[width=\linewidth]{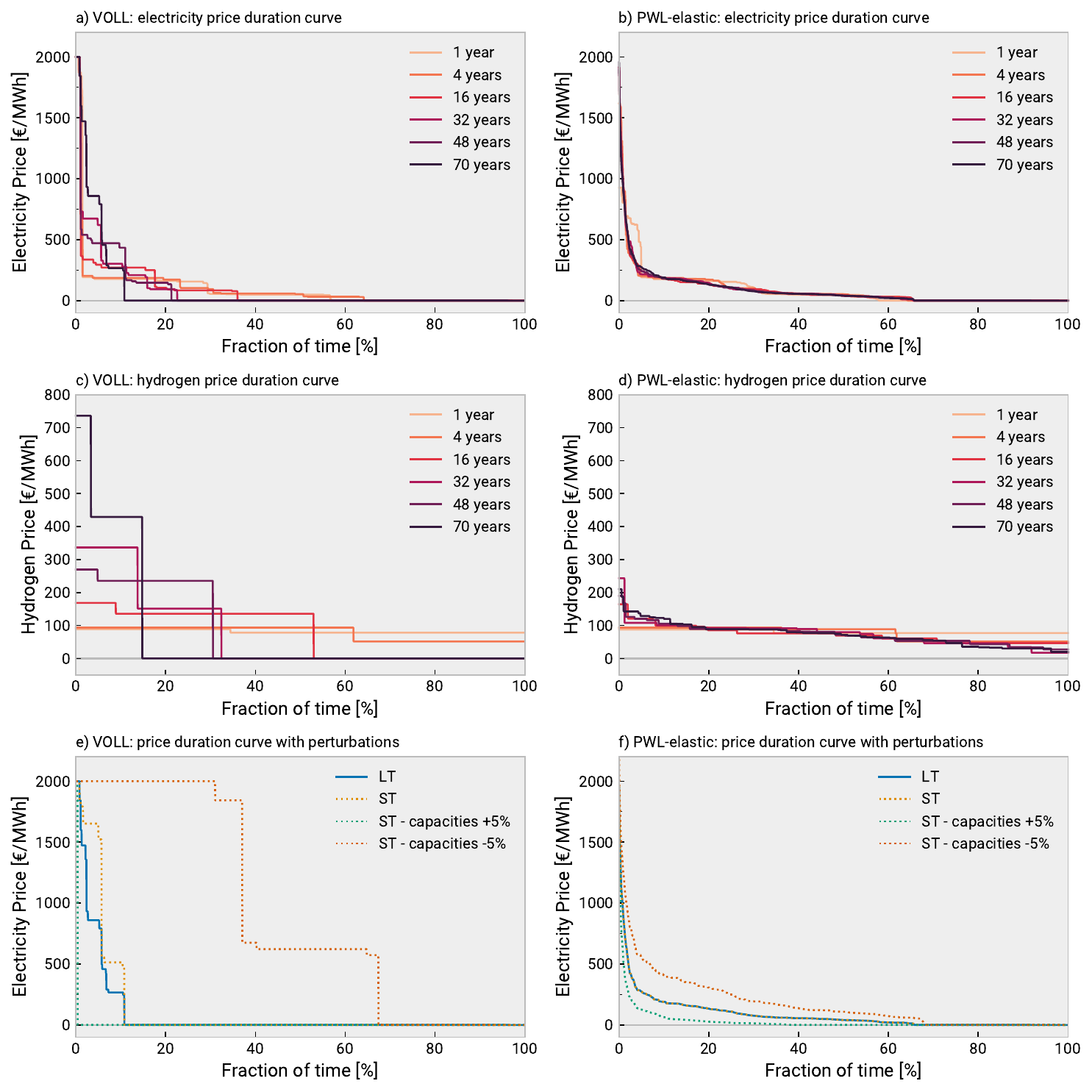}
  \caption{ Variant of Fig.~3 for Spain showing electricity and hydrogen price
    duration curves in long-term and short-term runs with perfect operational
    foresight.}
  \label{fig:si:price_duration_es}
\end{figure*}

\begin{figure*}[!htb]
  \centering
  \includegraphics[width=\linewidth]{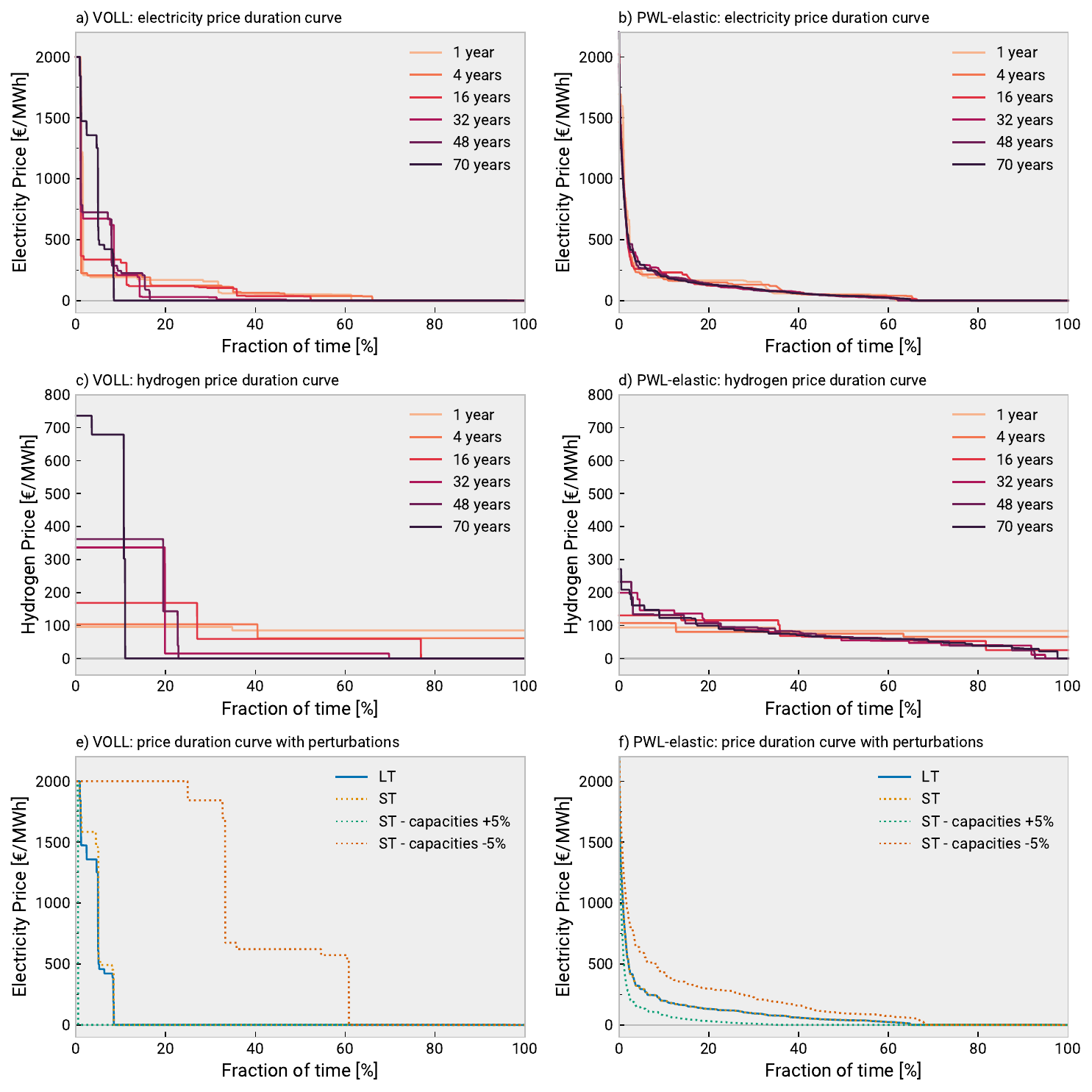}
  \caption{ Variant of Fig.~3 for United Kingdom showing electricity and
    hydrogen price duration curves in long-term and short-term runs with perfect
    operational foresight.}
  \label{fig:si:price_duration_uk}
\end{figure*}

\begin{figure*}[!htb]
  \small
  \begin{tabular}{cc}
    a) Spain                                                   & b) United Kingdom \\
    \includegraphics[width=0.45\linewidth]{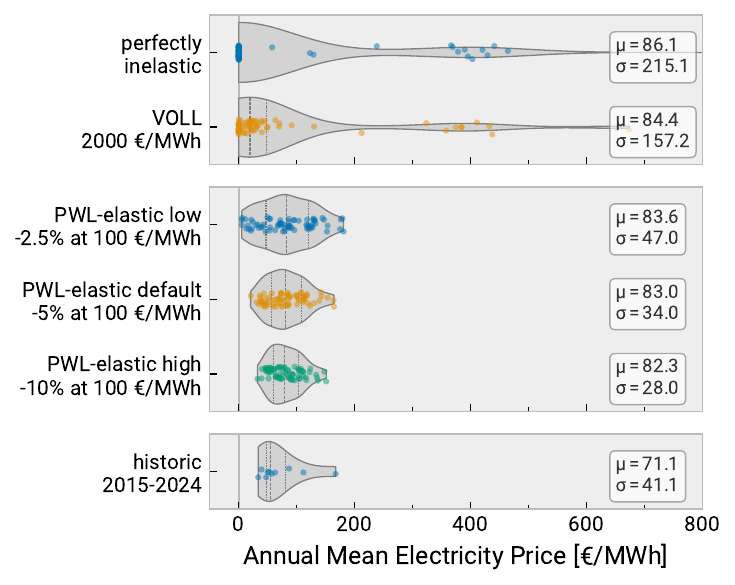} &
    \includegraphics[width=0.45\linewidth]{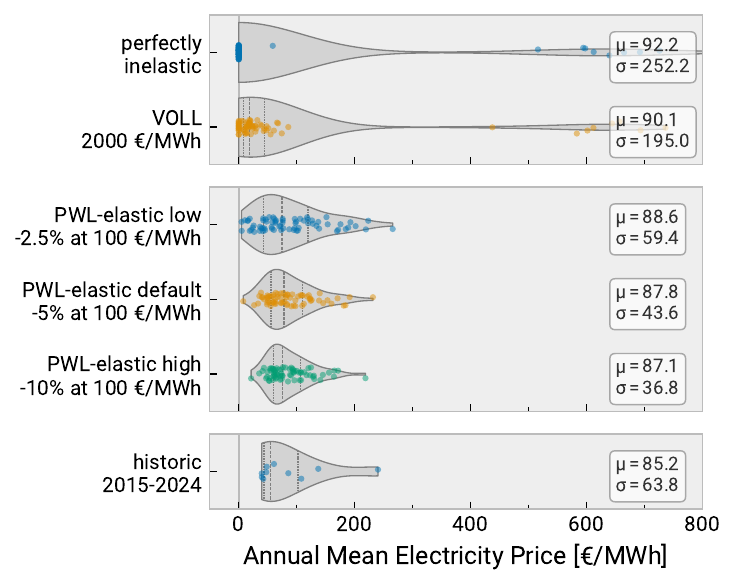}                     \\
  \end{tabular}
  \caption{Variant of Fig.~4 for different countries showing distribution of annual average baseload prices.}
  \label{fig:si:boxplots}
\end{figure*}

\begin{figure*}[!htb]
  \small
  \begin{tabular}{cc}
    a) Spain                                                  & b) United Kingdom \\
    \includegraphics[width=0.5\linewidth]{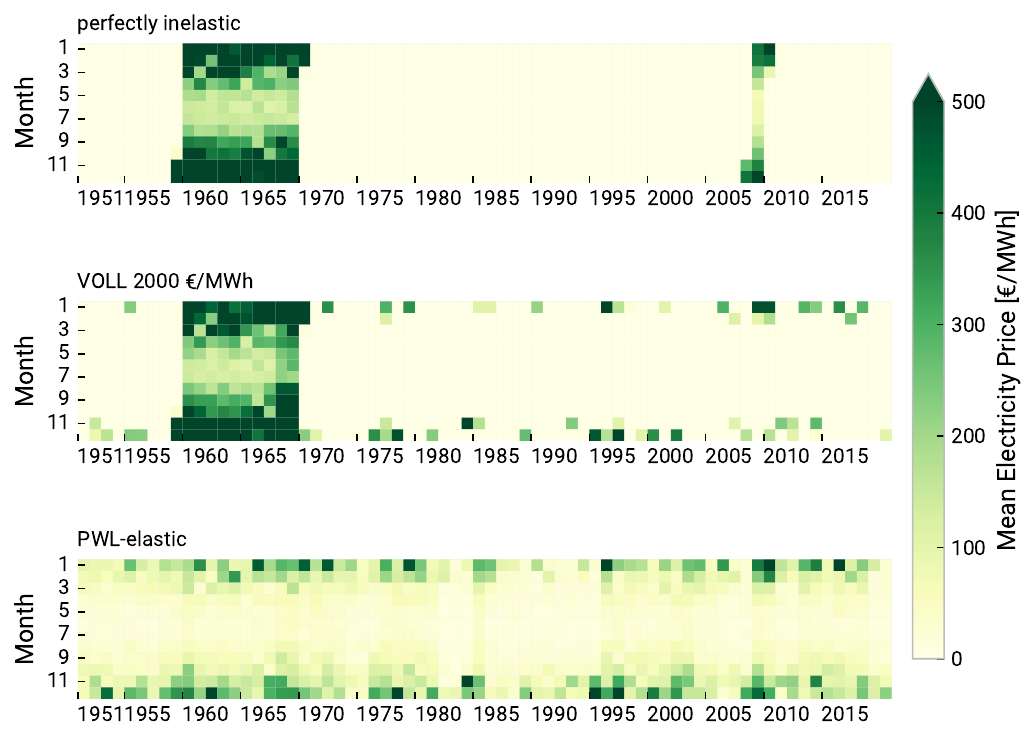} &
    \includegraphics[width=0.5\linewidth]{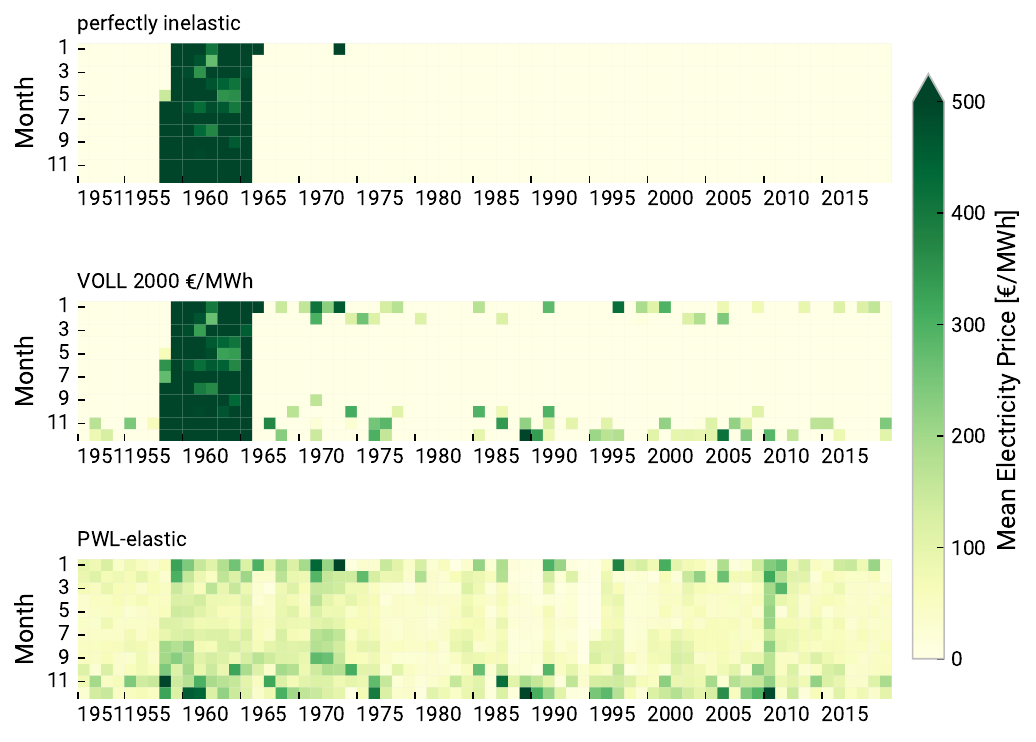}                     \\
  \end{tabular}
  \caption{ Variant of Fig.~5 for different countries showing heatmap of monthly
    average baseload prices.}
  \label{fig:si:heatmap}
\end{figure*}

\begin{figure*}[!htb]
  \small
  \centering
  a) Spain\\
  \includegraphics[width=0.8\linewidth]{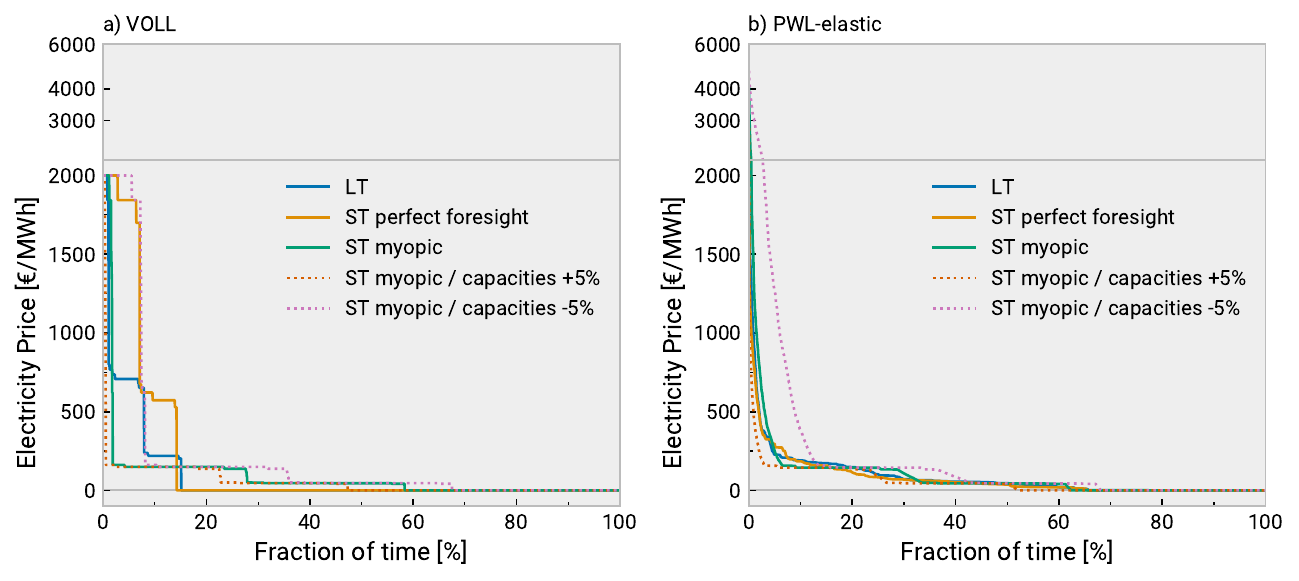}\\
  b) United Kingdom\\
  \includegraphics[width=0.8\linewidth]{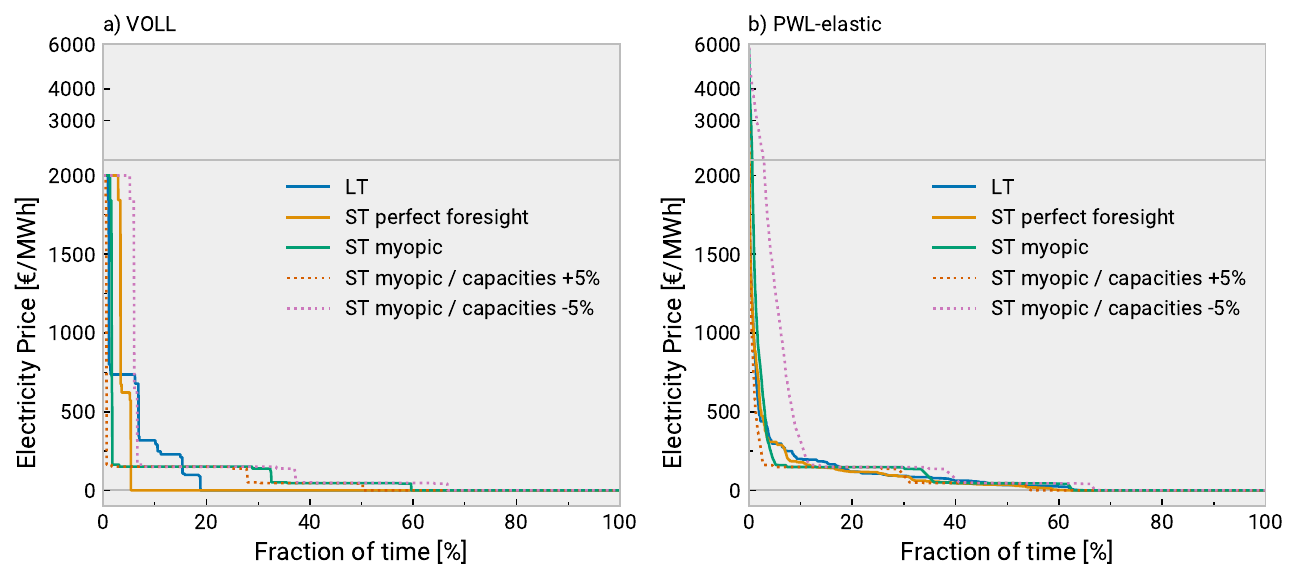}\\
  \caption{Variant of Fig.~6 for different countries showing price duration curves for runs with myopic foresight.}
  \label{fig:si:myopic}
\end{figure*}

\begin{figure*}[!htb]
  \small
  \centering
  a) Spain\\
  \includegraphics[width=0.66\linewidth]{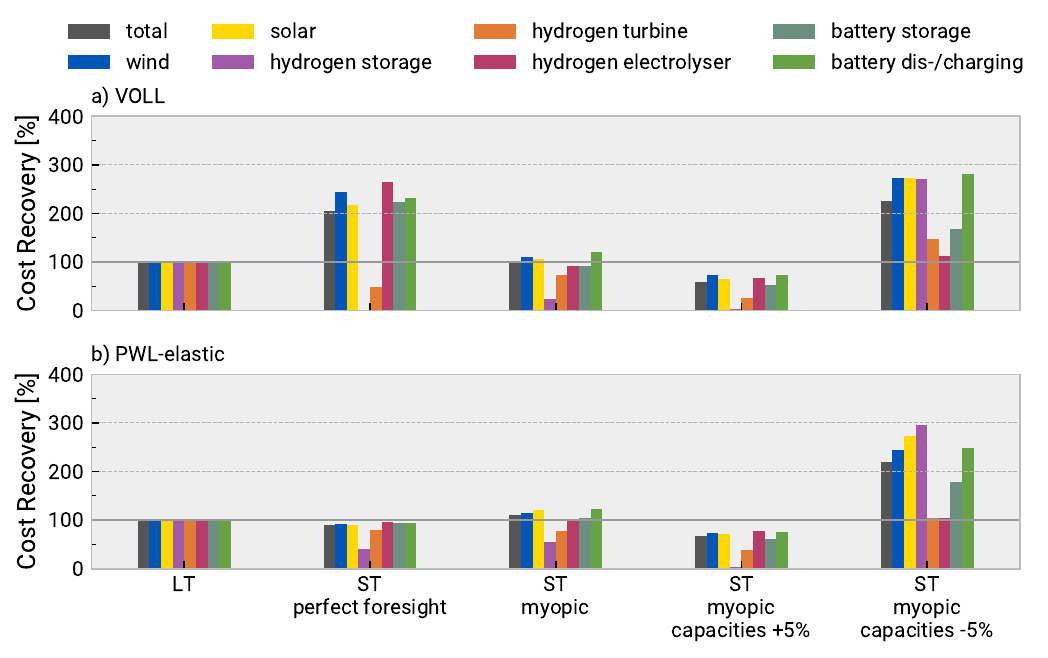}\\
  b) United Kingdom\\
  \includegraphics[width=0.66\linewidth]{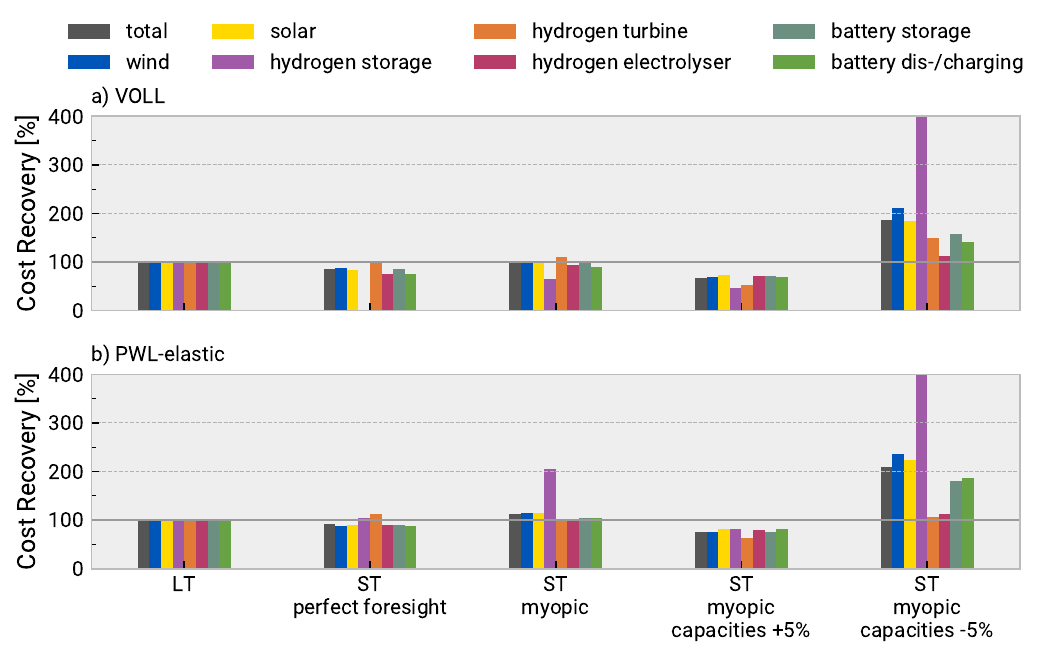}\\
  \caption{ Variant of Fig.~7 for different countries showing cost recovery
    factors.}
  \label{fig:si:crf}
\end{figure*}

\begin{figure*}[!htb]
  \includegraphics[width=\linewidth]{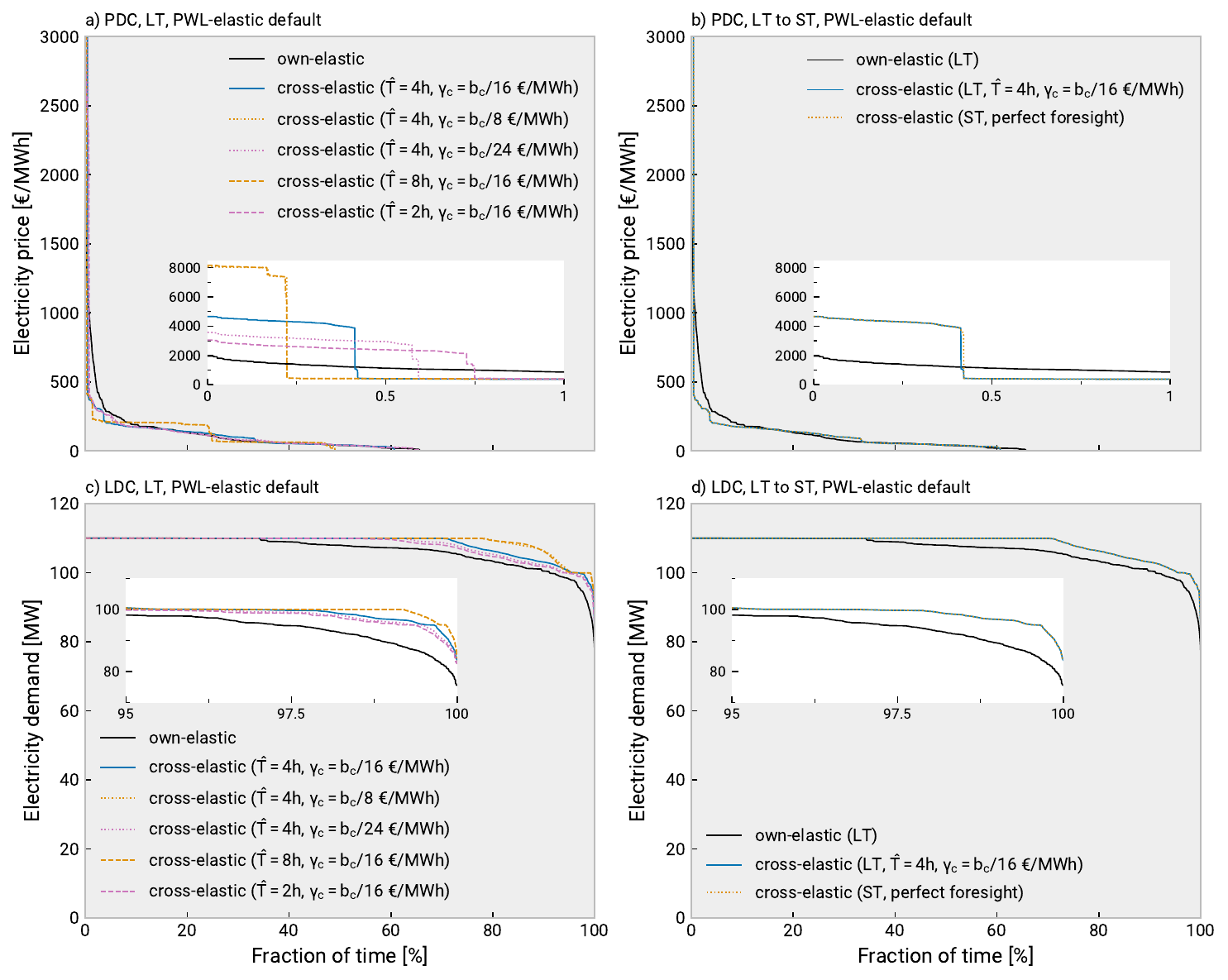}
  \caption{Variant of Fig.~S1 for Spain showing price and load duration curves for scenarios with cross-elasticity terms.}
  \label{fig:cross-elasticity_es}
\end{figure*}

\begin{figure*}[!htb]
  \includegraphics[width=\linewidth]{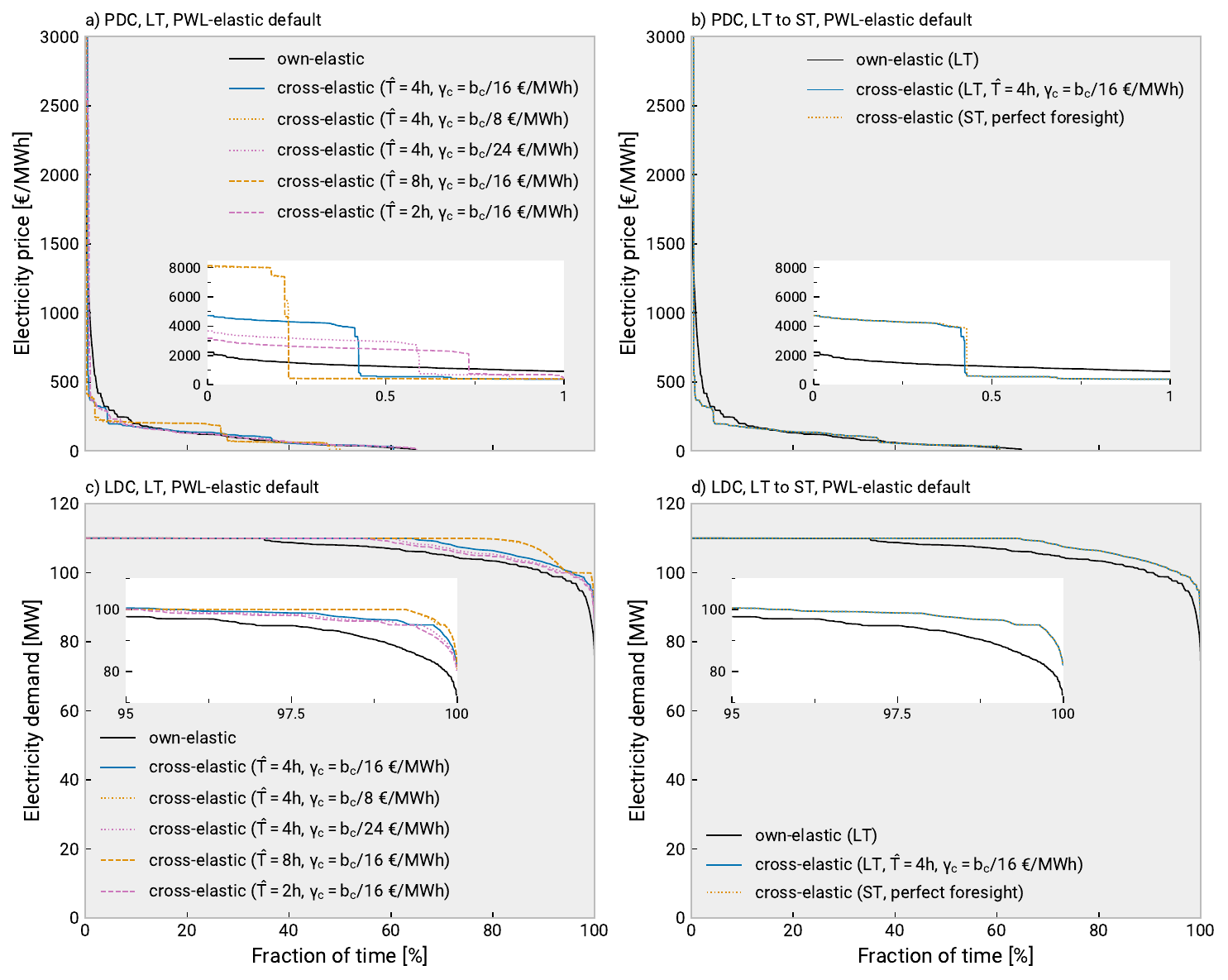}
  \caption{Variant of Fig.~S1 for the United Kingdom showing price and load duration curves for scenarios with cross-elasticity terms.}
  \label{fig:cross-elasticity_uk}
\end{figure*}

\begin{figure*}[!htb]
  \small
  \centering
  a) Spain\\
  \includegraphics[width=\linewidth]{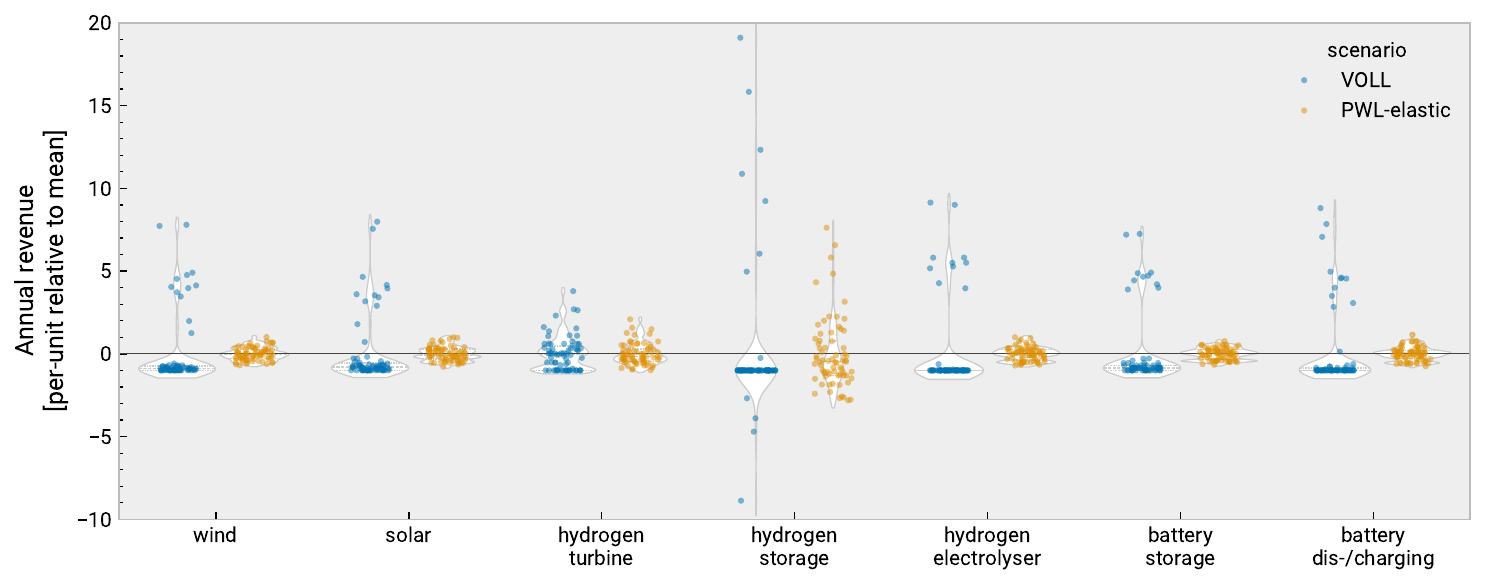}\\
  b) United Kingdom\\
  \includegraphics[width=\linewidth]{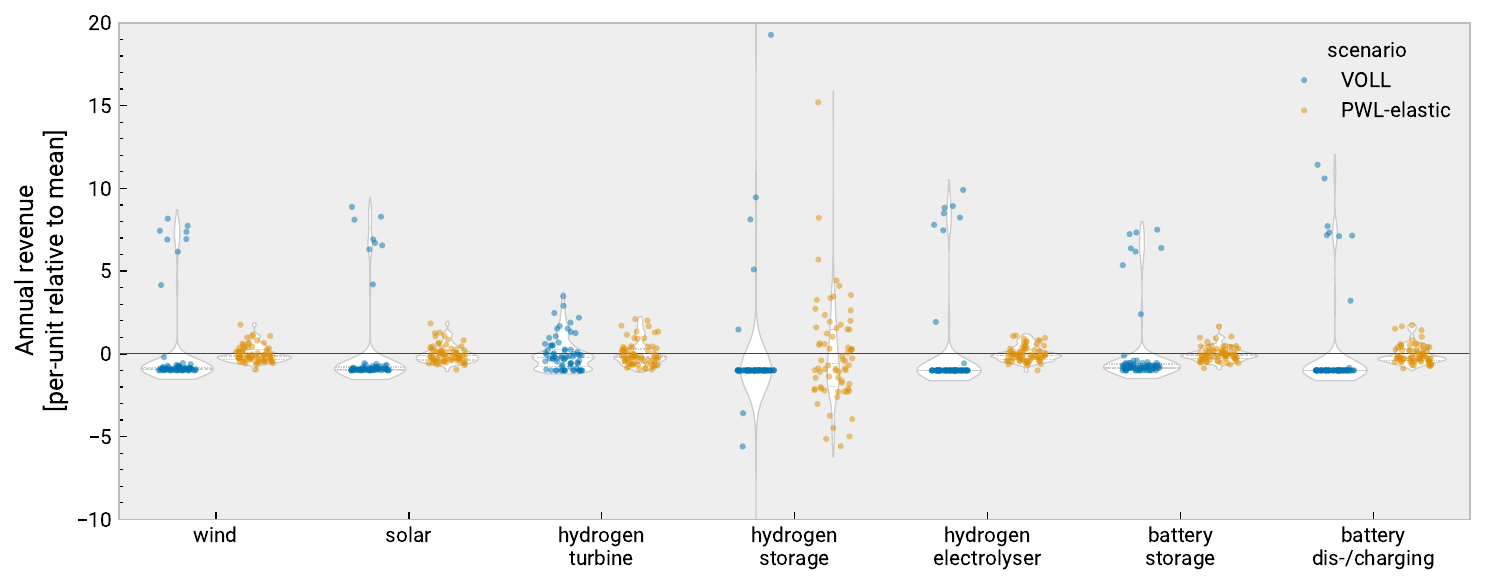}
\caption{Variant of Fig.~S9 showing the distribution of annual revenues per technology for Spain and the United Kingdom.}
\label{fig:annual-revenue-es-uk}
\end{figure*}

In this section, we replicate the paper's main figures for different countries,
namely Spain and the United Kingdom.
\cref{fig:si:price_duration_es,fig:si:price_duration_uk} correspond to the price
duration curves shown for Germany in \cref{fig:price_duration}.
\cref{fig:si:boxplots,fig:si:heatmap} replicate the distribution of average
annual and monthly baseload prices shown for Germany in
\cref{fig:interannual,fig:heatmaps}. The comparison of price duration curves and
cost recovery factors in different short-term models shown in
\cref{fig:si:myopic,fig:si:crf} for Spain and the United Kingdom correspond to
the comparisons shown for Germany in \cref{fig:myopic,fig:cost_recovery}.
\cref{fig:cross-elasticity_es,fig:cross-elasticity_uk} show the price and load
duration curves for scenarios with cross-elasticity terms.
\cref{fig:annual-revenue-es-uk} shows the distribution of annual revenues per
technology.

\FloatBarrier
\section{Additional tables}

The tables shown in this section collect a range of metrics for the different
cases examined. These metrics include, among other aspects, statistics on
prices, load served, system cost, energy mix and capacities built. The tables
cover the full 70-year optimisations for Germany (\cref{tab:metrics-70a-de}),
Spain (\cref{tab:metrics-70a-es}), and the United Kingdom
(\cref{tab:metrics-70a-uk}), as well as the model runs where the long-term model
was run on 35 different weather years than the 35 weather years the subsequent
short-term model was run on for Germany (\cref{tab:metrics-35a-35a-de}), Spain
(\cref{tab:metrics-35a-35a-es}), and the United Kingdom
(\cref{tab:metrics-35a-35a-uk}). Additionally, \cref{tab:metrics-reserve}
provides statistics for the cases where reserve capacity was forced into the
system in \cref{fig:si:reserve}

\begin{landscape}
  \begin{table}
    \begin{footnotesize}
      \begin{tabular}{lrrrrrrrrrrr}
\toprule
 & \shortstack[r]{perfectly\\inelastic,\\LT,\\70a} & \shortstack[r]{VOLL,\\LT,\\70a} & \shortstack[r]{PWL-elastic\\default,\\LT,\\70a} & \shortstack[r]{PWL-elastic\\higher,\\LT,\\70a} & \shortstack[r]{PWL-elastic\\lower,\\LT,\\70a} & \shortstack[r]{VOLL,\\ST,\\70a,\\PF} & \shortstack[r]{VOLL,\\ST,\\70a,\\PF,\\C+5\%} & \shortstack[r]{VOLL,\\ST,\\70a,\\PF,\\C-5\%} & \shortstack[r]{PWL-elastic,\\ST,\\70a,\\PF} & \shortstack[r]{PWL-elastic,\\ST,\\70a,\\PF,\\C+5\%} & \shortstack[r]{PWL-elastic,\\ST,\\70a,\\PF,\\C-5\%} \\
\midrule
system costs (bn€/period) & 6.31 & 6.15 & 5.92 & 5.59 & 6.16 & 6.15 & 6.46 & 5.84 & 5.92 & 6.21 & 5.62 \\
utility (bn€/period) & -- & 122.67 & 468.53 & 234.03 & 937.41 & 122.67 & 122.71 & 121.91 & 468.53 & 468.72 & 468.09 \\
welfare (bn€/period) & -- & 116.52 & 462.61 & 228.43 & 931.25 & 116.52 & 116.25 & 116.07 & 462.61 & 462.50 & 462.47 \\
average system costs (€/MWh) & 102.82 & 100.30 & 91.02 & 88.37 & 93.16 & 100.30 & 105.26 & 95.89 & 91.02 & 93.31 & 89.18 \\
average load served (MW) & 100.00 & 99.95 & 105.94 & 103.14 & 107.77 & 99.95 & 99.99 & 99.34 & 105.94 & 108.51 & 102.73 \\
peak load shedding (MW) & 0.00 & 29.44 & 33.92 & 41.53 & 27.84 & 29.85 & 20.67 & 47.08 & 33.92 & 30.04 & 37.80 \\
primary energy (TWh/period) & 72.46 & 72.43 & 76.62 & 73.82 & 78.56 & 72.41 & 70.95 & 73.03 & 76.62 & 78.86 & 73.78 \\
wind share (\%) & 45.37 & 46.88 & 50.51 & 50.81 & 50.31 & 46.85 & 43.34 & 49.68 & 50.51 & 49.76 & 50.93 \\
solar share (\%) & 54.63 & 53.12 & 49.49 & 49.19 & 49.69 & 53.15 & 56.66 & 50.32 & 49.49 & 50.24 & 49.07 \\
wind market value (€/MWh) & 66.50 & 66.95 & 59.69 & 59.40 & 59.95 & 63.01 & 2.90 & 393.28 & 59.69 & 19.38 & 126.40 \\
solar market value (€/MWh) & 42.20 & 41.96 & 41.58 & 41.51 & 41.65 & 39.70 & 1.84 & 262.22 & 41.58 & 13.11 & 89.86 \\
wind capacity factor (\%) & 21.21 & 21.21 & 21.21 & 21.21 & 21.21 & 21.21 & 21.21 & 21.21 & 21.21 & 21.21 & 21.21 \\
solar capacity factor (\%) & 11.67 & 11.67 & 11.67 & 11.67 & 11.67 & 11.67 & 11.67 & 11.67 & 11.67 & 11.67 & 11.67 \\
hydrogen consumed (TWh/period) & 9.16 & 9.19 & 9.54 & 8.59 & 10.26 & 9.18 & 7.83 & 10.03 & 9.54 & 10.12 & 8.78 \\
curtailment (\%) & 17.11 & 17.55 & 12.59 & 12.33 & 12.83 & 17.58 & 23.08 & 12.49 & 12.59 & 14.33 & 11.41 \\
wind capacity (MW) & 351.56 & 365.62 & 371.43 & 358.19 & 380.94 & 365.62 & 383.91 & 347.34 & 371.43 & 390.00 & 352.86 \\
solar capacity (MW) & 581.65 & 562.10 & 548.98 & 524.76 & 566.08 & 562.10 & 590.21 & 534.00 & 548.98 & 576.43 & 521.53 \\
electrolyser capacity (MW) & 63.00 & 62.88 & 55.62 & 50.41 & 59.31 & 62.88 & 66.03 & 59.74 & 55.62 & 58.40 & 52.84 \\
fuel cell capacity (MW) & 69.54 & 57.15 & 46.41 & 40.15 & 51.55 & 57.15 & 60.01 & 54.29 & 46.41 & 48.73 & 44.09 \\
battery inverter capacity (MW) & 111.54 & 107.77 & 107.57 & 102.79 & 110.63 & 107.77 & 113.16 & 102.38 & 107.57 & 112.95 & 102.19 \\
battery storage capacity (GWh) & 0.81 & 0.79 & 0.80 & 0.76 & 0.82 & 0.79 & 0.83 & 0.75 & 0.80 & 0.84 & 0.76 \\
hydrogen storage capacity (GWh) & 196.68 & 181.00 & 112.22 & 95.02 & 135.45 & 181.00 & 190.05 & 171.95 & 112.22 & 117.83 & 106.61 \\
mean electricity price (€/MWh) & 102.82 & 101.20 & 98.02 & 97.34 & 98.74 & 96.37 & 8.67 & 570.98 & 98.02 & 35.85 & 199.60 \\
mean hydrogen price (€/MWh) & 87.22 & 85.48 & 82.81 & 82.22 & 83.43 & 80.71 & 0.32 & 568.06 & 82.81 & 21.99 & 183.71 \\
STD electricity price (€/MWh) & 1587.53 & 338.16 & 166.02 & 145.49 & 192.35 & 318.84 & 127.95 & 792.10 & 166.02 & 109.00 & 255.33 \\
STD hydrogen price (€/MWh) & 214.16 & 211.13 & 49.72 & 43.10 & 66.48 & 192.59 & 0.00 & 485.59 & 49.72 & 31.32 & 84.41 \\
mean hydrogen MSV (€/MWh) & 87.22 & 85.48 & 82.81 & 82.22 & 83.43 & 80.71 & 0.32 & 568.06 & 82.81 & 21.99 & 183.71 \\
mean battery MSV (€/MWh) & 103.30 & 101.86 & 98.87 & 98.24 & 99.56 & 96.97 & 8.47 & 577.10 & 98.87 & 35.82 & 202.06 \\
STD hydrogen MSV (€/MWh) & 214.16 & 211.13 & 49.72 & 43.10 & 66.48 & 192.59 & 0.00 & 485.59 & 49.72 & 31.32 & 84.41 \\
STD battery MSV (€/MWh) & 1548.41 & 333.05 & 161.81 & 141.67 & 187.83 & 313.69 & 124.84 & 773.25 & 161.81 & 106.56 & 248.13 \\
\bottomrule
\end{tabular}

      \caption{ Metrics for 70-year optimisations for Germany. \tablelegend }
      \label{tab:metrics-70a-de}
    \end{footnotesize}
  \end{table}
\end{landscape}

\begin{landscape}
  \begin{table}
    \begin{footnotesize}
      \begin{tabular}{lrrrrrrrrrrrr}
\toprule
 & \shortstack[r]{VOLL,\\LT,\\35a} & \shortstack[r]{VOLL,\\ST,\\35a-35a,\\PF} & \shortstack[r]{VOLL,\\ST,\\35a-35a,\\myopic\\48/24} & \shortstack[r]{VOLL,\\ST,\\35a-35a,\\myopic\\96/48} & \shortstack[r]{VOLL,\\ST,\\35a-35a,\\myopic\\96/48,\\C+5\%} & \shortstack[r]{VOLL,\\ST,\\35a-35a,\\myopic\\96/48,\\C-5\%} & \shortstack[r]{PWL-elastic\\default,\\LT,\\35a} & \shortstack[r]{PWL-elastic\\default,\\ST,\\35a-35a,\\PF} & \shortstack[r]{PWL-elastic\\default,\\ST,\\35a-35a,\\myopic\\48/24} & \shortstack[r]{PWL-elastic\\default,\\ST,\\35a-35a,\\myopic\\96/48} & \shortstack[r]{PWL-elastic\\default,\\ST,\\35a-35a,\\myopic\\96/48,\\C+5\%} & \shortstack[r]{PWL-elastic\\default,\\ST,\\35a-35a,\\myopic\\96/48,\\C-5\%} \\
\midrule
system costs (bn€/period) & 3.04 & 3.04 & 3.04 & 3.04 & 3.19 & 2.89 & 2.96 & 2.96 & 2.96 & 2.96 & 3.11 & 2.81 \\
utility (bn€/period) & 61.34 & 61.07 & 61.04 & 61.06 & 61.33 & 60.11 & 233.02 & 233.04 & 232.83 & 232.87 & 233.07 & 232.18 \\
welfare (bn€/period) & 58.29 & 58.03 & 58.00 & 58.02 & 58.14 & 57.22 & 230.06 & 230.08 & 229.87 & 229.91 & 229.96 & 229.37 \\
average system costs (€/MWh) & 99.22 & 99.65 & 99.70 & 99.67 & 104.18 & 96.19 & 91.17 & 90.81 & 91.53 & 91.46 & 95.03 & 88.54 \\
average load served (MW) & 99.96 & 99.51 & 99.47 & 99.50 & 99.94 & 97.94 & 105.91 & 106.32 & 105.47 & 105.56 & 106.67 & 103.60 \\
peak load shedding (MW) & 23.96 & 77.55 & 88.31 & 85.60 & 71.19 & 93.69 & 29.90 & 58.09 & 71.59 & 66.89 & 55.49 & 83.30 \\
primary energy (TWh/period) & 36.67 & 36.11 & 36.58 & 36.61 & 36.31 & 35.54 & 38.20 & 38.45 & 37.74 & 37.82 & 37.86 & 37.03 \\
wind share (\%) & 48.54 & 48.58 & 49.43 & 49.45 & 48.09 & 50.10 & 49.94 & 51.07 & 50.51 & 50.53 & 49.22 & 51.51 \\
solar share (\%) & 51.46 & 51.42 & 50.57 & 50.55 & 51.91 & 49.90 & 50.06 & 48.93 & 49.49 & 49.47 & 50.78 & 48.49 \\
wind market value (€/MWh) & 61.82 & 232.08 & 78.52 & 79.79 & 51.07 & 151.40 & 60.92 & 55.51 & 74.68 & 74.56 & 51.55 & 131.31 \\
solar market value (€/MWh) & 41.77 & 155.51 & 50.63 & 51.01 & 31.47 & 91.78 & 41.36 & 39.36 & 52.01 & 50.45 & 33.56 & 86.85 \\
wind capacity factor (\%) & 20.73 & 21.69 & 21.69 & 21.69 & 21.69 & 21.69 & 20.73 & 21.69 & 21.69 & 21.69 & 21.69 & 21.69 \\
solar capacity factor (\%) & 11.78 & 11.56 & 11.56 & 11.56 & 11.56 & 11.56 & 11.78 & 11.56 & 11.56 & 11.56 & 11.56 & 11.56 \\
hydrogen consumed (TWh/period) & 4.95 & 4.61 & 4.59 & 4.61 & 4.16 & 4.50 & 4.67 & 4.78 & 4.28 & 4.34 & 4.02 & 4.26 \\
curtailment (\%) & 13.20 & 15.89 & 14.80 & 14.74 & 19.46 & 12.87 & 12.34 & 13.26 & 14.85 & 14.67 & 18.66 & 12.07 \\
wind capacity (MW) & 353.33 & 353.33 & 353.33 & 353.33 & 370.99 & 335.66 & 373.18 & 373.18 & 373.18 & 373.18 & 391.84 & 354.52 \\
solar capacity (MW) & 547.19 & 547.19 & 547.19 & 547.19 & 574.55 & 519.83 & 549.18 & 549.18 & 549.18 & 549.18 & 576.64 & 521.72 \\
electrolyser capacity (MW) & 55.60 & 55.60 & 55.60 & 55.60 & 58.38 & 52.82 & 54.67 & 54.67 & 54.67 & 54.67 & 57.41 & 51.94 \\
fuel cell capacity (MW) & 57.29 & 57.29 & 57.29 & 57.29 & 60.16 & 54.43 & 45.51 & 45.51 & 45.51 & 45.51 & 47.78 & 43.23 \\
battery inverter capacity (MW) & 107.09 & 107.09 & 107.09 & 107.09 & 112.44 & 101.73 & 108.72 & 108.72 & 108.72 & 108.72 & 114.16 & 103.28 \\
battery storage capacity (GWh) & 0.79 & 0.79 & 0.79 & 0.79 & 0.83 & 0.75 & 0.81 & 0.81 & 0.81 & 0.81 & 0.85 & 0.77 \\
hydrogen storage capacity (GWh) & 317.68 & 317.68 & 317.68 & 317.68 & 333.56 & 301.79 & 98.61 & 98.61 & 98.61 & 98.61 & 103.54 & 93.67 \\
mean electricity price (€/MWh) & 99.93 & 338.29 & 136.56 & 131.32 & 76.89 & 235.93 & 97.80 & 95.39 & 131.73 & 126.23 & 79.41 & 236.87 \\
mean hydrogen price (€/MWh) & 83.23 & 348.63 & 95.00 & 98.89 & 68.79 & 158.86 & 83.04 & 75.35 & 99.25 & 97.32 & 67.20 & 170.19 \\
STD electricity price (€/MWh) & 270.09 & 676.68 & 356.55 & 343.06 & 197.83 & 532.42 & 158.09 & 192.05 & 368.50 & 340.12 & 179.46 & 651.32 \\
STD hydrogen price (€/MWh) & 133.64 & 476.34 & 120.83 & 130.99 & 56.45 & 252.28 & 45.96 & 69.05 & 234.24 & 192.18 & 78.31 & 391.13 \\
mean hydrogen MSV (€/MWh) & 83.23 & 348.63 & 95.00 & 98.89 & 68.79 & 158.86 & 83.04 & 75.35 & 99.25 & 97.32 & 67.20 & 170.19 \\
mean battery MSV (€/MWh) & 100.54 & 342.01 & 136.15 & 131.19 & 77.15 & 234.52 & 98.65 & 95.94 & 131.92 & 126.52 & 79.80 & 236.19 \\
STD hydrogen MSV (€/MWh) & 133.64 & 476.34 & 120.83 & 130.99 & 56.45 & 252.28 & 45.96 & 69.05 & 234.24 & 192.18 & 78.31 & 391.13 \\
STD battery MSV (€/MWh) & 264.70 & 664.97 & 347.52 & 334.82 & 193.29 & 519.41 & 153.88 & 187.77 & 360.79 & 333.22 & 176.12 & 637.72 \\
\bottomrule
\end{tabular}

      \caption{Metrics for 35-year LT plus 35-year ST optimisations for Germany. \tablelegend}
      \label{tab:metrics-35a-35a-de}
    \end{footnotesize}
  \end{table}
\end{landscape}

\begin{landscape}
  \begin{table}
    \begin{footnotesize}
      \begin{tabular}{lrrrrrrrrrrr}
\toprule
 & \shortstack[r]{perfectly\\inelastic,\\LT,\\70a} & \shortstack[r]{VOLL,\\LT,\\70a} & \shortstack[r]{PWL-elastic\\default,\\LT,\\70a} & \shortstack[r]{PWL-elastic\\higher,\\LT,\\70a} & \shortstack[r]{PWL-elastic\\lower,\\LT,\\70a} & \shortstack[r]{VOLL,\\ST,\\70a,\\PF} & \shortstack[r]{VOLL,\\ST,\\70a,\\PF,\\C+5\%} & \shortstack[r]{VOLL,\\ST,\\70a,\\PF,\\C-5\%} & \shortstack[r]{PWL-elastic,\\ST,\\70a,\\PF} & \shortstack[r]{PWL-elastic,\\ST,\\70a,\\PF,\\C+5\%} & \shortstack[r]{PWL-elastic,\\ST,\\70a,\\PF,\\C-5\%} \\
\midrule
system costs (bn€/period) & 5.28 & 5.13 & 5.03 & 4.76 & 5.23 & 5.13 & 5.38 & 4.87 & 5.03 & 5.28 & 4.78 \\
utility (bn€/period) & -- & 122.67 & 468.58 & 234.09 & 937.46 & 122.67 & 122.71 & 120.80 & 468.58 & 468.74 & 468.20 \\
welfare (bn€/period) & -- & 117.55 & 463.55 & 229.33 & 932.22 & 117.55 & 117.33 & 115.93 & 463.55 & 463.46 & 463.42 \\
average system costs (€/MWh) & 86.08 & 83.59 & 76.93 & 74.56 & 78.87 & 83.59 & 87.72 & 80.64 & 76.93 & 79.06 & 75.23 \\
average load served (MW) & 100.00 & 99.96 & 106.51 & 104.05 & 108.11 & 99.96 & 99.99 & 98.43 & 106.51 & 108.82 & 103.48 \\
peak load shedding (MW) & 0.00 & 26.93 & 34.45 & 41.56 & 28.65 & 25.38 & 15.85 & 48.31 & 34.45 & 30.59 & 38.30 \\
primary energy (TWh/period) & 71.28 & 71.22 & 74.95 & 72.44 & 76.72 & 71.22 & 69.89 & 69.97 & 74.95 & 76.90 & 72.32 \\
wind share (\%) & 19.86 & 20.35 & 21.22 & 21.29 & 21.21 & 20.35 & 15.67 & 22.12 & 21.22 & 20.27 & 21.52 \\
solar share (\%) & 80.14 & 79.65 & 78.78 & 78.71 & 78.79 & 79.65 & 84.33 & 77.88 & 78.78 & 79.73 & 78.48 \\
wind market value (€/MWh) & 68.20 & 68.87 & 62.44 & 62.14 & 62.70 & 107.37 & 3.33 & 771.76 & 62.44 & 19.54 & 136.77 \\
solar market value (€/MWh) & 30.67 & 30.92 & 30.86 & 30.79 & 30.91 & 45.98 & 2.15 & 343.25 & 30.86 & 10.48 & 65.50 \\
wind capacity factor (\%) & 21.86 & 21.86 & 21.86 & 21.86 & 21.86 & 21.86 & 21.86 & 21.86 & 21.86 & 21.86 & 21.86 \\
solar capacity factor (\%) & 17.79 & 17.79 & 17.79 & 17.79 & 17.79 & 17.79 & 17.79 & 17.79 & 17.79 & 17.79 & 17.79 \\
hydrogen consumed (TWh/period) & 7.19 & 7.18 & 6.76 & 5.91 & 7.43 & 7.18 & 5.89 & 6.96 & 6.76 & 7.17 & 6.15 \\
curtailment (\%) & 18.40 & 19.20 & 17.29 & 17.05 & 17.46 & 19.20 & 24.49 & 16.43 & 17.29 & 19.18 & 15.99 \\
wind capacity (MW) & 155.29 & 160.57 & 159.67 & 154.09 & 164.04 & 160.57 & 168.60 & 152.54 & 159.67 & 167.66 & 151.69 \\
solar capacity (MW) & 609.38 & 610.12 & 633.95 & 610.61 & 649.88 & 610.12 & 640.62 & 579.61 & 633.95 & 665.65 & 602.26 \\
electrolyser capacity (MW) & 43.04 & 42.59 & 37.40 & 32.96 & 40.92 & 42.59 & 44.72 & 40.46 & 37.40 & 39.27 & 35.53 \\
fuel cell capacity (MW) & 60.75 & 47.83 & 35.44 & 29.90 & 40.19 & 47.83 & 50.22 & 45.43 & 35.44 & 37.21 & 33.66 \\
battery inverter capacity (MW) & 163.63 & 160.49 & 169.64 & 163.85 & 173.27 & 160.49 & 168.51 & 152.47 & 169.64 & 178.12 & 161.16 \\
battery storage capacity (GWh) & 1.19 & 1.16 & 1.22 & 1.17 & 1.24 & 1.16 & 1.22 & 1.10 & 1.22 & 1.28 & 1.16 \\
hydrogen storage capacity (GWh) & 157.50 & 153.86 & 101.03 & 85.88 & 121.12 & 153.86 & 161.55 & 146.17 & 101.03 & 106.08 & 95.98 \\
mean electricity price (€/MWh) & 86.08 & 84.41 & 82.97 & 82.35 & 83.57 & 123.86 & 7.69 & 920.32 & 82.97 & 29.18 & 173.21 \\
mean hydrogen price (€/MWh) & 76.02 & 73.93 & 72.15 & 71.66 & 72.82 & 122.22 & 0.32 & 1000.00 & 72.15 & 16.03 & 167.91 \\
STD electricity price (€/MWh) & 1863.44 & 302.45 & 155.24 & 134.53 & 180.90 & 406.96 & 120.37 & 846.20 & 155.24 & 102.50 & 235.92 \\
STD hydrogen price (€/MWh) & 171.61 & 183.59 & 33.42 & 28.53 & 48.49 & 292.90 & 0.00 & 0.00 & 33.42 & 20.18 & 57.18 \\
mean hydrogen MSV (€/MWh) & 76.02 & 73.93 & 72.15 & 71.66 & 72.82 & 122.22 & 0.32 & 1000.00 & 72.15 & 16.03 & 167.91 \\
mean battery MSV (€/MWh) & 86.91 & 85.32 & 83.95 & 83.36 & 84.54 & 125.57 & 7.57 & 939.07 & 83.95 & 29.24 & 175.91 \\
STD hydrogen MSV (€/MWh) & 171.61 & 183.59 & 33.42 & 28.53 & 48.49 & 292.90 & 0.00 & 0.00 & 33.42 & 20.18 & 57.18 \\
STD battery MSV (€/MWh) & 1819.54 & 298.92 & 151.94 & 131.67 & 177.17 & 403.39 & 117.94 & 815.61 & 151.94 & 100.52 & 230.33 \\
\bottomrule
\end{tabular}

      \caption{Metrics for 70-year optimisations for Spain. \tablelegend}
      \label{tab:metrics-70a-es}
    \end{footnotesize}
  \end{table}
\end{landscape}

\begin{landscape}
  \begin{table}
    \begin{footnotesize}
      \begin{tabular}{lrrrrrrrrrrrr}
\toprule
 & \shortstack[r]{VOLL,\\LT,\\35a} & \shortstack[r]{VOLL,\\ST,\\35a-35a,\\PF} & \shortstack[r]{VOLL,\\ST,\\35a-35a,\\myopic\\48/24} & \shortstack[r]{VOLL,\\ST,\\35a-35a,\\myopic\\96/48} & \shortstack[r]{VOLL,\\ST,\\35a-35a,\\myopic\\96/48,\\C+5\%} & \shortstack[r]{VOLL,\\ST,\\35a-35a,\\myopic\\96/48,\\C-5\%} & \shortstack[r]{PWL-elastic\\default,\\LT,\\35a} & \shortstack[r]{PWL-elastic\\default,\\ST,\\35a-35a,\\PF} & \shortstack[r]{PWL-elastic\\default,\\ST,\\35a-35a,\\myopic\\48/24} & \shortstack[r]{PWL-elastic\\default,\\ST,\\35a-35a,\\myopic\\96/48} & \shortstack[r]{PWL-elastic\\default,\\ST,\\35a-35a,\\myopic\\96/48,\\C+5\%} & \shortstack[r]{PWL-elastic\\default,\\ST,\\35a-35a,\\myopic\\96/48,\\C-5\%} \\
\midrule
system costs (bn€/period) & 2.58 & 2.58 & 2.58 & 2.58 & 2.71 & 2.45 & 2.53 & 2.53 & 2.53 & 2.53 & 2.66 & 2.40 \\
utility (bn€/period) & 61.33 & 61.27 & 61.24 & 61.27 & 61.35 & 60.63 & 233.04 & 233.10 & 232.97 & 233.01 & 233.13 & 232.46 \\
welfare (bn€/period) & 58.75 & 58.69 & 58.66 & 58.69 & 58.64 & 58.18 & 230.51 & 230.57 & 230.44 & 230.48 & 230.48 & 230.06 \\
average system costs (€/MWh) & 84.15 & 84.23 & 84.27 & 84.24 & 88.32 & 80.88 & 77.44 & 77.19 & 77.55 & 77.52 & 80.59 & 75.17 \\
average load served (MW) & 99.95 & 99.85 & 99.80 & 99.84 & 99.97 & 98.79 & 106.46 & 106.80 & 106.30 & 106.35 & 107.39 & 104.19 \\
peak load shedding (MW) & 21.37 & 59.26 & 87.81 & 89.76 & 63.61 & 91.42 & 33.58 & 34.76 & 69.07 & 61.80 & 44.91 & 68.80 \\
primary energy (TWh/period) & 35.56 & 35.29 & 35.41 & 35.44 & 34.94 & 35.12 & 37.47 & 37.59 & 37.19 & 37.23 & 37.27 & 36.27 \\
wind share (\%) & 18.65 & 17.76 & 19.62 & 19.64 & 18.24 & 20.58 & 21.06 & 21.05 & 20.73 & 20.76 & 19.63 & 21.35 \\
solar share (\%) & 81.35 & 82.24 & 80.38 & 80.36 & 81.76 & 79.42 & 78.94 & 78.95 & 79.27 & 79.24 & 80.37 & 78.65 \\
wind market value (€/MWh) & 70.09 & 181.32 & 76.95 & 73.38 & 55.92 & 166.32 & 62.75 & 58.16 & 71.35 & 73.47 & 52.15 & 148.04 \\
solar market value (€/MWh) & 31.42 & 68.04 & 36.76 & 33.92 & 21.91 & 84.43 & 31.08 & 28.13 & 38.34 & 37.40 & 22.97 & 83.55 \\
wind capacity factor (\%) & 21.80 & 21.93 & 21.93 & 21.93 & 21.93 & 21.93 & 21.80 & 21.93 & 21.93 & 21.93 & 21.93 & 21.93 \\
solar capacity factor (\%) & 17.82 & 17.76 & 17.76 & 17.76 & 17.76 & 17.76 & 17.82 & 17.76 & 17.76 & 17.76 & 17.76 & 17.76 \\
hydrogen consumed (TWh/period) & 3.51 & 3.30 & 3.27 & 3.30 & 2.75 & 3.45 & 3.38 & 3.39 & 3.10 & 3.13 & 2.82 & 2.96 \\
curtailment (\%) & 19.99 & 20.52 & 20.27 & 20.19 & 25.06 & 16.76 & 17.62 & 17.30 & 18.18 & 18.07 & 21.89 & 15.99 \\
wind capacity (MW) & 149.31 & 149.31 & 149.31 & 149.31 & 156.78 & 141.85 & 158.97 & 158.97 & 158.97 & 158.97 & 166.92 & 151.03 \\
solar capacity (MW) & 630.50 & 630.50 & 630.50 & 630.50 & 662.02 & 598.97 & 637.65 & 637.65 & 637.65 & 637.65 & 669.53 & 605.76 \\
electrolyser capacity (MW) & 41.89 & 41.89 & 41.89 & 41.89 & 43.99 & 39.80 & 37.32 & 37.32 & 37.32 & 37.32 & 39.18 & 35.45 \\
fuel cell capacity (MW) & 48.34 & 48.34 & 48.34 & 48.34 & 50.75 & 45.92 & 36.13 & 36.13 & 36.13 & 36.13 & 37.94 & 34.33 \\
battery inverter capacity (MW) & 168.33 & 168.33 & 168.33 & 168.33 & 176.74 & 159.91 & 171.02 & 171.02 & 171.02 & 171.02 & 179.57 & 162.47 \\
battery storage capacity (GWh) & 1.19 & 1.19 & 1.19 & 1.19 & 1.25 & 1.13 & 1.22 & 1.22 & 1.22 & 1.22 & 1.28 & 1.16 \\
hydrogen storage capacity (GWh) & 141.25 & 141.25 & 141.25 & 141.25 & 148.31 & 134.19 & 104.02 & 104.02 & 104.02 & 104.02 & 109.22 & 98.82 \\
mean electricity price (€/MWh) & 85.07 & 175.72 & 104.68 & 86.12 & 52.91 & 204.42 & 83.65 & 75.26 & 102.14 & 96.54 & 56.81 & 211.47 \\
mean hydrogen price (€/MWh) & 74.75 & 193.73 & 72.79 & 75.08 & 53.36 & 137.41 & 72.64 & 68.86 & 80.88 & 82.42 & 56.90 & 149.71 \\
STD electricity price (€/MWh) & 266.05 & 496.81 & 310.38 & 248.72 & 134.62 & 499.34 & 159.22 & 138.26 & 276.88 & 247.37 & 104.54 & 539.77 \\
STD hydrogen price (€/MWh) & 144.29 & 381.63 & 90.53 & 99.94 & 54.64 & 232.64 & 31.27 & 52.44 & 144.62 & 126.70 & 42.51 & 310.30 \\
mean hydrogen MSV (€/MWh) & 74.75 & 193.73 & 72.79 & 75.08 & 53.36 & 137.41 & 72.64 & 68.86 & 80.88 & 82.42 & 56.90 & 149.71 \\
mean battery MSV (€/MWh) & 85.95 & 178.21 & 105.30 & 87.59 & 53.78 & 207.89 & 84.62 & 76.21 & 103.17 & 97.96 & 57.73 & 214.18 \\
STD hydrogen MSV (€/MWh) & 144.29 & 381.63 & 90.53 & 99.94 & 54.64 & 232.64 & 31.27 & 52.44 & 144.62 & 126.70 & 42.51 & 310.30 \\
STD battery MSV (€/MWh) & 262.33 & 492.92 & 305.23 & 247.56 & 132.94 & 496.70 & 155.85 & 135.76 & 273.00 & 245.24 & 102.95 & 534.58 \\
\bottomrule
\end{tabular}

      \caption{Metrics for 35-year LT plus 35-year ST optimisations for Spain. \tablelegend}
      \label{tab:metrics-35a-35a-es}
    \end{footnotesize}
  \end{table}
\end{landscape}

\begin{landscape}
  \begin{table}
    \begin{footnotesize}
      \begin{tabular}{lrrrrrrrrrrr}
\toprule
 & \shortstack[r]{perfectly\\inelastic,\\LT,\\70a} & \shortstack[r]{VOLL,\\LT,\\70a} & \shortstack[r]{PWL-elastic\\default,\\LT,\\70a} & \shortstack[r]{PWL-elastic\\higher,\\LT,\\70a} & \shortstack[r]{PWL-elastic\\lower,\\LT,\\70a} & \shortstack[r]{VOLL,\\ST,\\70a,\\PF} & \shortstack[r]{VOLL,\\ST,\\70a,\\PF,\\C+5\%} & \shortstack[r]{VOLL,\\ST,\\70a,\\PF,\\C-5\%} & \shortstack[r]{PWL-elastic,\\ST,\\70a,\\PF} & \shortstack[r]{PWL-elastic,\\ST,\\70a,\\PF,\\C+5\%} & \shortstack[r]{PWL-elastic,\\ST,\\70a,\\PF,\\C-5\%} \\
\midrule
system costs (bn€/period) & 5.66 & 5.47 & 5.30 & 5.00 & 5.52 & 5.47 & 5.74 & 5.20 & 5.30 & 5.56 & 5.04 \\
utility (bn€/period) & -- & 122.66 & 468.55 & 234.06 & 937.43 & 122.66 & 122.70 & 121.52 & 468.55 & 468.72 & 468.16 \\
welfare (bn€/period) & -- & 117.19 & 463.25 & 229.06 & 931.91 & 117.19 & 116.96 & 116.33 & 463.25 & 463.16 & 463.13 \\
average system costs (€/MWh) & 92.19 & 89.16 & 81.21 & 78.58 & 83.27 & 89.16 & 93.57 & 85.50 & 81.21 & 83.40 & 79.40 \\
average load served (MW) & 100.00 & 99.95 & 106.34 & 103.76 & 108.01 & 99.95 & 99.98 & 99.02 & 106.34 & 108.72 & 103.33 \\
peak load shedding (MW) & 0.00 & 29.62 & 37.56 & 45.45 & 31.15 & 26.41 & 21.56 & 44.13 & 37.56 & 33.87 & 41.26 \\
primary energy (TWh/period) & 74.87 & 74.16 & 78.42 & 75.61 & 80.26 & 74.16 & 72.67 & 74.13 & 78.42 & 80.52 & 75.64 \\
wind share (\%) & 67.43 & 66.27 & 67.01 & 66.94 & 66.87 & 66.27 & 63.88 & 67.92 & 67.01 & 66.29 & 67.48 \\
solar share (\%) & 32.57 & 33.73 & 32.99 & 33.06 & 33.13 & 33.73 & 36.12 & 32.08 & 32.99 & 33.71 & 32.52 \\
wind market value (€/MWh) & 46.82 & 47.18 & 44.11 & 43.86 & 44.32 & 53.15 & 2.32 & 477.35 & 44.11 & 13.55 & 94.93 \\
solar market value (€/MWh) & 44.12 & 44.15 & 44.16 & 44.15 & 44.16 & 50.49 & 1.89 & 500.91 & 44.16 & 12.73 & 98.07 \\
wind capacity factor (\%) & 29.11 & 29.11 & 29.11 & 29.11 & 29.11 & 29.11 & 29.11 & 29.11 & 29.11 & 29.11 & 29.11 \\
solar capacity factor (\%) & 10.63 & 10.63 & 10.63 & 10.63 & 10.63 & 10.63 & 10.63 & 10.63 & 10.63 & 10.63 & 10.63 \\
hydrogen consumed (TWh/period) & 11.78 & 11.13 & 11.29 & 10.18 & 12.02 & 11.13 & 9.78 & 11.55 & 11.29 & 11.88 & 10.47 \\
curtailment (\%) & 18.85 & 19.05 & 15.17 & 14.81 & 15.43 & 19.05 & 24.46 & 14.83 & 15.17 & 17.04 & 13.86 \\
wind capacity (MW) & 379.67 & 372.40 & 372.19 & 356.49 & 381.94 & 372.40 & 391.02 & 353.78 & 372.19 & 390.80 & 353.58 \\
solar capacity (MW) & 374.61 & 384.54 & 397.80 & 384.23 & 408.89 & 384.54 & 403.77 & 365.31 & 397.80 & 417.69 & 377.91 \\
electrolyser capacity (MW) & 72.16 & 68.00 & 61.97 & 56.08 & 65.65 & 68.00 & 71.40 & 64.60 & 61.97 & 65.07 & 58.87 \\
fuel cell capacity (MW) & 77.67 & 61.67 & 49.91 & 42.98 & 55.63 & 61.67 & 64.76 & 58.59 & 49.91 & 52.40 & 47.41 \\
battery inverter capacity (MW) & 62.01 & 67.63 & 70.86 & 69.19 & 72.53 & 67.63 & 71.01 & 64.25 & 70.86 & 74.40 & 67.31 \\
battery storage capacity (GWh) & 0.48 & 0.53 & 0.56 & 0.55 & 0.57 & 0.53 & 0.56 & 0.51 & 0.56 & 0.59 & 0.53 \\
hydrogen storage capacity (GWh) & 164.12 & 155.33 & 95.58 & 80.36 & 113.57 & 155.33 & 163.10 & 147.56 & 95.58 & 100.36 & 90.80 \\
mean electricity price (€/MWh) & 92.19 & 90.14 & 87.84 & 87.12 & 88.60 & 100.21 & 9.61 & 820.48 & 87.84 & 31.83 & 179.66 \\
mean hydrogen price (€/MWh) & 77.35 & 75.63 & 73.95 & 73.38 & 74.51 & 86.82 & 0.32 & 879.28 & 73.95 & 18.14 & 167.44 \\
STD electricity price (€/MWh) & 1315.41 & 338.78 & 165.33 & 142.35 & 194.57 & 370.80 & 134.94 & 841.26 & 165.33 & 113.37 & 246.63 \\
STD hydrogen price (€/MWh) & 219.81 & 215.94 & 42.29 & 37.18 & 58.75 & 246.69 & 0.00 & 308.17 & 42.29 & 26.39 & 72.39 \\
mean hydrogen MSV (€/MWh) & 77.35 & 75.63 & 73.95 & 73.38 & 74.51 & 86.82 & 0.32 & 879.28 & 73.95 & 18.14 & 167.44 \\
mean battery MSV (€/MWh) & 92.51 & 90.82 & 88.77 & 88.11 & 89.46 & 101.08 & 9.38 & 832.24 & 88.77 & 31.86 & 182.14 \\
STD hydrogen MSV (€/MWh) & 219.81 & 215.94 & 42.29 & 37.18 & 58.75 & 246.69 & 0.00 & 308.17 & 42.29 & 26.39 & 72.39 \\
STD battery MSV (€/MWh) & 1283.92 & 334.27 & 161.34 & 138.89 & 190.07 & 366.33 & 131.47 & 815.92 & 161.34 & 110.79 & 240.44 \\
\bottomrule
\end{tabular}

      \caption{Metrics for 70-year optimisations for the United Kingdom.  \tablelegend}
      \label{tab:metrics-70a-uk}
    \end{footnotesize}
  \end{table}
\end{landscape}

\begin{landscape}
  \begin{table}
    \begin{footnotesize}
      \begin{tabular}{lrrrrrrrrrrrr}
\toprule
 & \shortstack[r]{VOLL,\\LT,\\35a} & \shortstack[r]{VOLL,\\ST,\\35a-35a,\\PF} & \shortstack[r]{VOLL,\\ST,\\35a-35a,\\myopic\\48/24} & \shortstack[r]{VOLL,\\ST,\\35a-35a,\\myopic\\96/48} & \shortstack[r]{VOLL,\\ST,\\35a-35a,\\myopic\\96/48,\\C+5\%} & \shortstack[r]{VOLL,\\ST,\\35a-35a,\\myopic\\96/48,\\C-5\%} & \shortstack[r]{PWL-elastic\\default,\\LT,\\35a} & \shortstack[r]{PWL-elastic\\default,\\ST,\\35a-35a,\\PF} & \shortstack[r]{PWL-elastic\\default,\\ST,\\35a-35a,\\myopic\\48/24} & \shortstack[r]{PWL-elastic\\default,\\ST,\\35a-35a,\\myopic\\96/48} & \shortstack[r]{PWL-elastic\\default,\\ST,\\35a-35a,\\myopic\\96/48,\\C+5\%} & \shortstack[r]{PWL-elastic\\default,\\ST,\\35a-35a,\\myopic\\96/48,\\C-5\%} \\
\midrule
system costs (bn€/period) & 2.73 & 2.73 & 2.73 & 2.73 & 2.86 & 2.59 & 2.66 & 2.66 & 2.66 & 2.66 & 2.79 & 2.52 \\
utility (bn€/period) & 61.33 & 61.27 & 61.25 & 61.27 & 61.35 & 60.64 & 233.03 & 233.07 & 232.89 & 232.92 & 233.10 & 232.27 \\
welfare (bn€/period) & 58.60 & 58.54 & 58.53 & 58.54 & 58.48 & 58.05 & 230.37 & 230.41 & 230.24 & 230.26 & 230.31 & 229.74 \\
average system costs (€/MWh) & 88.92 & 89.01 & 89.03 & 89.01 & 93.32 & 85.45 & 81.47 & 81.19 & 81.71 & 81.68 & 85.05 & 79.08 \\
average load served (MW) & 99.95 & 99.84 & 99.81 & 99.84 & 99.97 & 98.81 & 106.35 & 106.70 & 106.02 & 106.06 & 106.93 & 104.07 \\
peak load shedding (MW) & 30.10 & 48.61 & 85.67 & 84.69 & 68.12 & 99.11 & 33.84 & 38.18 & 91.05 & 81.65 & 51.10 & 87.39 \\
primary energy (TWh/period) & 37.16 & 36.99 & 37.26 & 37.27 & 36.65 & 36.98 & 39.09 & 39.31 & 38.79 & 38.83 & 38.69 & 37.93 \\
wind share (\%) & 65.97 & 66.25 & 66.55 & 66.53 & 64.42 & 67.90 & 66.76 & 67.37 & 66.95 & 66.97 & 65.27 & 67.85 \\
solar share (\%) & 34.03 & 33.75 & 33.45 & 33.47 & 35.58 & 32.10 & 33.24 & 32.63 & 33.05 & 33.03 & 34.73 & 32.15 \\
wind market value (€/MWh) & 46.87 & 40.88 & 47.45 & 46.25 & 35.15 & 92.07 & 44.88 & 39.32 & 51.69 & 51.86 & 36.95 & 101.88 \\
solar market value (€/MWh) & 44.01 & 37.10 & 45.36 & 44.26 & 32.49 & 82.14 & 44.01 & 40.55 & 52.01 & 51.19 & 36.29 & 99.87 \\
wind capacity factor (\%) & 28.78 & 29.44 & 29.44 & 29.44 & 29.44 & 29.44 & 28.78 & 29.44 & 29.44 & 29.44 & 29.44 & 29.44 \\
solar capacity factor (\%) & 10.70 & 10.56 & 10.56 & 10.56 & 10.56 & 10.56 & 10.70 & 10.56 & 10.56 & 10.56 & 10.56 & 10.56 \\
hydrogen consumed (TWh/period) & 5.63 & 5.51 & 5.50 & 5.51 & 4.90 & 5.67 & 5.54 & 5.63 & 5.25 & 5.27 & 4.89 & 5.10 \\
curtailment (\%) & 17.84 & 19.27 & 18.68 & 18.65 & 23.81 & 15.05 & 15.39 & 15.99 & 17.10 & 17.02 & 21.26 & 14.68 \\
wind capacity (MW) & 368.54 & 368.54 & 368.54 & 368.54 & 386.96 & 350.11 & 375.61 & 375.61 & 375.61 & 375.61 & 394.39 & 356.83 \\
solar capacity (MW) & 386.40 & 386.40 & 386.40 & 386.40 & 405.72 & 367.08 & 396.95 & 396.95 & 396.95 & 396.95 & 416.80 & 377.11 \\
electrolyser capacity (MW) & 66.79 & 66.79 & 66.79 & 66.79 & 70.13 & 63.45 & 60.94 & 60.94 & 60.94 & 60.94 & 63.98 & 57.89 \\
fuel cell capacity (MW) & 61.38 & 61.38 & 61.38 & 61.38 & 64.45 & 58.31 & 49.10 & 49.10 & 49.10 & 49.10 & 51.56 & 46.65 \\
battery inverter capacity (MW) & 68.15 & 68.15 & 68.15 & 68.15 & 71.56 & 64.74 & 71.53 & 71.53 & 71.53 & 71.53 & 75.11 & 67.96 \\
battery storage capacity (GWh) & 0.53 & 0.53 & 0.53 & 0.53 & 0.56 & 0.51 & 0.57 & 0.57 & 0.57 & 0.57 & 0.60 & 0.54 \\
hydrogen storage capacity (GWh) & 157.72 & 157.72 & 157.72 & 157.72 & 165.61 & 149.84 & 87.49 & 87.49 & 87.49 & 87.49 & 91.86 & 83.12 \\
mean electricity price (€/MWh) & 89.92 & 78.82 & 98.93 & 91.86 & 63.82 & 181.91 & 88.35 & 82.02 & 110.15 & 107.83 & 70.17 & 223.67 \\
mean hydrogen price (€/MWh) & 75.57 & 61.81 & 72.48 & 72.72 & 55.75 & 123.70 & 74.45 & 65.41 & 88.52 & 86.12 & 60.45 & 163.10 \\
STD electricity price (€/MWh) & 267.78 & 363.86 & 271.10 & 250.11 & 162.37 & 458.81 & 167.48 & 175.47 & 339.43 & 319.51 & 150.22 & 623.97 \\
STD hydrogen price (€/MWh) & 135.66 & 240.18 & 64.77 & 75.24 & 41.62 & 205.31 & 47.23 & 50.88 & 235.75 & 180.71 & 50.86 & 394.59 \\
mean hydrogen MSV (€/MWh) & 75.57 & 61.81 & 72.48 & 72.72 & 55.75 & 123.70 & 74.45 & 65.41 & 88.52 & 86.12 & 60.45 & 163.10 \\
mean battery MSV (€/MWh) & 90.61 & 78.90 & 98.99 & 92.13 & 64.26 & 181.00 & 89.28 & 82.75 & 110.87 & 108.50 & 70.97 & 224.46 \\
STD hydrogen MSV (€/MWh) & 135.66 & 240.18 & 64.77 & 75.24 & 41.62 & 205.31 & 47.23 & 50.88 & 235.75 & 180.71 & 50.86 & 394.59 \\
STD battery MSV (€/MWh) & 262.56 & 357.87 & 263.76 & 243.55 & 158.53 & 447.10 & 163.49 & 171.30 & 333.21 & 313.40 & 147.81 & 613.99 \\
\bottomrule
\end{tabular}

      \caption{Metrics for 35-year LT plus 35-year ST optimisations for the United Kingdom. \tablelegend}
      \label{tab:metrics-35a-35a-uk}
    \end{footnotesize}
  \end{table}
\end{landscape}

\begin{landscape}
  \begin{table}
    \begin{footnotesize}
      \begin{tabular}{lrrrrrrrrrrrr}
\toprule
 & \shortstack[r]{VOLL,\\0 MW} & \shortstack[r]{VOLL,\\10 MW} & \shortstack[r]{VOLL,\\30 MW} & \shortstack[r]{VOLL,\\50 MW} & \shortstack[r]{VOLL,\\70 MW} & \shortstack[r]{VOLL,\\90 MW} & \shortstack[r]{linear-elastic,\\0 MW} & \shortstack[r]{linear-elastic,\\10 MW} & \shortstack[r]{linear-elastic,\\30 MW} & \shortstack[r]{linear-elastic,\\50 MW} & \shortstack[r]{linear-elastic,\\70 MW} & \shortstack[r]{linear-elastic,\\90 MW} \\
\midrule
system costs (bn€/period) & 1.71 & 1.71 & 1.71 & 1.71 & 1.75 & 1.82 & 1.47 & 1.47 & 1.47 & 1.55 & 1.62 & 1.68 \\
utility (bn€/period) & 35.05 & 35.05 & 35.05 & 35.05 & 35.06 & 35.06 & 17.42 & 17.42 & 17.42 & 17.47 & 17.47 & 17.47 \\
welfare (bn€/period) & 33.34 & 33.34 & 33.34 & 33.34 & 33.31 & 33.25 & 15.96 & 15.96 & 15.96 & 15.92 & 15.85 & 15.79 \\
average system costs (€/MWh) & 97.54 & 97.54 & 97.54 & 97.54 & 99.76 & 103.61 & 88.00 & 88.00 & 88.00 & 92.24 & 96.29 & 100.32 \\
average load served (MW) & 99.97 & 99.97 & 99.97 & 99.97 & 100.00 & 100.00 & 95.19 & 95.19 & 95.19 & 95.66 & 95.71 & 95.71 \\
peak load shedding (MW) & 17.57 & 17.26 & 18.21 & 20.55 & 0.00 & 0.00 & 37.65 & 37.65 & 37.65 & 19.50 & 14.96 & 14.96 \\
primary energy (TWh/period) & 20.84 & 20.84 & 20.84 & 20.84 & 21.07 & 21.07 & 19.37 & 19.37 & 19.37 & 19.73 & 19.77 & 19.77 \\
wind share (\%) & 49.55 & 49.55 & 49.55 & 49.55 & 51.02 & 51.02 & 50.37 & 50.37 & 50.37 & 50.98 & 51.04 & 51.04 \\
solar share (\%) & 50.45 & 50.45 & 50.45 & 50.45 & 48.98 & 48.98 & 49.63 & 49.63 & 49.63 & 49.02 & 48.96 & 48.96 \\
wind market value (€/MWh) & 63.14 & 63.14 & 63.14 & 63.14 & 60.83 & 60.83 & 61.86 & 61.86 & 61.86 & 60.33 & 60.23 & 60.22 \\
solar market value (€/MWh) & 40.71 & 40.71 & 40.71 & 40.71 & 40.56 & 40.56 & 40.70 & 40.70 & 40.70 & 40.49 & 40.47 & 40.47 \\
wind capacity factor (\%) & 20.74 & 20.74 & 20.74 & 20.74 & 20.74 & 20.74 & 20.74 & 20.74 & 20.74 & 20.74 & 20.74 & 20.74 \\
solar capacity factor (\%) & 12.04 & 12.04 & 12.04 & 12.04 & 12.04 & 12.04 & 12.04 & 12.04 & 12.04 & 12.04 & 12.04 & 12.04 \\
hydrogen consumed (TWh/period) & 2.74 & 2.74 & 2.74 & 2.74 & 2.96 & 2.96 & 2.18 & 2.18 & 2.18 & 2.44 & 2.46 & 2.46 \\
curtailment (\%) & 13.81 & 13.81 & 13.81 & 13.81 & 12.12 & 12.12 & 12.96 & 12.96 & 12.96 & 11.65 & 11.55 & 11.55 \\
wind capacity (MW) & 364.89 & 364.89 & 364.89 & 364.89 & 366.13 & 366.12 & 337.88 & 337.88 & 337.88 & 339.72 & 340.09 & 340.09 \\
solar capacity (MW) & 516.67 & 516.67 & 516.67 & 516.67 & 505.41 & 505.40 & 472.46 & 472.46 & 472.46 & 472.88 & 472.84 & 472.84 \\
electrolyser capacity (MW) & 57.53 & 57.53 & 57.53 & 57.53 & 60.95 & 60.94 & 44.43 & 44.43 & 44.43 & 51.69 & 52.34 & 52.35 \\
fuel cell capacity (MW) & 58.10 & 58.10 & 58.10 & 58.10 & 70.00 & 90.00 & 32.00 & 32.00 & 32.00 & 50.00 & 70.00 & 90.00 \\
battery inverter capacity (MW) & 104.84 & 104.84 & 104.84 & 104.84 & 100.11 & 100.11 & 95.95 & 95.95 & 95.95 & 94.40 & 94.19 & 94.19 \\
battery storage capacity (GWh) & 0.79 & 0.79 & 0.79 & 0.79 & 0.73 & 0.73 & 0.72 & 0.72 & 0.72 & 0.69 & 0.69 & 0.69 \\
hydrogen storage capacity (GWh) & 131.02 & 131.02 & 131.02 & 131.02 & 195.40 & 195.45 & 80.84 & 80.84 & 80.84 & 90.99 & 90.65 & 90.66 \\
mean electricity price (€/MWh) & 98.21 & 98.21 & 98.21 & 98.21 & 86.32 & 86.28 & 96.12 & 96.12 & 96.12 & 86.85 & 85.76 & 85.76 \\
mean hydrogen price (€/MWh) & 84.65 & 84.65 & 84.65 & 84.65 & 89.46 & 89.48 & 80.63 & 80.63 & 80.63 & 88.31 & 88.89 & 88.89 \\
STD electricity price (€/MWh) & 238.02 & 238.02 & 238.02 & 238.02 & 151.01 & 150.86 & 124.29 & 124.29 & 124.29 & 85.22 & 82.35 & 82.35 \\
STD hydrogen price (€/MWh) & 98.77 & 98.77 & 98.77 & 98.77 & 103.89 & 103.89 & 28.14 & 28.14 & 28.14 & 28.46 & 27.69 & 27.70 \\
mean hydrogen MSV (€/MWh) & 84.65 & 84.65 & 84.65 & 84.65 & 89.46 & 89.48 & 80.63 & 80.63 & 80.63 & 88.31 & 88.89 & 88.89 \\
mean battery MSV (€/MWh) & 98.86 & 98.86 & 98.86 & 98.86 & 87.08 & 87.01 & 97.09 & 97.09 & 97.09 & 87.87 & 86.58 & 86.57 \\
STD hydrogen MSV (€/MWh) & 98.77 & 98.77 & 98.77 & 98.77 & 103.89 & 103.89 & 28.14 & 28.14 & 28.14 & 28.46 & 27.69 & 27.70 \\
STD battery MSV (€/MWh) & 232.31 & 232.31 & 232.31 & 232.31 & 147.82 & 147.62 & 120.70 & 120.70 & 120.70 & 82.29 & 79.15 & 79.14 \\
\bottomrule
\end{tabular}

      \caption{Metrics for 20-year LT optimisations (2001-2020) for Germany with forced reserve capacity (hydrogen turbine). VOLL with 2000~\euro/MWh. Linear-elastic demand with $a=2000$ and $b=10$.}
      \label{tab:metrics-reserve}
    \end{footnotesize}
  \end{table}
\end{landscape}

\end{document}